\pdfoutput=1
\documentclass[11pt,twoside,a4paper,cmspaper,final,collab]{cms-tdr}
\def\svnVersion{3809499}\def\svnDate{2022/05/14}\def\cmsCernNoTag{CERN-EP-2021-214}\def\cmsCernDate{\today}\def\cmsMessage{Published in the Journal of High Energy Physics as \href{http://dx.doi.org/10.1007/JHEP05(2022)014}{\doi{10.1007/JHEP05(2022)014}.}}
\begin{document}\cmsNoteHeader{SUS-20-004}

\newlength\cmsTabSkip\setlength{\cmsTabSkip}{1ex}

\newcommand{\met}{\ptmiss}
\newcommand{\mettr}{\ensuremath{p_{\mathrm{T,\,trig}}^\text{miss}}\xspace}
\newcommand{\kapa}{\ensuremath{\kappa}\xspace}
\newcommand{\kapap}{\ensuremath{\kappa^\prime}\xspace}
\newcommand{\dR}{\ensuremath{\Delta R}\xspace}
\newcommand{\imini}{\ensuremath{I}\xspace}
\newcommand{\drmax}{\ensuremath{\Delta R_{\text{max}}}\xspace}
\newcommand{\amjj}{\ensuremath{\langle m_{\botq\botq}\rangle}\xspace}
\newcommand{\dmjj}{\ensuremath{\Delta m_{\botq\botq}}\xspace}
\newcommand{\njetsisr}{\ensuremath{N_{\text{jet}}^{\text{ISR}}}\xspace}
\newcommand{\nb}{\ensuremath{N_\PQb}\xspace}
\newcommand{\nbl}{\ensuremath{N_{\PQb,\mathrm{L}}}\xspace}
\newcommand{\nbm}{\ensuremath{N_{\PQb,\mathrm{M}}}\xspace}
\newcommand{\nbt}{\ensuremath{N_{\PQb,\mathrm{T}}}\xspace}
\newcommand{\dphi}{\ensuremath{\Delta\phi}\xspace}
\newcommand{\mt}{\mT}
\newcommand{\nH}{\ensuremath{N_{\mathrm{\PH}}}\xspace}
\newcommand{\mjet}{\ensuremath{m_{\mathrm{J}}}\xspace}
\newcommand{\Dbb}{\ensuremath{D_{\operatorname{\PQb\PQb}}}\xspace}
\newcommand{\nsr}{\ensuremath{N_{\text{SR}}}\xspace}
\newcommand{\nsrp}{\ensuremath{N_{\text{SR}}^{\text{pred}}}\xspace}
\newcommand{\nsrpi}{\ensuremath{N_{\text{SR},i}^{\text{pred}}}\xspace}
\newcommand{\nsrf}{\ensuremath{N_{\text{SR}}^{\text{fit}}}\xspace}
\newcommand{\nsro}{\ensuremath{N_{\text{SR}}^{\text{obs}}}\xspace}
\newcommand{\nsb}{\ensuremath{N_{\text{SB}}}\xspace}
\newcommand{\ncsr}{\ensuremath{N_{\text{CSR}}}\xspace}
\newcommand{\ncsb}{\ensuremath{N_{\text{CSB}}}\xspace}
\newcommand{\crb}{\ensuremath{\text{0H+b}}\xspace}
\newcommand{\tbkg}{\ensuremath{N_{\text{SR, tot}}^{\text{pred}}}\xspace}
\newcommand{\fbkg}{\ensuremath{f_{\text{pTmiss}}}\xspace}
\newcommand{\fbkgi}{\ensuremath{f_{\text{pTmiss},i}}\xspace}
\newcommand{\rscale}{\ensuremath{\mu_{\mathrm{R}}}\xspace}
\newcommand{\fscale}{\ensuremath{\mu_{\mathrm{F}}}\xspace}

\newcommand{\gluino}{\PSg}
\newcommand{\chizone}{\PSGczDo}
\newcommand{\chiztwo}{\PSGczDt}
\newcommand{\chizthree}{\PSGczDT}
\newcommand{\chipmone}{\PSGcpmDo}
\newcommand{\chimpone}{\ensuremath{\PSGc_1^\mp}\xspace}
\newcommand{\LSP}{\PSGczDo}
\newcommand{\lsp}{\LSP}
\newcommand{\botq}{\PQb}

\newcommand{\pp}{\ensuremath{{\Pp\Pp}}\xspace}
\newcommand{\znn}{{\ensuremath{\PZ\to\PGn\PAGn}}\xspace}
\newcommand{\zll}{{\ensuremath{\PZ\to\ell^+\ell^-}}\xspace}
\newcommand{\zjets}{{{\PZ}\!+jets}\xspace}
\newcommand{\wjets}{{{\PW}\!+jets}\xspace}
\newcommand{\vjets}{{\text{V}\!+jets}\xspace}
\newcommand{\Hbb}{\ensuremath{\PH\to\bbbar}\xspace}

\cmsNoteHeader{SUS-20-004}
\title{Search for higgsinos decaying to two Higgs bosons and missing transverse momentum in proton-proton collisions at \texorpdfstring{$\sqrt{s}=13\TeV$}{sqrt(s) = 13 TeV}}

\date{\today}

\abstract{
Results are presented from a search for physics beyond the standard model in proton-proton collisions at $\sqrt{s}=13\TeV$ in channels with two Higgs bosons, each decaying via the process $\PH\to\bbbar$, and large missing transverse momentum. The search uses a data sample corresponding to an integrated luminosity of 137\fbinv collected by the CMS experiment at the CERN LHC. The search is motivated by models of supersymmetry that predict the production of neutralinos, the neutral partners of the electroweak gauge and Higgs bosons. The observed event yields in the signal regions are found to be consistent with the standard model background expectations. The results are interpreted using simplified models of supersymmetry. For the electroweak production of nearly mass-degenerate higgsinos, each of whose decay chains yields a neutralino ({\chizone}) that in turn decays to a massless goldstino and a Higgs boson, \chizone masses in the range 175 to 1025\GeV are excluded at 95\% confidence level. For the strong production of gluino pairs decaying via a slightly lighter \chiztwo to \PH and a light \chizone, gluino masses below 2330\GeV are excluded.
}

\hypersetup{%
pdfauthor={CMS Collaboration},%
pdftitle={Search for higgsinos decaying to two Higgs bosons and missing transverse momentum in proton-proton collisions at sqrt(s) = 13 TeV},%
pdfsubject={CMS},%
pdfkeywords={CMS, SUSY, Higgs, LHC}}

\maketitle

\section{Introduction}
\label{sec:introduction}
The discovery of the Higgs boson (\PH) at the CERN Large Hadron Collider (LHC) \cite{Aad:2012tfa,Chatrchyan:2012ufa,Chatrchyan:2013lba,Aad:2015zhl} provides a new tool for probing physics beyond the standard model (SM). Higgs boson production is expected to be prominent in the decays of a variety of new, heavy particles, such as those arising in theories based on supersymmetry (SUSY)~\cite{Ramond:1971gb,Golfand:1971iw,Neveu:1971rx,Volkov:1972jx,Wess:1973kz,Wess:1974tw,Fayet:1974pd,Fayet:1976cr,Nilles:1983ge,Martin:1997ns}. For example, Higgs bosons would be produced in the decays of
higgsinos, their supersymmetric partners.
A key motivation to search for higgsinos is that in certain models they are generically expected to be among the lightest supersymmetric particles, with masses kinematically accessible at the LHC.
This is particularly the case in so-called natural SUSY models~\cite{Dimopoulos:1995mi,Barbieri:2009ev,Papucci:2011wy}, which are designed to stabilize the electroweak mass scale.
Searches for higgsinos are nevertheless challenging, not only because of the wide range of possible physics scenarios, but also because their electroweak coupling to other particles leads to small production cross sections. Higgsino searches are therefore expected to be a long-term part of the LHC physics program. 

We search for processes in proton-proton (\pp) collisions leading to the production of two Higgs bosons, together with missing momentum transverse to the beam direction (\ptvecmiss) and possible additional jets. The Higgs bosons are reconstructed in the decay $\PH\to\bbbar$, which has a branching fraction of 58\%~\cite{deFlorian:2016spz}. To provide sensitivity to the wide range of kinematic configurations that can arise, the analysis searches both for events containing pairs of $\PH\to\bbbar$ decays in which the \botq-tagged jets are separately resolved, and for events in which the \botq jets from a given \PH boson decay are merged together into a single, wider jet. We refer to the unmerged and merged kinematic configurations as the resolved signature and the boosted signature, respectively, because the merging of \botq jets typically occurs when the parent Higgs boson has a high momentum. 

The search uses an event sample of \pp collision data at $\sqrt{s}=13\TeV$, corresponding to an integrated luminosity of 137\fbinv, collected in 2016--2018 by the CMS experiment at the
LHC. Searches for this and related decay scenarios have been performed by
ATLAS~\cite{Aad:2012jva,Aad:2015jqa,Aaboud:2017dmy,Aaboud:2017hdf,Aaboud:2017bac,Aaboud:2017vwy,Aaboud:2017leg,Aaboud:2018doq,Aaboud:2018jiw,Aaboud:2018sua,Aaboud:2018htj,Aaboud:2018ngk,Aad:2019vnb,Aad:2019vvf,Aad:2019qnd,Aad:2019vvi,Aad:2020qnn,ATLAS:2021yyr,ATLAS:2021yqv} and CMS~\cite{Khachatryan:2014qwa,Chatrchyan:2013mya,Khachatryan:2014mma,Sirunyan:2017uyt,Sirunyan:2017zss,Sirunyan:2017yse,Sirunyan:2017eie,Sirunyan:2017obz,Sirunyan:2018ubx,Sirunyan:2017qaj,Sirunyan:2017qaj,Sirunyan:2017mrs,Sirunyan:2017hvp,Sirunyan:2017nyt,Sirunyan:2017bsh,Sirunyan:2018iwl,Sirunyan:2018lul,Sirunyan:2018psa,Sirunyan:2019hzr,Sirunyan:2019mbp,Sirunyan:2019sem,Sirunyan:2019iwo,Sirunyan:2019xwh,Sirunyan:2020ztc,Sirunyan:2020zzv,Sirunyan:2020eab,CMS:2021cox,CMS:2021few}
using 7, 8, and 13\TeV data. The data for the results reported here include the smaller samples of Ref.~\cite{Sirunyan:2017obz} for the resolved signature and Ref.~\cite{Sirunyan:2017bsh} for the boosted signature.
Tabulated results are provided in the HEPData record for this analysis~\cite{hepdata}.  

The simplest SUSY framework is the minimal supersymmetric standard model (MSSM), which includes an additional doublet of complex scalar Higgs fields~\cite{Fayet:1976cr,Nilles:1983ge,Martin:1997ns} beyond the doublet in the SM. The SUSY partners of the gauge and Higgs bosons, referred to as gauginos and higgsinos, respectively, are all spin $J=1/2$ fermions.  
In the MSSM, there are four higgsinos, two of which are charged ($\mathrm{\widetilde{H}^\pm}$) and two of which are neutral ($\mathrm{\widetilde{h}^0}$, $\mathrm{\widetilde{H}^0}$).
However, the phenomenology of the electroweak sector of the MSSM is complicated by the fact that the partners of the electroweak gauge and
Higgs bosons can mix. The resulting physical states are then the electrically neutral mass eigenstates, $\PSGc_{\text{1-4}}^0$ (where the particles are ordered in mass, with \chizone corresponding to the lightest), referred to as neutralinos, and the electrically charged mass eigenstates, $\PSGc_{1,2}^\pm$~\cite{PDG2020}, referred to as charginos. 
In models that conserve $R$-parity~\cite{Fayet:1974pd,Farrar:1978}, the lightest supersymmetric particle (LSP) is stable, and because of its weak interactions would escape experimental detection.
The \chizone would be a candidate for the LSP, as would the goldstino (\sGra), a Nambu--Goldstone particle associated with the spontaneous breaking of global supersymmetry.  

\begin{figure}[tb!]
\centering
\includegraphics[width=0.32\textwidth]{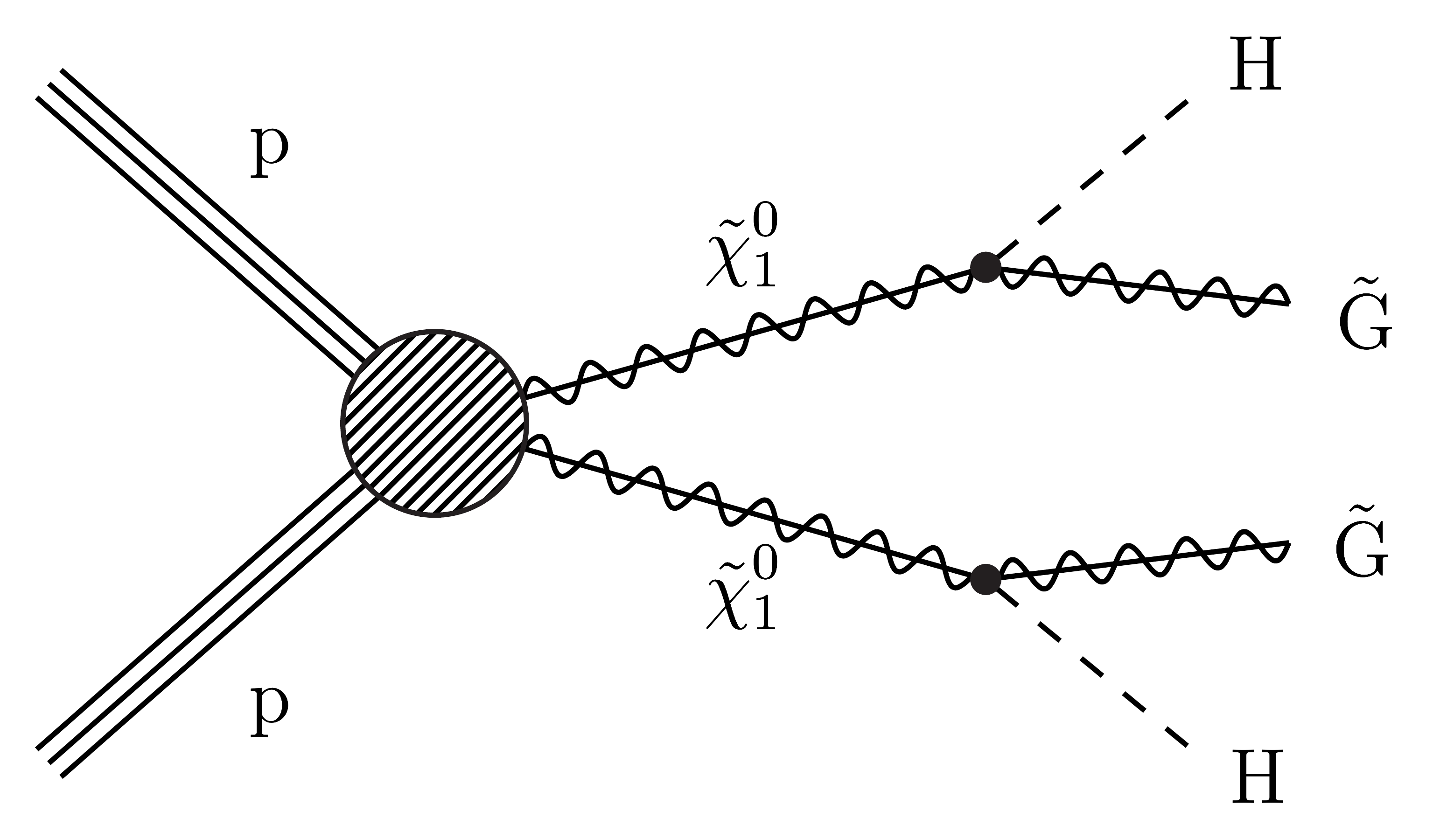}
\includegraphics[width=0.32\textwidth]{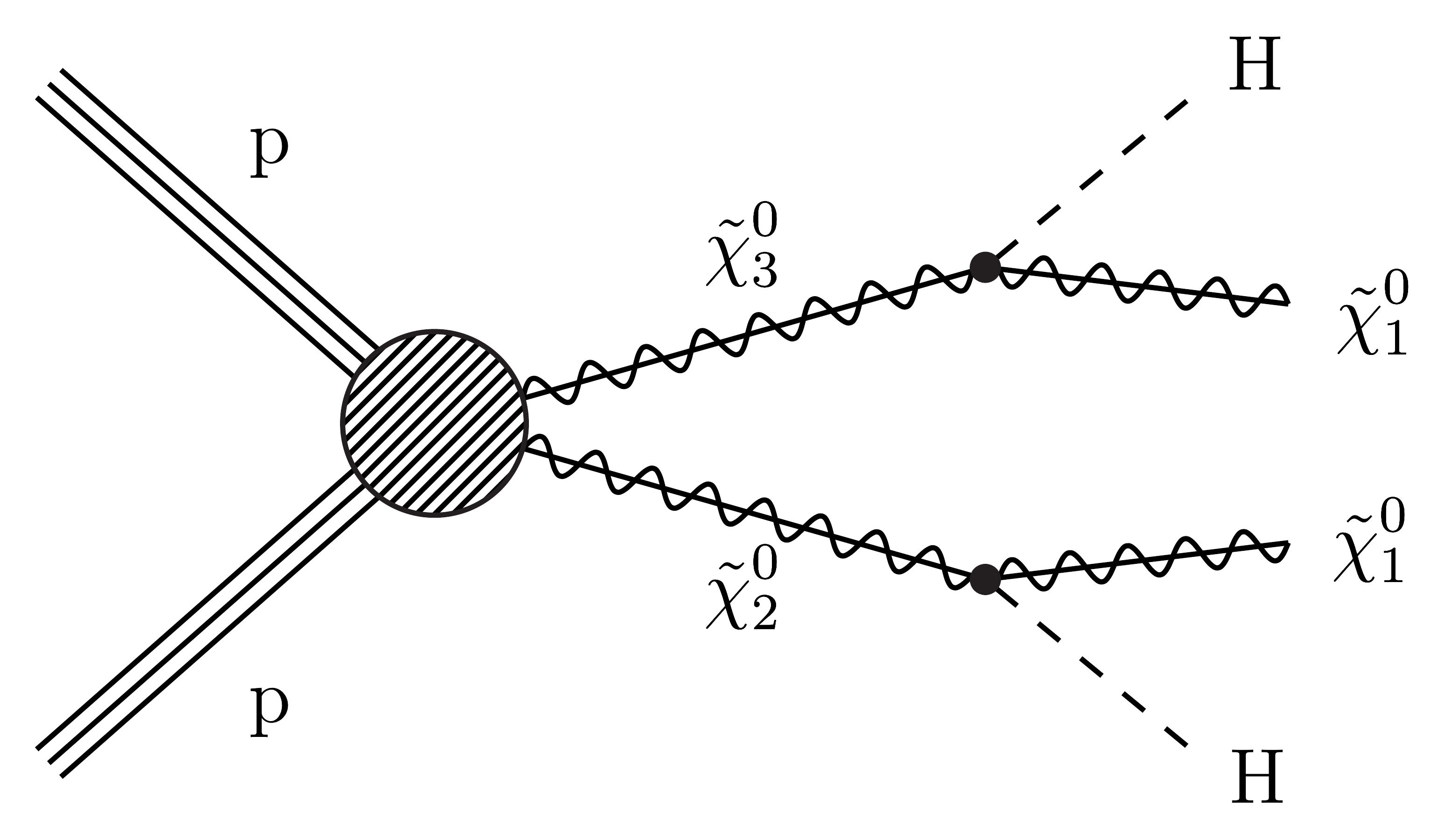}
\includegraphics[width=0.32\textwidth]{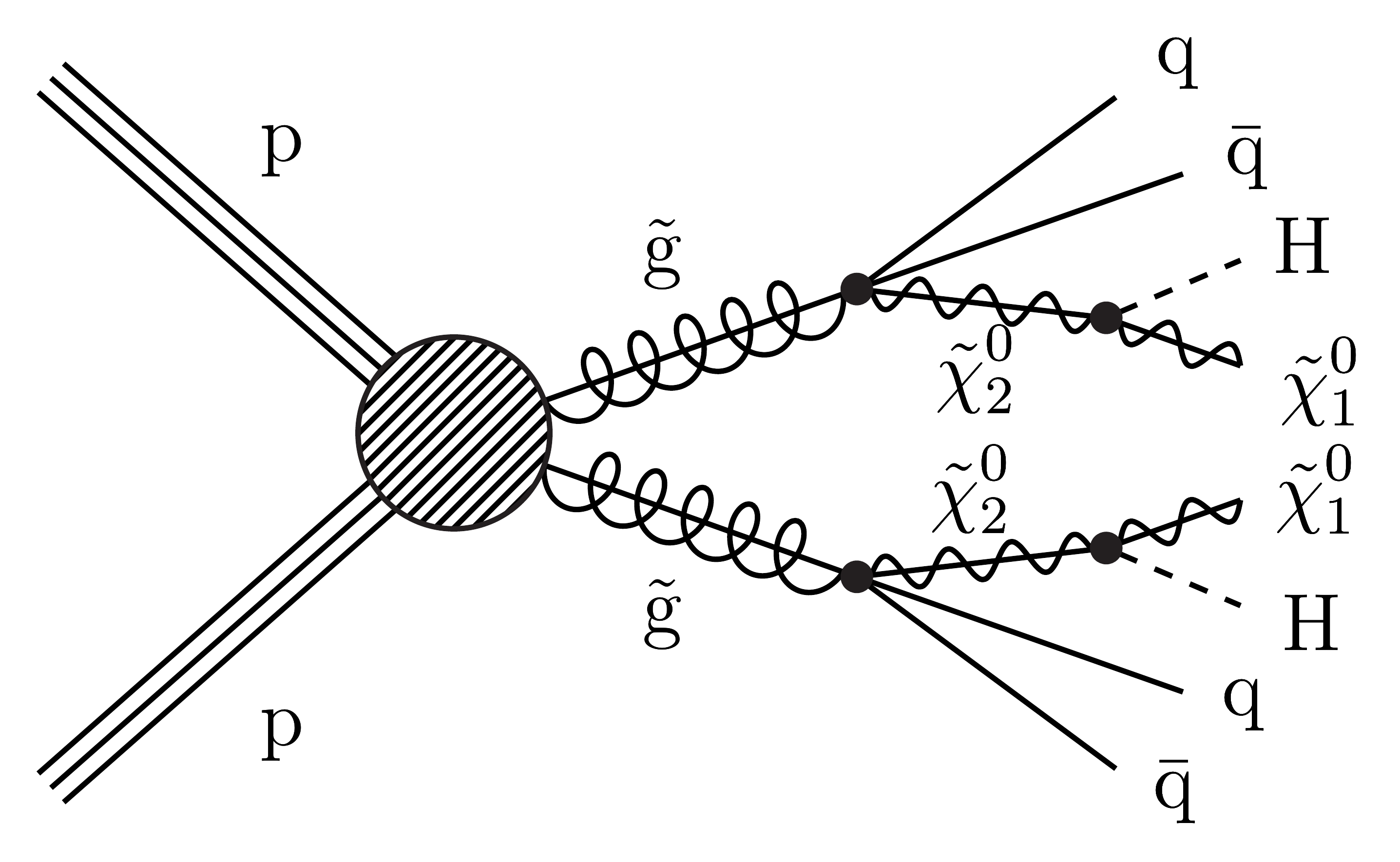}
\caption{Diagrams for
(left) the TChiHH-G signal model,
  $\lsp\lsp\to\PH\PH\sGra\sGra$,
in which the \chizone NLSPs are produced indirectly through the cascade decays of several combinations of neutralinos and charginos, as described in the text;
(center) TChiHH, in which the electroweak production of two neutralinos leads to two Higgs bosons and two neutralinos ($\LSP$);
(right) T5HH, the strong production of a pair of gluinos, each of which decays via a three-body process to quarks and a neutralino, with
the neutralino subsequently decaying to a Higgs boson and a \chizone LSP.
In each diagram, the hatched circle represents the sum of processes that can lead to the SUSY particles shown.
}
\label{fig:susy_models}
\end{figure}

Figure~\ref{fig:susy_models} shows three models that can lead to the experimental signature considered in this analysis.
These scenarios
are described using the framework of simplified SUSY models~\cite{bib-sms-1,bib-sms-2,bib-sms-3,bib-sms-4}, in which many of the SUSY particles are assumed to be decoupled or otherwise irrelevant to the process under consideration. 
In the TChiHH-G simplified model (Fig.~\ref{fig:susy_models}, left), the LSP is the goldstino.
In a broad range of scenarios in which SUSY breaking is mediated at a low scale, such as gauge-mediated supersymmetry breaking (GMSB) models~\cite{Dimopoulos:1996vz,Matchev:1999ft}, the goldstino is
nearly massless on the scale of the other particles and is the LSP. The \chizone is then the next-to-lightest supersymmetric particle (NLSP)~\cite{Ruderman:2011vv}.
The NLSPs are produced in the cascade decays of several different combinations of neutralinos and charginos, and the goldstino is taken to be approximately massless.
An important case arises when the lighter neutralinos $\widetilde\chi_{1,2}^0$ and charginos $\chipmone$ are dominated by their higgsino content and, as a consequence,
are nearly mass degenerate.  In this case, all of their cascade decays can lead to the production of the NLSP and soft particles.
The NLSP subsequently decays to the goldstino and a Higgs boson, with a coupling that is suppressed by the SUSY breaking scale~\cite{Dimopoulos:1996vz}.
Integrating over the contributions
from the allowed combinations of produced charginos and neutralinos ($\lsp\PSGczDt$, $\lsp\chipmone$,
$\PSGczDt\chipmone$, $\chipmone\chimpone$) leads to an effective rate for $\lsp\lsp$ production~\cite{Fuks:2012qx,Fuks:2013vua} that
is significantly larger than that for any of the individual primary pairs. We assume a branching fraction of 100\% for $\chizone\to\PH\sGra$.

In the TChiHH simplified model (Fig.~\ref{fig:susy_models}, center),
two higgsinos \chiztwo and \chizthree are produced. The \chizone is the LSP, assumed to be a bino (the superpartner of the SM boson corresponding to the $U(1)$
weak hypercharge gauge field $B$), while \chiztwo and \chizthree, nearly degenerate in mass, are the NLSPs.  The other higgsinos have allowed decay channels, such as $\chipmone\to\PWpm\chizone$, that do not lead to Higgs bosons.
This type of mass hierarchy can arise in various scenarios, as discussed in Refs.~\cite{Baer:2015tva,Kang:2015nga}.
Unlike in the TChiHH-G simplified model described above, where the heavier higgsino states were assumed to decay directly to the lightest one, here only
the exclusive $\chiztwo\chizthree$ higgsino production cross section contributes; we are not sensitive to the other higgsino production channels
$\chipmone\chimpone$, $\chipmone\chizthree$, and $\chipmone\chiztwo$.
(The identical-particle combinations $\chiztwo\chiztwo$ and $\chizthree\chizthree$ are not produced because their couplings to the \PZ boson are suppressed~\cite{Kang:2015nga}.)
The cross section for $\chiztwo\chizthree$ alone
is about 17\% of that for the sum of all six higgsino cross sections~\cite{Fuks:2012qx,Fuks:2013vua}.

Some SUSY models~\cite{Kribs:2010hp,Gori:2011hj} predict production rates for energetic Higgs bosons that are greater in gluino cascade decays than in direct higgsino
production, because of the much larger strong production cross section for gluinos.
Figure~\ref{fig:susy_models} (right) corresponds to a model (T5HH) in which two gluinos are produced, each of which decays via a three-body process to
a quark, an antiquark, and a \chiztwo.
The \chiztwo
decays to a Higgs boson and a \chizone, which is taken to be the LSP.
The same signature also arises in a recently proposed model~\cite{Arganda:2021lpg,Delgado:2020url}, motivated by dark-matter considerations,
where a bino-like \chizthree becomes the NLSP, decaying via a Higgs boson to the \chizone LSP.

This paper is organized as follows. Section~\ref{sec:analysis} provides an overview of the analysis strategy, while Sections~\ref{sec:detector} and \ref{sec:simulation} describe the
CMS detector and the simulated event samples, respectively. The event triggers and reconstruction of the data are presented in Section~\ref{sec:evtreco}, while event selection and the
reconstruction of Higgs boson candidates are discussed in Section~\ref{sec:evtsel}. The background estimation is described in Section~\ref{sec:background}, and the
systematic uncertainties are discussed in Section~\ref{sec:systematics}. Section~\ref{sec:results} presents the results and interpretation,
and Section~\ref{sec:summary} gives a summary. 

\section{Analysis strategy}
\label{sec:analysis}
As discussed in Section~\ref{sec:introduction}, the common signature of the SUSY processes targeted by this search is the appearance of two Higgs bosons, each decaying via $\PH\to\bbbar$, and accompanied by $\ptmiss=\abs{\ptvecmiss}$.
Additional jets can be present, either from initial-state radiation, or, as in the case of strong SUSY production, from additional particles in the decay chains. Experimentally, this signature can be manifested in a number of different jet configurations.
The analysis is designed to address not only the simplest case, in which the \PQb quarks from the Higgs boson decays appear as jets (\PQb jets) that are individually resolved, but also the case in which \PQb jets resulting from the decay of a high-momentum Higgs boson merge into a single wide jet. Furthermore, to provide sensitivity to a broad range of experimental b-tagging outcomes, including those arising from reconstruction inefficiencies and detector acceptance, we do not restrict the search to precisely the nominal four resolved b-tagged jets (for the resolved signature), or to the nominal two merged double b-tagged jets (for the boosted signature).

The momentum of the Higgs boson largely determines whether the resolved or boosted signature contributes more to the sensitivity; this momentum is in turn determined by the masses of
the particles in the decay chain, especially the mass splitting between the higgsino and its daughter SUSY particle.
 For both types of signatures, the basic analysis procedure involves five steps:
\begin{enumerate}
\item Baseline event selection requirements are applied, narrowing the event sample to an \textit{analysis region}, which includes signal, sideband, and control regions. These requirements include a
veto of events containing isolated leptons (either electrons or muons). A loose \ptmiss requirement is also applied for both signatures. 
\item Higgs-boson candidates are identified, and their masses are reconstructed.
\item Binning is applied in certain analysis variables. This binning has two roles: first, to define the signal, sideband, and control regions within the analysis region for a robust, data-driven background estimation,
and second, using additional variables (such as \ptmiss), to enhance the sensitivity to the different SUSY models across the large range of kinematically accessible SUSY particle masses. 
\item \textit{Control samples} are defined and used to validate the background estimation procedures and to estimate systematic uncertainties.  These samples are selected with modifications of the baseline criteria (for example, by requiring one or more leptons) so that they do not overlap with the analysis region.  These samples contain signal, sideband, and control regions paralleling those in the analysis region.
\item As a test of the null hypothesis (no signal), the background predictions based on yields in the sideband and control regions are compared with the observed yields in the signal regions.
For each test of a specific SUSY model, a fit is performed that allows 
for the expected level of signal contamination across all parts of the overall analysis region. 
\end{enumerate}
The detailed event selection requirements and background estimation methods used are somewhat different for the resolved and boosted signatures, but they have certain basic features in common. The most important of these is the use of two key types of variables to define the signal, sideband, and control regions: (i) the number of single- or double-\PQb-tagged jets and (ii) the measured masses of the Higgs-boson candidates.

The composition of the SM backgrounds in this analysis is relatively simple, with the dominant contribution arising from \ttbar events that contain 
a single leptonic \PW boson decay, together with additional \PQb-tagged jets from the misidentification of lighter-quark and gluon jets, or from gluon jets that split into b quark pairs. The leptonic \PW boson decay produces a neutrino that
is the source of the observed \ptmiss. Most such events are vetoed by the baseline selection requirement that excludes events with a single isolated lepton. However, \ttbar single-lepton events can sometimes escape this veto if the lepton is ``lost'' in some manner,
or is a \Pgt lepton that decays hadronically.
Such events, which escape the isolated-lepton veto applied in the baseline selections, are referred to as \textit{lost-lepton events}.
Sub-dominant backgrounds arise from \wjets and \zjets events, as well as from multijet events, produced by quantum chromodynamics (QCD) processes, in which \ptmiss arises from jet mismeasurement.
However, the QCD contribution is highly suppressed 
by requiring that the \ptvecmiss vector not be aligned with any of the highest transverse-momentum jets in the event.

\section{The CMS detector}
\label{sec:detector}
A detailed description of the CMS detector,
along with a definition of the coordinate system and
pertinent kinematic variables,
is given in Ref.~\cite{Chatrchyan:2008aa}.
Briefly,
a cylindrical superconducting solenoid with an inner diameter of 6\unit{m}
provides a 3.8\unit{T} axial magnetic field.
Within the cylindrical volume
are a silicon pixel and strip tracker,
a lead tungstate crystal electromagnetic calorimeter (ECAL),
and a brass and scintillator hadron calorimeter (HCAL).
The tracking detectors cover the range $\abs{\eta}<2.5$,
where $\eta$ is the pseudorapidity.
The ECAL and HCAL,
each composed of a barrel and two endcap sections,
cover $\abs{\eta}<3.0$.
Forward calorimeters extend the coverage to $3.0<\abs{\eta}<5.2$.
Muons are measured within $\abs{\eta}<2.4$ by gas-ionization detectors
embedded in the steel flux-return yoke outside the solenoid.
The detector is nearly hermetic,
permitting accurate measurements of~\ptmiss.

Events of interest are selected using a two-tiered trigger system. The first level (L1), composed of custom hardware processors, uses information from the calorimeters and muon detectors to select events at a rate of around 100\unit{kHz} within a fixed latency of about 4\mus~\cite{Sirunyan:2020zal}. The second level, known as the high-level trigger (HLT), consists of a farm of processors running a version of the full event reconstruction software optimized for fast processing, and reduces the event rate to around 1\unit{kHz} before data storage~\cite{Khachatryan:2016bia}.

In addition to the \pp collision of interest selected by the trigger, the event record generally contains tracks and calorimeter energy deposits from a number of other collisions (pileup) occurring in the same bunch crossing, or in adjacent crossings within the time resolution of the data acquisition.  The number of collisions varies with the instantaneous luminosity, averaging about 29 over the data set~\cite{CMS:2020ebo}.

\section{Simulated event samples}
\label{sec:simulation}
While the evaluation of the SM background, described in Section~\ref{sec:background}, is based primarily on observed event yields in multiple control regions in data, Monte Carlo (MC) simulated event samples nevertheless play an important role in the analysis. Such samples are used to optimize the event selection requirements and background prediction methods without biasing the result; to evaluate correction factors, typically near unity, to the background predictions; and to evaluate the acceptance and efficiency for signal processes.

We use the
{\MGvATNLO}\,2.2.2 (2.4.2)~\cite{Alwall:2014hca,Alwall:2007fs}
event generator with leading order (LO) precision for simulation of
the SM production of \ttbar, \wjets, \zjets,
and QCD processes,
for the 2016 (2017--2018) data taking periods.
The \ttbar events are generated with
up to three additional partons in the matrix element calculations.
The \wjets and \zjets events are generated
with up to four additional partons.
For the \MADGRAPH LO samples, MLM parton matching is used~\cite{Alwall:2007fs}.
Single top quark events produced through the $s$ channel,
diboson events such as those originating from
$\PW\PW$, $\PZ\PZ$, or $\PZ\PH$ production,
and rare events such as those from $\ttbar\PW$,
$\ttbar\PZ$, and $\PW\PW\PZ$ production,
are generated with {\MGvATNLO}
at next-to-leading order (NLO), with FXFX matching~\cite{Frederix:2012ps},
except that $\PW\PW$ events in which both {\PW} bosons decay leptonically
are generated using the
{\POWHEG}\,v2.0~\cite{Nason:2004rx,Frixione:2007vw,Alioli:2010xd,Alioli:2009je,Re:2010bp}
program at NLO.
This same \POWHEG generator is used to describe both $\PQt\PW$ production and
$t$-channel production of single top quark events at NLO precision.

Samples of simulated signal events are generated at LO using {\MGvATNLO}\,2.4.2,
with up to two additional partons included in the matrix element calculations.
The production cross sections for the electroweak models are computed at NLO plus next-to-leading-log (NLL) accuracy~\cite{Fuks:2012qx,Fuks:2013vua}, while those for strong production are
determined with approximate next-to-NLO (NNLO) plus next-to-next-to-leading logarithmic (NNLL)
accuracy~\cite{bib-nlo-nll-01,bib-nlo-nll-02,bib-nlo-nll-03,bib-nlo-nll-04,bib-nlo-nll-05,
Beenakker:2016lwe,Beenakker:2011sf,Beenakker:2013mva,Beenakker:2014sma,
Beenakker:1997ut,Beenakker:2010nq,Beenakker:2016gmf}.
For the TChiHH-G model, these samples are generated by performing
a scan of $m(\chizone)$,
with $m(\sGra)=1\GeV$.
The TChiHH model is represented by a two-dimensional scan
of $m(\chizone)$ and
$m(\widetilde\chi_{2,3}^0)$.
The TChiHH-G and TChiHH simulated samples are generated with both Higgs bosons constrained to decay via \Hbb, and each event is weighted by the square of the standard model branching fraction.  Auxiliary simulations show that inclusion of the non-$\bbbar$ decay modes of the Higgs boson would slightly improve the sensitivity, mainly for the boosted signature, but these additional modes are not included here.
Additional details on the simulation of electroweak production models are given in Ref.~\cite{Sirunyan:2017obz}.
Events with gluino pair production (T5HH) are generated
for a range of $m(\gluino)$ values.
The gluino decay is simulated with a
three-body phase space model~\cite{Abdullin:2011zz}.
The mass of the intermediate \chiztwo is set $50\GeV$ below $m(\gluino)$,
ensuring that the daughter Higgs bosons have a large Lorentz boost, $m(\chizone)$ is set to $1\GeV$, and all decay modes of the Higgs bosons are included.

All simulated samples make use of the
{\PYTHIA}\,8.205~\cite{Sjostrand:2014zea} program
to describe parton showering and hadronization.
The CUETP8M1~\cite{ Khachatryan:2015pea} (CP5~\cite{CMS:2019csb})
{\PYTHIA} tune was used to produce the SM background samples
for the analysis of the 2016 (2017 and 2018) data,
with signal samples based on the CUETP8M1 tune for 2016
and on the CP2 tune~\cite{CMS:2019csb} for 2017 and 2018.
Simulated samples generated at LO (NLO) with the CUETP8M1 tune use the
NNPDF3.0LO (NNPDF3.0NLO)~\cite{Ball:2013hta}
parton distribution functions (PDFs),
while those using the CP2 or CP5 tune use the NNPDF3.1LO
(NNPDF3.1NNLO)~\cite{Ball:2017nwa} PDFs.

The detector response is modeled with
\GEANTfour~\cite{Agostinelli:2002hh}.
Normalization of the simulated background samples is performed using
the most accurate cross section calculations
available~\cite{Alioli:2009je,Re:2010bp,Alwall:2014hca,Melia:2011tj,Beneke:2011mq,
Cacciari:2011hy,Baernreuther:2012ws,Czakon:2012zr,Czakon:2012pz,Czakon:2013goa,
Gavin:2012sy,Gavin:2010az},
which generally correspond to NLO or NNLO precision.
Because scans over numerous mass points are required for the signal models, the detector response for such events (except for those of the T5HH model) is described using the CMS fast simulation program~\cite{Abdullin:2011zz,Giammanco:2014bza},
which yields results that are generally
consistent with those from the simulation based on {\GEANTfour}.
Pileup \pp interactions are superimposed on the generated events, with a number distribution that is adjusted to match
the pileup distribution measured in data.

\section{Triggers and event reconstruction}
\label{sec:evtreco}
The data sample for the analysis region was obtained using triggers~\cite{Khachatryan:2016bia,Sirunyan:2020zal} that require \mettr to exceed a threshold that varied between 90 and $140\GeV$, depending on the LHC instantaneous luminosity.
Single-lepton control samples, defined below, are collected with single-electron, single-muon, and \mettr triggers, while
dilepton control samples are collected with single-electron and single-muon triggers.
The value of \mettr is computed with trigger-level quantities and therefore has somewhat poorer resolution than \ptmiss.  The trigger efficiency for the analysis sample is measured as a function of \ptmiss,
and of \HT, the scalar sum of the transverse momentum (\pt) of jets, separately for each year of data taking. For the single-lepton control sample, it is measured as a function of \ptmiss, \HT, and lepton \pt, and for the dilepton control sample, as a function of the leading-lepton \pt.
For the analysis sample,
the efficiency at $\ptmiss=150\GeV$ is 30--70\%, depending on the trigger thresholds (which varied with the instantaneous luminosity); the efficiency rises with \ptmiss and is over 99\% for $\ptmiss>260\GeV$.
The analysis of the boosted signature requires $\ptmiss>300\GeV$ and thus operates on the efficiency plateau; 
the resolved-signature analysis operates partly on the turn-on section of the efficiency curve, with a lower \ptmiss threshold of $150\GeV$.
These measured efficiencies are applied as weights to the simulated SM and signal events, for the analysis region and the control samples.

The reconstruction of physics objects in an event proceeds from the candidate particles identified by the particle-flow (PF) algorithm~\cite{CMS-PRF-14-001}, which uses information from the tracker, calorimeters, and muon systems to identify the candidates as charged or neutral hadrons, photons, electrons, or muons.

The primary \pp interaction vertex is taken to be the reconstructed vertex with the largest value of summed physics-object $\pt^2$, as described in Section 9.4.1 of Ref.~\cite{CMS-TDR-15-02}.
The physics objects for this purpose are the jets
(reconstructed using the anti-\kt jet finding algorithm~\cite{Cacciari:2008gp,Cacciari:2011ma}
with the charged particle tracks assigned to the vertex),
isolated tracks (including those identified as leptons),
and the associated missing transverse momentum
(computed as the negative vector sum of the \ptvec values of those objects).
Charged particle tracks associated with vertices other than
the primary vertex are removed from further consideration.
The primary vertex is required to lie within 24\unit{cm} of the
center of the detector in the direction along the beam axis
and within 2\unit{cm} in the plane transverse to that axis.

Jets from the primary \pp interaction
are formed from the charged PF candidates associated with the primary vertex, together with the neutral PF candidates, again using
the anti-$\kt$ algorithm~\cite{Cacciari:2008gp},
as implemented in the \FASTJET package~\cite{Cacciari:2011ma}.
The clustering is performed
with distance parameter $R=0.4$ (AK4 jets)
when optimized for jets containing the fragmentation products of a single parton,
and with distance parameter $R=0.8$ (AK8 jets) for
jets containing multiple partons.
No removal of overlap jets between the AK4 and AK8 collections is applied.
Jet quality criteria~\cite{CMS-PAS-JME-10-003,CMS-PAS-JME-16-003}
are imposed to select quark and gluon jets while rejecting those from 
spurious sources, such as electronics noise and detector malfunctions.
The jet energies are corrected for the nonlinear response of the
detector~\cite{Khachatryan:2016kdb}.
The estimated contribution to the jet $\pt$ of neutral PF candidates
from pileup
is removed with a correction based on the
area of the jet and the average energy density of the event~\cite{Cacciari:2007fd}, for AK4 jets, and with the PUPPI technique~\cite{Bertolini:2014bba}, for AK8 jets.
Jets are required to have $\abs{\eta}<2.4$; AK4 and AK8 jets are required to have $\pt>30\GeV$ and $\pt>300\GeV$, respectively.
Finally, AK4 jets that have PF constituents matched to an isolated electron or muon are removed.
To improve the consistency of the fast simulation description
with respect to that based on \GEANTfour,
we apply a small correction, about 1\%, to account for differences in the efficiency
of the jet quality requirements~\cite{CMS-PAS-JME-10-003,CMS-PAS-JME-16-003}.

To improve the modeling of jets associated with initial-state radiation (ISR),
the prediction of \MGvATNLO is compared with data in a control
sample enriched in \ttbar events by the requirement of
two light charged leptons ($\Pe\Pe$, $\Pgm\Pgm$, or $\Pe\Pgm$)
and two {\PQb}-tagged jets.
The number of all remaining jets in the event is denoted \njetsisr.
A correction factor is applied to simulated \ttbar and strong-production signal events
so that the \njetsisr distribution agrees with that in data.
The central value of the correction ranges from 0.92 for $\njetsisr = 1$
to 0.51 for $\njetsisr \geq 6$.
The correction is found not to be necessary for \ttbar samples that are
generated with the CP5 tune.
For the electroweak signal samples, we account for the effect of ISR on the \pt distribution of the system of SUSY particles, $\pt^{\text{ISR}}$, by reweighting the distribution based on studies of the transverse momentum of \zjets events~\cite{Chatrchyan:2013xna}.
The reweighting factors range between 1.18 at $\pt^{\text{ISR}}\sim125\GeV$ and
0.78 for $\pt^{\text{ISR}}>600\GeV$.

The identification of jets originating from {\PQb} quarks ({\PQb} tagging) is integral to the reconstruction of Higgs bosons in the analysis and is performed separately for AK4 and AK8 jets.  For AK4 jets we use a version of the combined secondary vertex algorithm based on deep neural networks (\textsc{DeepCSV}~\cite{Sirunyan:2017ezt}), making use of Loose (L), Medium (M), and Tight (T) working points, as discussed in Section~\ref{sec:evtsel}.  The detection efficiencies for true \PQb quark jets, as well as the mistag rates for jets originating from up, down, or strange quarks or from gluons
(light-parton jets), and for charm-quark jets, are measured with data samples enriched in, respectively, \ttbar, QCD, and a combination of $\PW\!+\!\PQc$ and semileptonic \ttbar events~\cite{Sirunyan:2017ezt}.
Typical {\PQb} jet efficiencies for the three \textsc{DeepCSV} working points are 85, 70, and 50\%, with corresponding light-parton jet  mistag rates of 10, 1, and 0.1\%, and charm jet mistag rates of 40, 12, and 2.5\%, for the L, M, and T working points, respectively.  The numbers of observed \PQb-tagged AK4 jets satisfying the L, M, and T requirements in an event are denoted \nbl, \nbm, and \nbt, respectively.

The identification of AK8 jets containing two {\PQb} quarks, used in the analysis of the boosted signature, is performed
with a deep-learning-based double-\PQb tagging algorithm (\textsc{DeepDoubleBvL}, mass-decorrelated~\cite{CMS-DP-2018-046})
at its loose working point.
The efficiencies for tagging AK8 jets from \Hbb and for those that contain only light partons are 90 and 5\%, respectively~\cite{CMS-DP-2018-046}.
The efficiency of the double-\PQb tagging is checked with a data sample enriched in $\Pg\to\bbbar$; the mistag rate is checked with data in a lower-\ptmiss sideband ($200<\ptmiss<300\GeV$) of the baseline selection for the boosted signature.
For both single and double \PQb tagging, the simulation is reweighted to compensate for differences with respect to data based on measurements of the {\PQb} tagging efficiency and mistag rate for each working point.
Additional corrections are applied for differences in the flavor tagging efficiencies between the fast simulation and {\GEANTfour}, where applicable.

The reconstructed \ptvecmiss is computed as the negative vector sum of the \ptvec of all PF candidates, adjusted for known detector effects by taking into account the jet energy corrections described in Ref.~\cite{Sirunyan:2019kia}. Filters are applied to reject events with well-defined anomalous sources of \ptmiss, such as calorimeter noise, nonfunctioning calorimeter cells, beam particles externally scattered into the detector, and other effects~\cite{Sirunyan:2019kia}.
For data generated with the fast simulation, corrections of 0--12\% are applied to account for differences in the modeling of \ptmiss.

Because the targeted signature is fully hadronic, we veto events with isolated charged lepton candidates as part of the definition of the analysis region.
Conversely, we require isolated electrons and muons in the selection of certain control samples.
Electrons are reconstructed from showers in the ECAL matched to tracks in the silicon detectors, taking into account the effects of bremsstrahlung in the material of the tracker.
Additional criteria are imposed on the shower shape in the ECAL and on the ratio of energies associated with the electron candidate
in the HCAL and ECAL~\cite{CMS:2020uim}.  We use the ``cut-based veto'' working point of the electron identification algorithm~\cite{CMS:2020uim}.
Muons are constructed from muon detector track segments matched to silicon detector tracks; we require candidates to satisfy the ``medium'' working point of the muon identification~\cite{Sirunyan:2018fpa}.
For both electrons and muons, the track is required to be consistent with the particle's origination at the primary vertex.  Electron and muon candidates must also satisfy $\abs{\eta}<2.5$ and 2.4, respectively.  Requirements on the lepton \pt are imposed that differ between veto leptons, where efficiency is prioritized, and control-sample leptons, where higher purity is desired;  we require $\pt>10\GeV$ for veto leptons, and $\pt>30\GeV$ for those used to select events in the single-lepton control samples.  For the dilepton control samples, the threshold \pt is 30 (20)\GeV for the leading (subleading) lepton.

To select electrons and muons that originate in the decay of \PW and \PZ bosons, while suppressing those from the decays of hadrons and from particles misidentified as \Pe or \PGm, both are required to be isolated from other PF candidates.
The isolation criterion is based on the variable~$\imini$,
which is the scalar \pt sum of charged-hadron,
neutral-hadron, and photon PF candidates within a cone
of radius $\dR\equiv\sqrt{\smash[b]{(\Delta\phi)^2+(\Delta\eta)^2}}$
around the lepton direction, after pileup subtraction,
divided by the lepton~\pt,
where $\phi$ is the azimuthal angle.
The radius of the cone depends on the lepton \pt:  $\dR = 0.2$ for  $\pt<50\GeV$, $10\GeV/\pt$ for $50\leq\pt\leq 200\GeV$,
and 0.05 for $\pt>200\GeV$~\cite{Rehermann:2010vq}.
The isolation requirement is $\imini<0.1$ (0.2) for electrons (muons).

The efficiency of electron and muon identification is measured with a tag-and-probe method applied to \zll samples in the data~\cite{CMS:2020uim,Sirunyan:2018fpa}.  Yields of these leptons in simulation are corrected to match the observations in data.

The efficiency of the charged lepton veto can be compromised if the charged lepton is lost, either because it is not reconstructed or fails the lepton selection requirements (including the isolation and \pt threshold requirements), or because it is a tau lepton that decays hadronically.
To recover some of this veto efficiency,
we also reject events with any additional isolated tracks corresponding to leptonic or hadronic PF candidates. To reduce the influence of tracks resulting from pileup, such isolated tracks are considered only if their distance of closest approach along the beam axis to a reconstructed vertex is smaller for the primary vertex than for any other vertex.
The tracks are required to satisfy $\abs{\eta}<2.5$,
as well as $\pt>10\GeV$, or ${>}5\GeV$ for a PF electron or muon.
The isolation requirement is that
the scalar \pt sum of all other charged particle tracks
within a cone of radius 0.3 around the track direction,
divided by the track~\pt,
must be less than 0.2 if the track is identified
as a PF electron or muon,
less than 0.1 otherwise.
To limit the veto tracks to those likely to originate in \PW boson leptonic decays,
we also restrict the transverse mass, requiring $\mt\equiv \sqrt{\smash[b]{2\pt\ptmiss(1-\cos{\Delta\phi})}}<100\GeV$.
Here \pt is the transverse momentum of the particle and $\Delta\phi$ is the azimuthal angle between the particle momentum and \ptvecmiss.

\section{Event selection and reconstruction of Higgs boson candidates}
\label{sec:evtsel}
In the first step of the analysis procedure, as outlined in Section~\ref{sec:analysis}, baseline selection requirements are applied to define the analysis regions for both the
resolved and boosted signatures. These baseline selection requirements are described below. Three of the baseline selection requirements are based on quantities that are in common to both
signatures and are defined as follows: 
\begin{enumerate}
\item A minimum requirement on \ptmiss is applied: $\ptmiss>150\GeV$ for the resolved signature and $\ptmiss>300\GeV$ for the boosted signature.
\item For both signatures, events are excluded if any veto leptons or isolated charged particle tracks are present. These veto objects are defined in Section~\ref{sec:evtreco}.
\item To suppress QCD background, events in either signature are rejected if any of the four highest \pt AK4 jets is approximately aligned with the missing momentum vector \ptvecmiss.
Such alignment is an indication that the observed \ptvecmiss in the event is a consequence of jet mismeasurement and is not genuine.
This set of requirements, which we refer to collectively as the \dphi requirement, can only be imposed in the plane transverse to the beam axis. An event is vetoed if any of the azimuthal separations is small, specifically, if $\Delta\phi_{\ptvecmiss,\,\text{j}_i}<\{0.5,0.5,0.3,0.3\}$ for the $i^{\text{th}}$-highest \pt jet $\text{j}_i$. 
\end{enumerate}
The remaining baseline selection requirements depend on the individual details of the Higgs boson mass reconstruction and the counting of \PQb jets, and we discuss these for the
two signatures in turn.  

\begin{figure}[htbp!]
\centering
\includegraphics[width=0.40\textwidth]{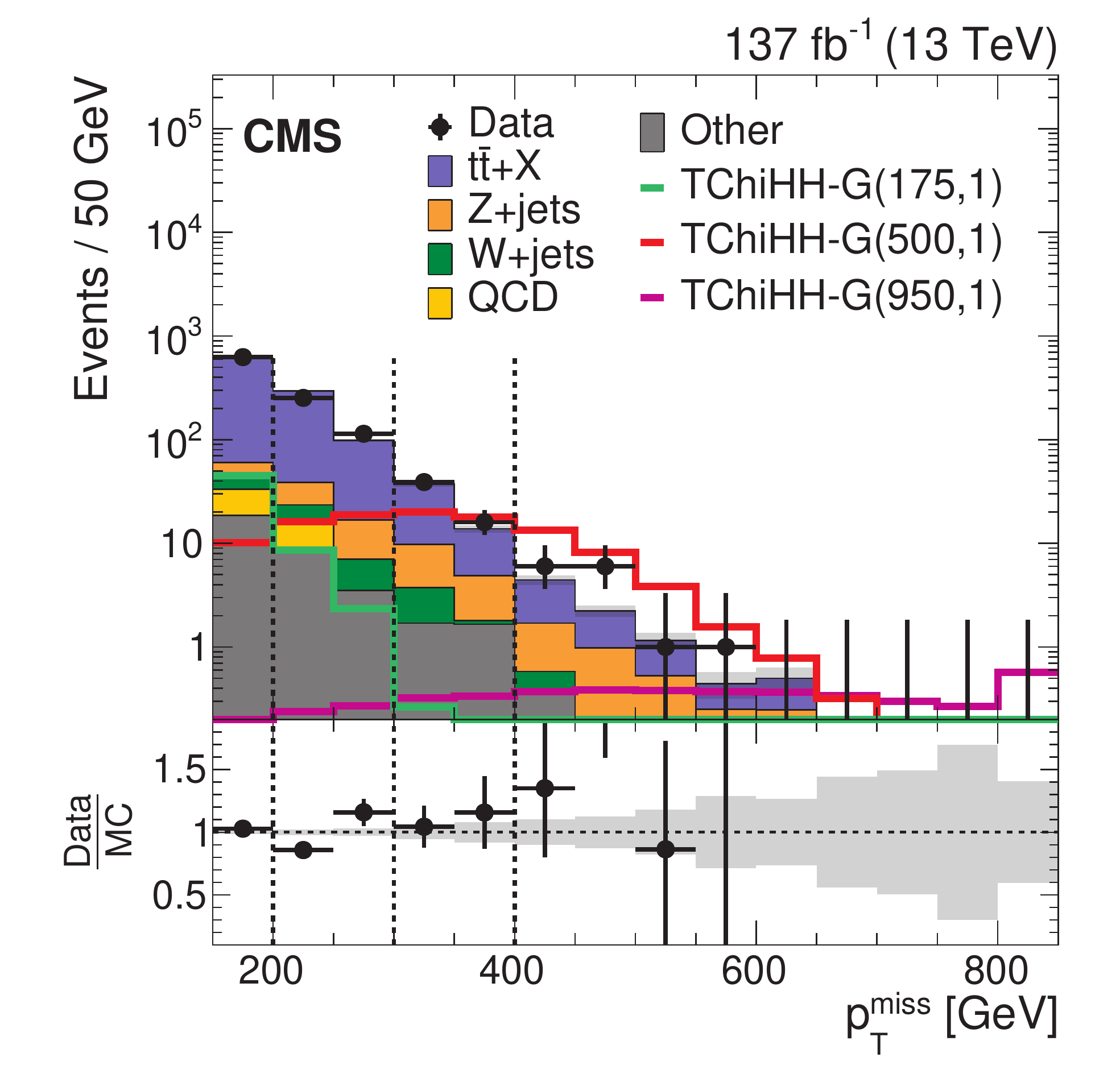}
\includegraphics[width=0.40\textwidth]{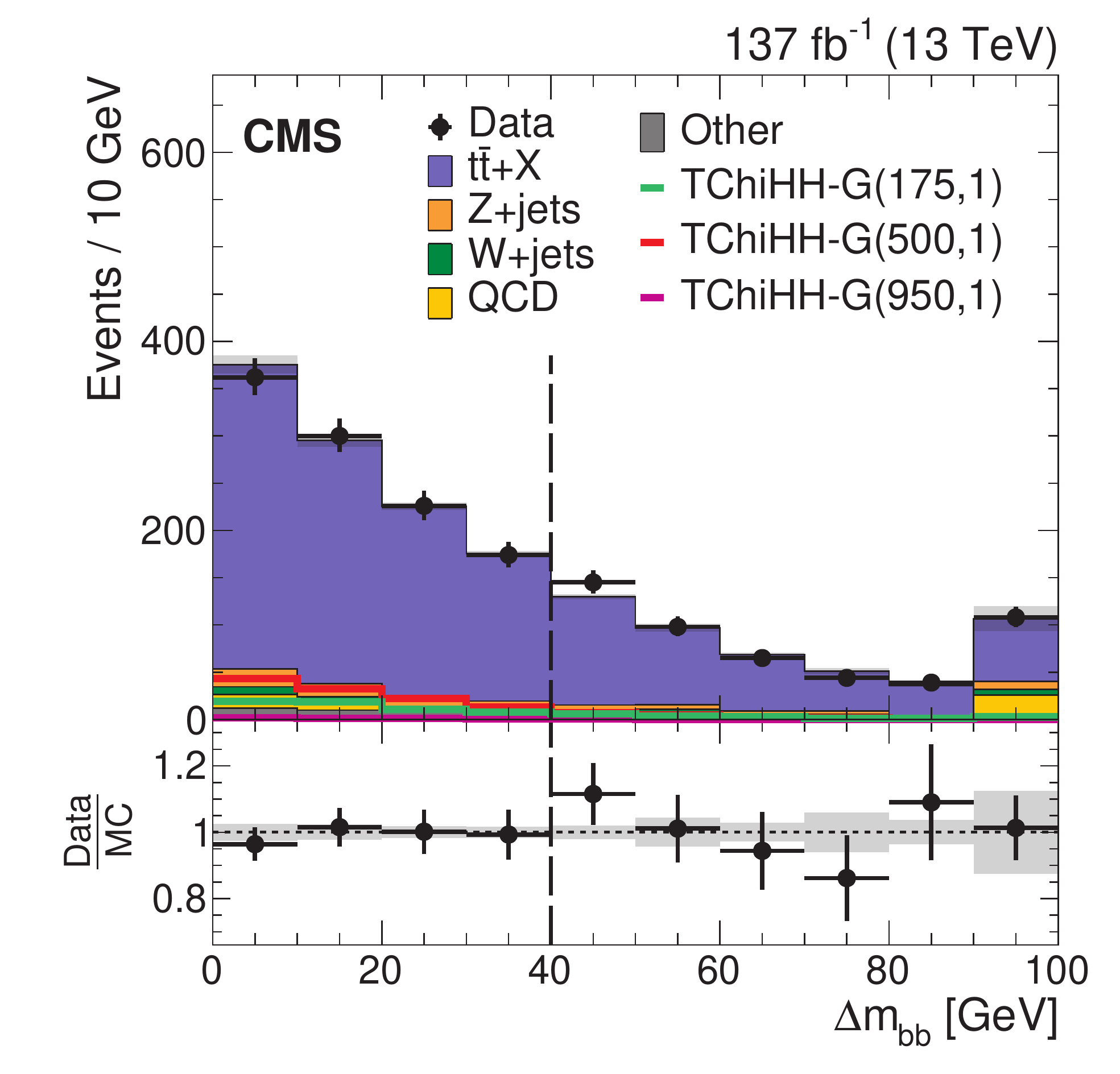}
\includegraphics[width=0.40\textwidth]{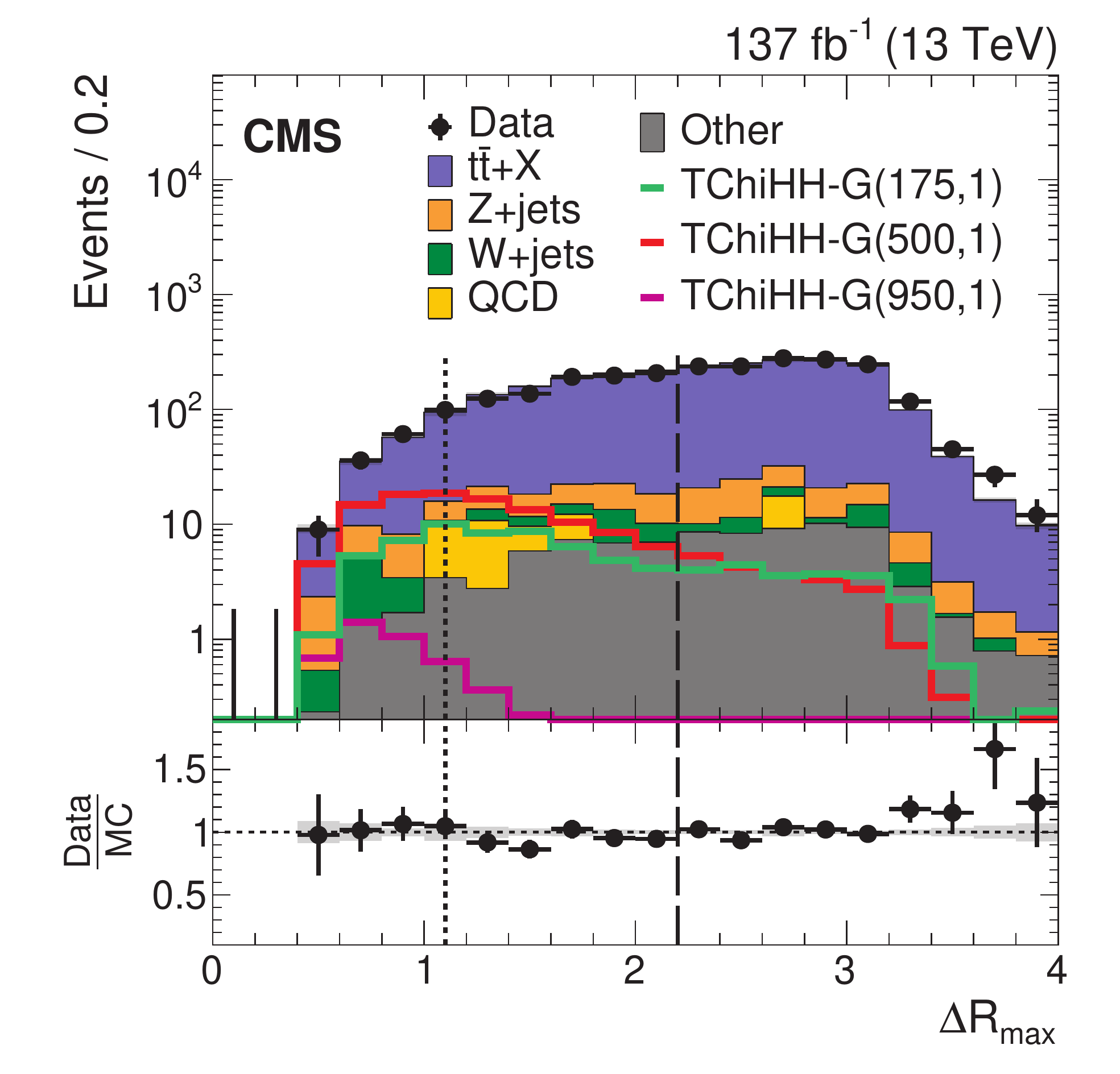}\\
\includegraphics[width=0.40\textwidth]{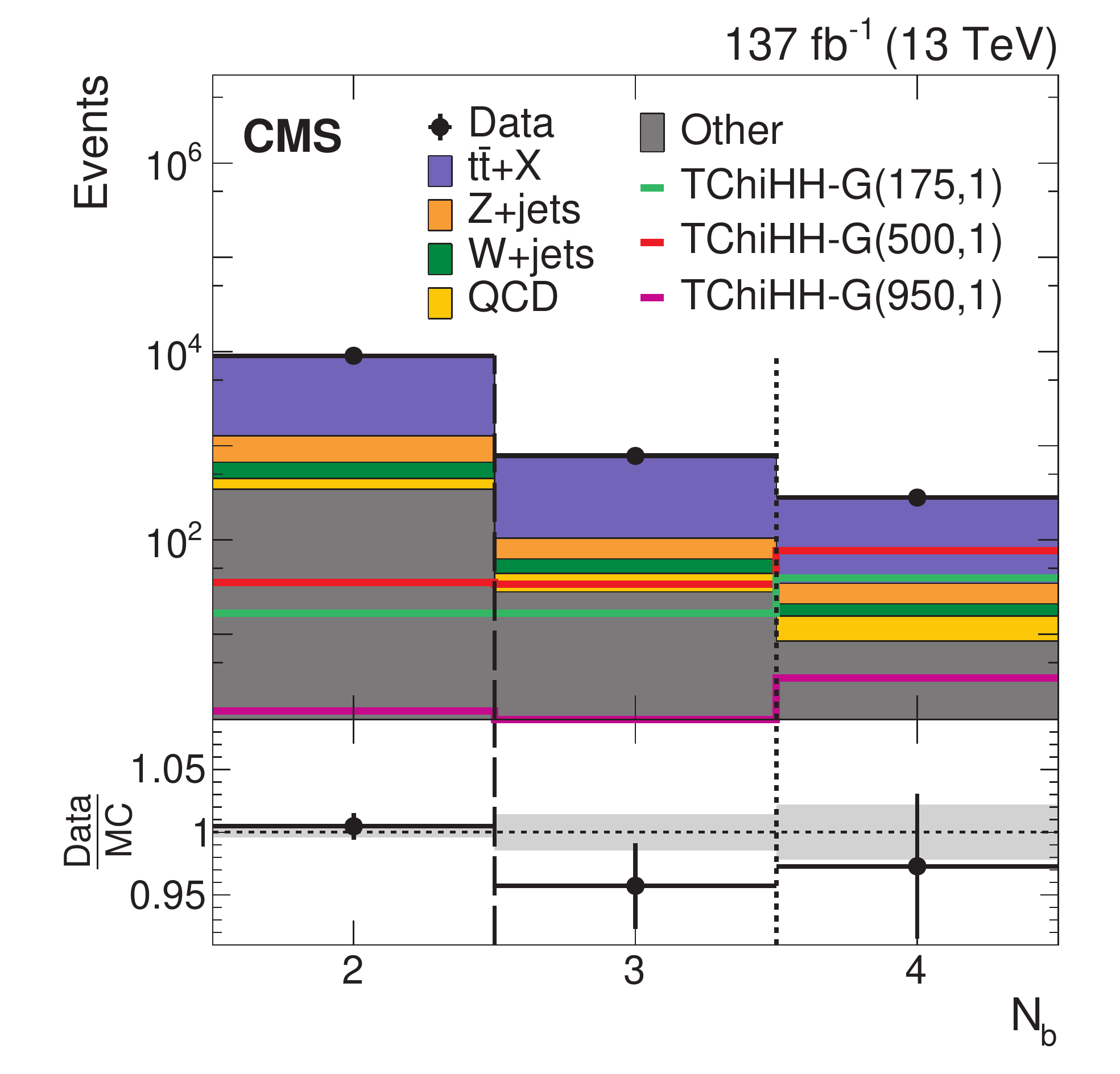}
\includegraphics[width=0.40\textwidth]{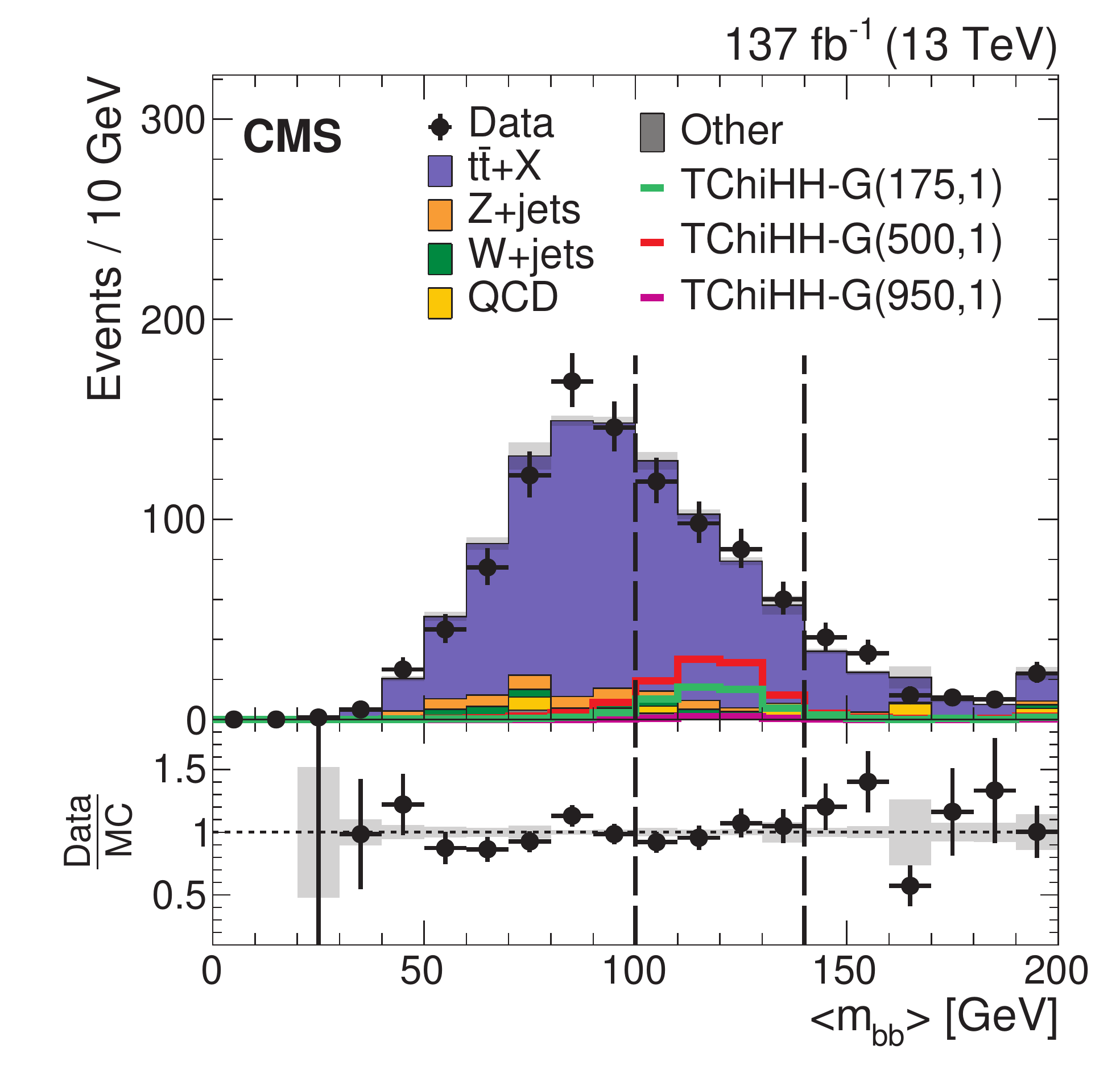}\\
\caption{
Distributions in key analysis variables of the resolved signature for events satisfying
the baseline requirements and $\nb=3$ or 4, except for those on the variable plotted:
(upper left) \ptmiss, (upper right) \dmjj, (middle) \drmax, (lower left) \nb, and (lower right) \amjj.
The rightmost bins include the overflow entries. The data are shown as black markers with error bars, simulated SM backgrounds by the stacked histograms (scaled by factors of 0.93--0.97 to match the total data yield in each plot), and simulations of signals $\text{TChiHH-G}(m(\chizone),m(\sGra)\,[\GeVns])$ by the open colored histograms.  Gray shading represents the statistical uncertainties of the simulation. Vertical lines indicate the boundaries for signal region selection (dashed), or for the signal region binning
(dotted).  The lower panels show the ratio of data to (scaled) simulation.}
\label{fig:resVarsBaseline}
\end{figure}

For the resolved signature, the following additional criteria are used to define the baseline requirements:
\begin{enumerate}
\item  To control combinatorics in the Higgs boson reconstruction, events must contain either 4 or 5 AK4 jets. Besides the jets associated with the Higgs boson candidates, this requirement
allows for one additional jet in the event, for example, from initial-state radiation.
\item Higgs boson reconstruction is performed as follows. The four jets with the highest \PQb-tag discriminator values are selected and used to form two Higgs boson candidates,
each one decaying into two \PQb jets. There are three pairings among these four jets that could form the two Higgs boson candidates. To select one of them without biasing the mass
distributions towards the true Higgs boson mass, we use the pair with the smallest absolute value of the mass difference, \dmjj, between the two candidate masses $m(\PQb_i\PQb_j)$ and $m(\PQb_k\PQb_l)$,
$\dmjj \equiv \abs{m(\PQb_i\PQb_j)-m(\PQb_k\PQb_l)}$,
and require $\dmjj<40\GeV$.
The average mass \amjj of these two Higgs boson candidates is required to satisfy the loose baseline requirement $\amjj<200\GeV$.  
This approach exploits the fact that signal events contain two particles of identical mass, without using the value of the Higgs boson mass itself. As such, it prevents the sculpting of an artificial peak in the background at $m(\PH)$,
so that a peak observed at that mass would be meaningful, and the combinatorial background in the \amjj spectrum can be investigated in the subsequent analysis.
\item To further suppress background from \ttbar lost-lepton events, we compute the angular separation, \dR, between the \PQb-tagged jets within each Higgs boson candidate. In \ttbar lost-lepton events, one of the \PQt quark decays produces a \PW boson that decays leptonically. The \PQb jet from that \PQt quark decay is then usually paired with a mistagged jet from the other \PQt quark decay in the event to form a Higgs boson candidate. The value of \dR for that candidate will then typically be large. This background is therefore reduced
by requiring $\drmax<2.2$, where \drmax is the larger of the \dR values associated with the two \PH boson candidates.
\item We define a \PQb-counting variable, \nb, which is the number of \PQb-tagged jets but using optimized selections based on tight, medium, and loose categories (see Section~\ref{sec:evtreco}) as follows:
\begin{itemize}
\item[$\nb=2$:] $\nbt=2$ and $\nbm=2$;
\item[$\nb=3$:] $\nbt\geq2$ and $\nbm=3$ and $\nbl=3$;
\item[$\nb=4$:] $\nbt\geq2$ and $\nbm\geq3$ and $\nbl\geq4$.
\end{itemize}
We refer to these categories as 2b, 3b, and 4b, but it should be remembered
that the \PQb tagging thresholds on individual jets become looser as \nb increases.  
The baseline selection requires $\nb\ge 2$. 
\end{enumerate}
Figure~\ref{fig:resVarsBaseline} shows the distributions in \ptmiss, \dmjj, \drmax, \nb, and \amjj, for both data and simulation. The last two quantities shown, 
\nb and \amjj, are used as the basis for the background estimation, while additional binning is performed in the quantities \ptmiss and \drmax
to improve the sensitivity of the analysis.  
In each plot, the baseline selection is applied to all variables other than the one shown, and, for all distributions except \nb,
an additional requirement $\nb\ge3$ is used to select the signal-like parts of the overall analysis region.
To facilitate comparison of the distributions between data and simulation, the distributions from simulation are scaled to match the area of the data, using factors listed in the caption.
The representative signals show the expected peaking in \amjj near the \PH boson mass, as well as toward lower values of \drmax.
The binning indicated by the vertical dotted lines is discussed in Section~\ref{sec:background}. While these plots provide an overall picture of the background shapes
and background composition in the analysis region, they do not give an accurate picture of the sensitivity of the analysis to the signal, because they do not yet reflect
the final binning used to extract the signal. 

The additional baseline criteria for the analysis of the boosted signature are
as follows:
\begin{enumerate}
\item At least two AK8 jets, each with $\pt>300\GeV$. The \pt threshold results in a high probability that all daughter particles from the Higgs boson decay are contained within the jet.  The kinematic merging of the Higgs boson decay products reduces the combinatorial challenge encountered in the resolved signature, and so the number of additional AK8 jets is unrestricted, as is the 
number of AK4 jets. Such additional jets can be produced in the signal processes involving strongly interacting particles, such as the T5HH model described in Section~\ref{sec:introduction}.  They can also arise from initial- or final-state radiation.
\item  The mass \mjet attributed to an AK8 jet is computed by the ``soft drop'' algorithm~\cite{Larkoski:2014wba,Dasgupta:2013ihk}, in which soft wide-angle radiation is recursively removed from a jet.  For signal events, \mjet is expected to show a peak near $m(\PH)$.  For the baseline selection,
the two highest \pt AK8 jets are required to satisfy the loose requirement $60<\mjet<260\GeV$.
\item We define the variable \nH as the number of AK8 jets passing the above criteria that are double-{\PQb} tagged, that is, for which the value of the double-{\PQb} tagging discriminator \Dbb exceeds the loose working point threshold of 0.7.  The value of \nH, 0, 1, or 2, is unrestricted in the baseline selection; it is used for the event classification discussed in Section~\ref{sec:background}.
\end{enumerate} 

\begin{figure}[tbp!]
\includegraphics[width=0.48\textwidth]{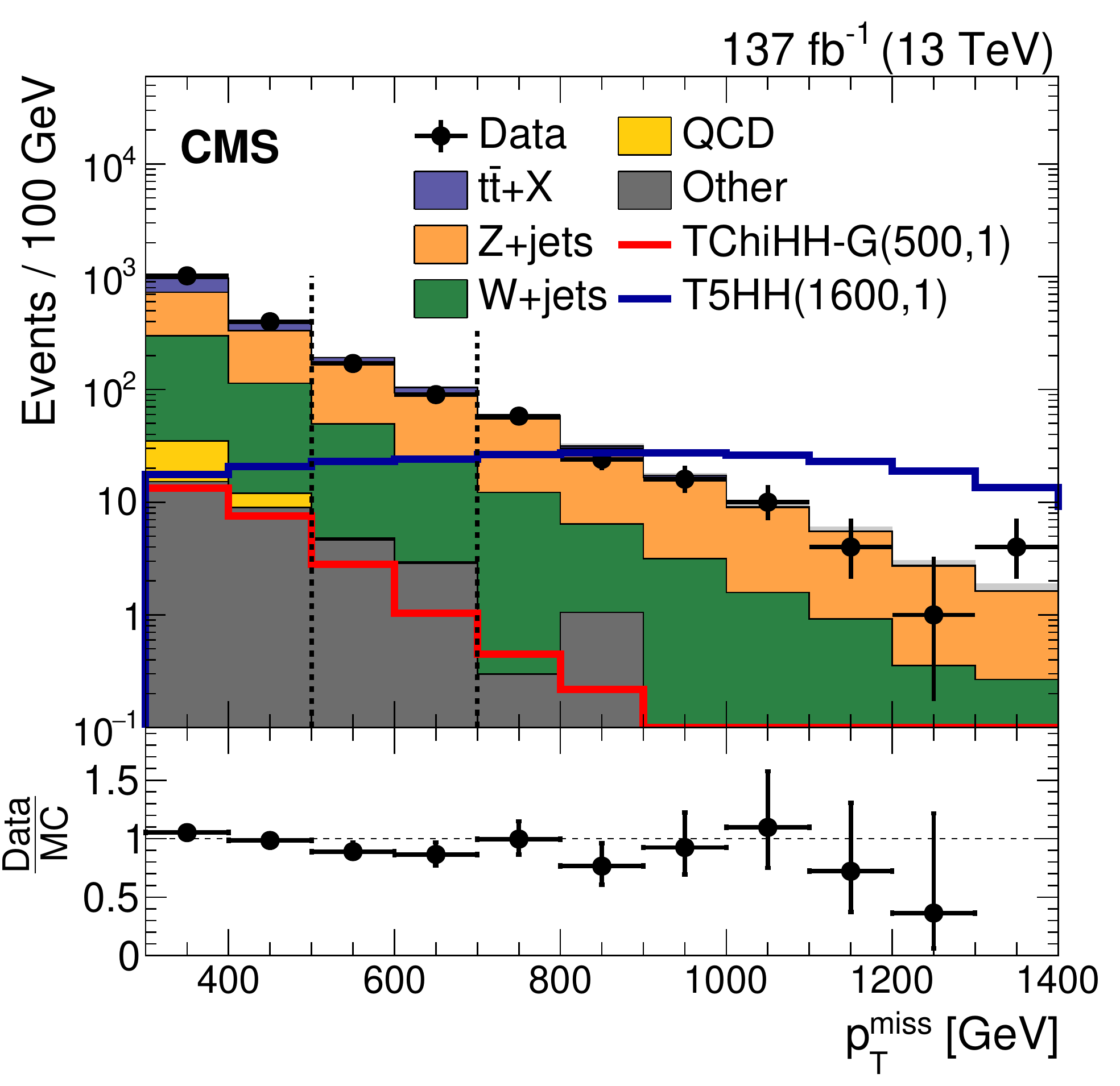}
\includegraphics[width=0.48\textwidth]{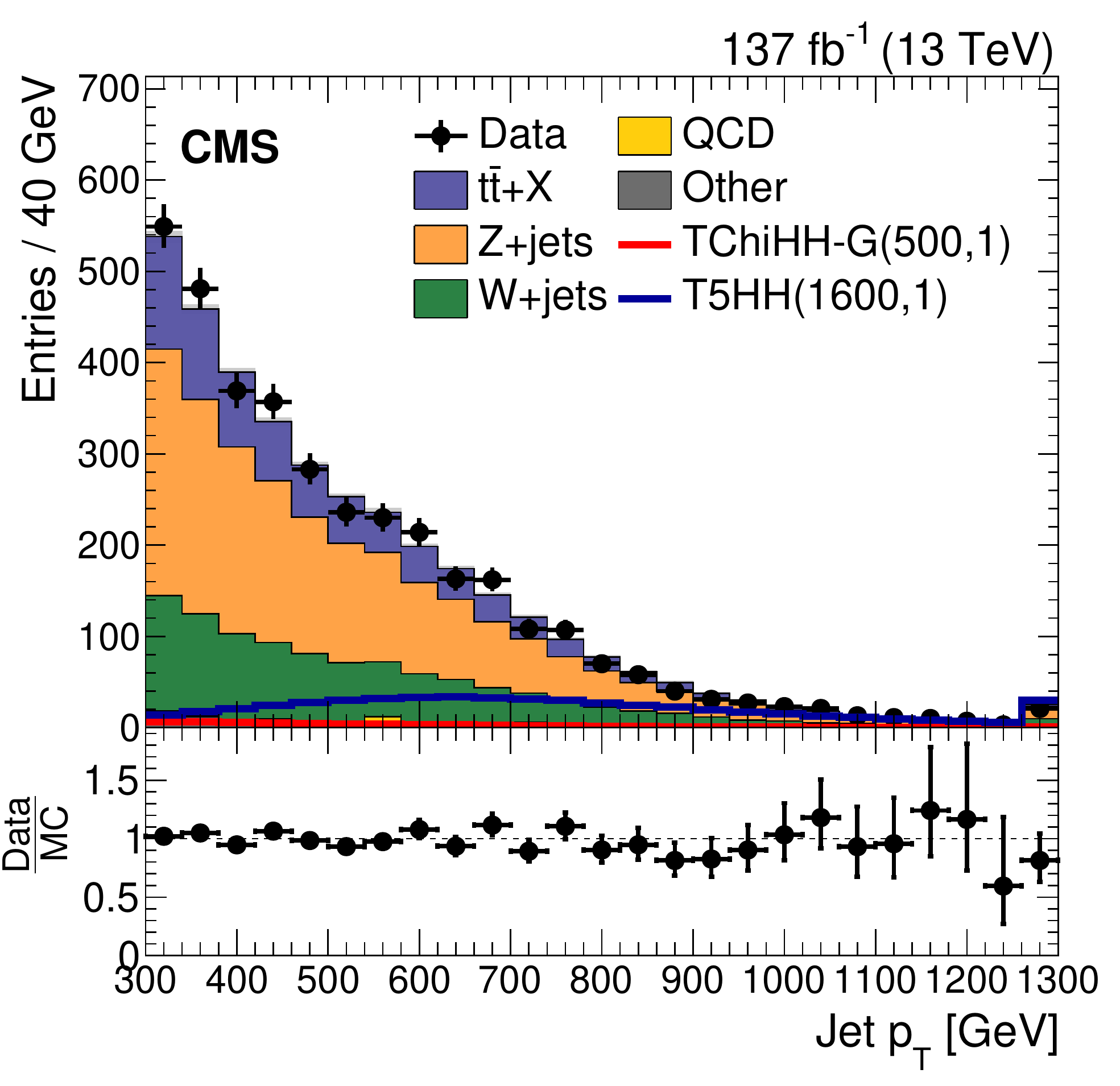}\\
\includegraphics[width=0.48\textwidth]{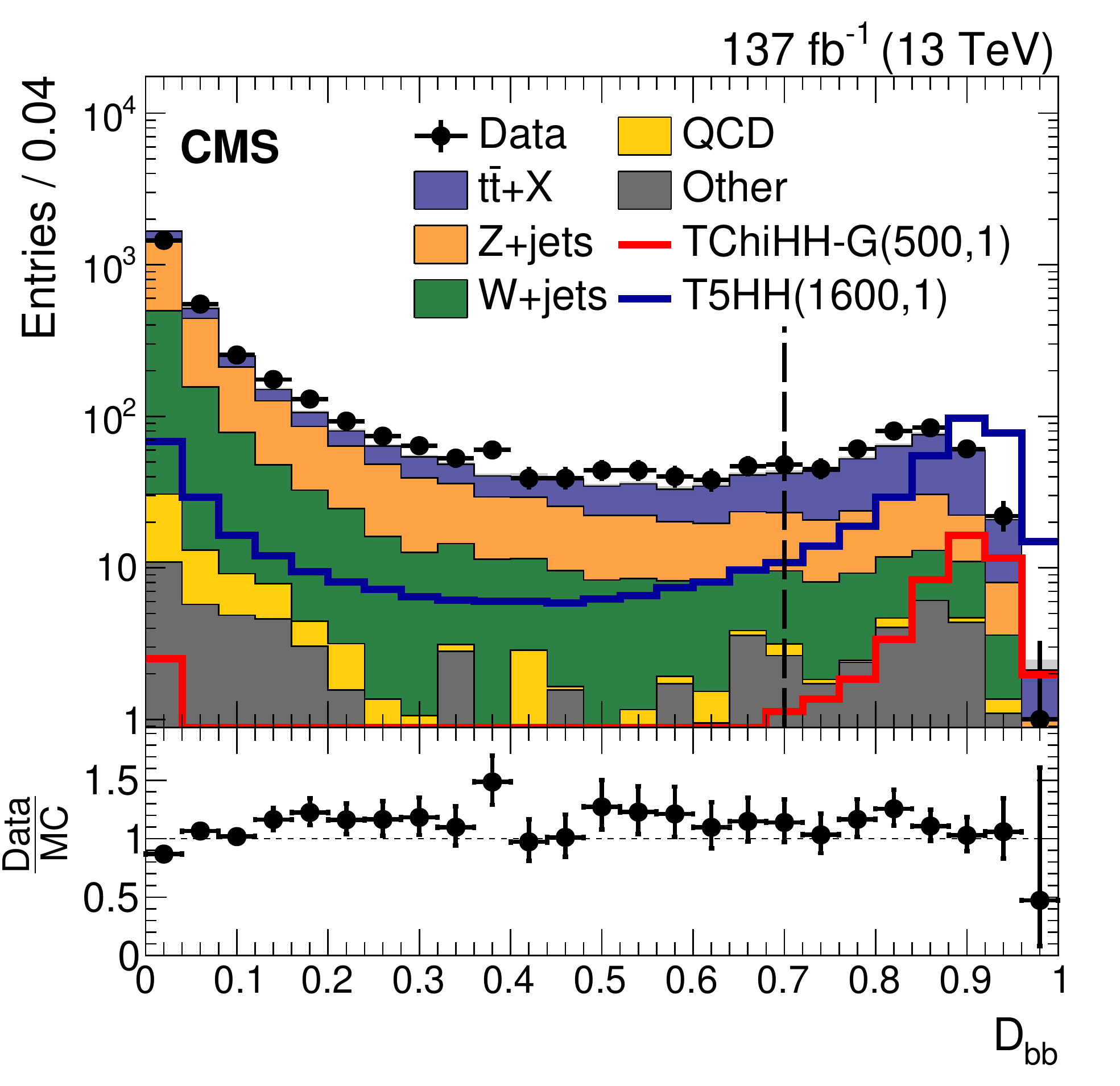}
\includegraphics[width=0.48\textwidth]{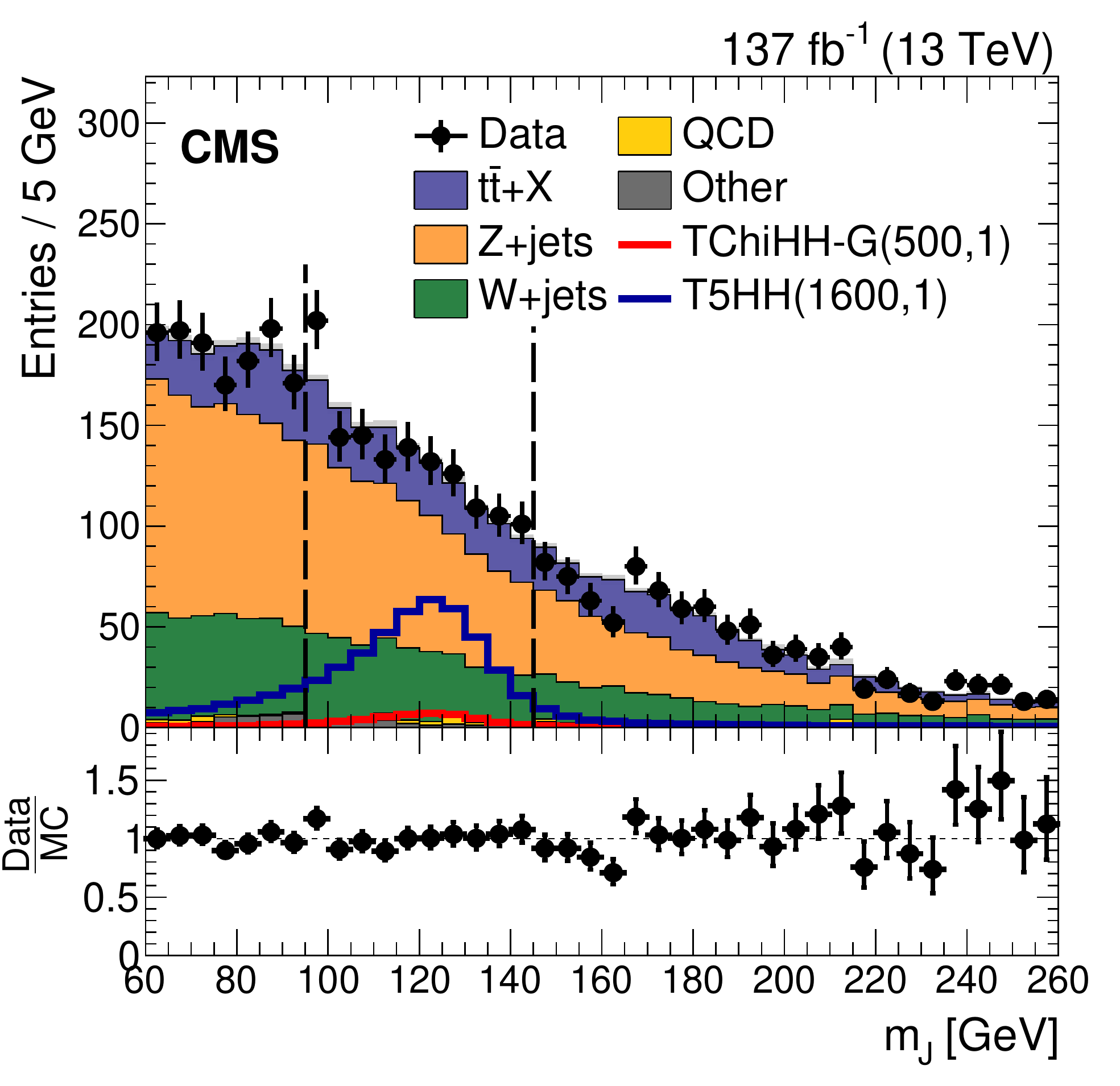}
\caption{Distributions in key analysis variables after the baseline requirements for the boosted signature, except for those on the variable plotted: (upper left) \ptmiss, (upper right) jet \pt, (lower left) \Dbb, and (lower right) \mjet.  Except for the \ptmiss distribution, each plot contains two entries per event, for each of the two \pt-leading AK8 jets.  The data are shown by the black markers with error bars, and SM backgrounds from simulation (scaled by 86\% to match the data integral) by the filled histograms.  Open colored histograms show simulations of the signals $\text{TChiHH-G}(m(\chizone),m(\sGra))$ or $\text{T5HH}(m(\gluino),m(\chizone)\,[\GeVns])$.
The vertical dashed lines show boundaries for the event classifications defined in Section~\ref{sec:background}: (upper left) the \ptmiss binning; (lower left) the \Dbb threshold for identification of a jet as an \Hbb candidate; (lower right) the boundaries of the \PH boson mass window.
The lower panel in each plot shows the ratio of observed to (scaled) simulated yields.
}
\label{fig:boostedBaselineVars}
\end{figure}

Figure~\ref{fig:boostedBaselineVars} shows the distributions for data and simulated event samples
with the boosted signature
in \ptmiss, and in \pt, \Dbb, and \mjet for the two leading AK8 jets.  The baseline requirements for the boosted signature are applied, except for those on the variable shown.  The simulated SM background yields, scaled by the factor noted in the caption, are seen to describe the shape of the data well.  Also shown are examples of SUSY models, which peak around the Higgs boson mass in \mjet, and at high values of \Dbb, for both jets. 

In a small region of parameter space a candidate may be selected by both signatures.  In this case it is assigned to the resolved selection.  Slightly higher expected sensitivity is achieved with this choice than with the alternative.
The efficiency of the combined resolved and boosted signatures for the targeted signal models ranges from 1 to 15\%, depending on the values of the mass parameters in the models.

\section{Background estimation}
\label{sec:background}
\begin{figure}[bt!h]
  \includegraphics[width=\textwidth]{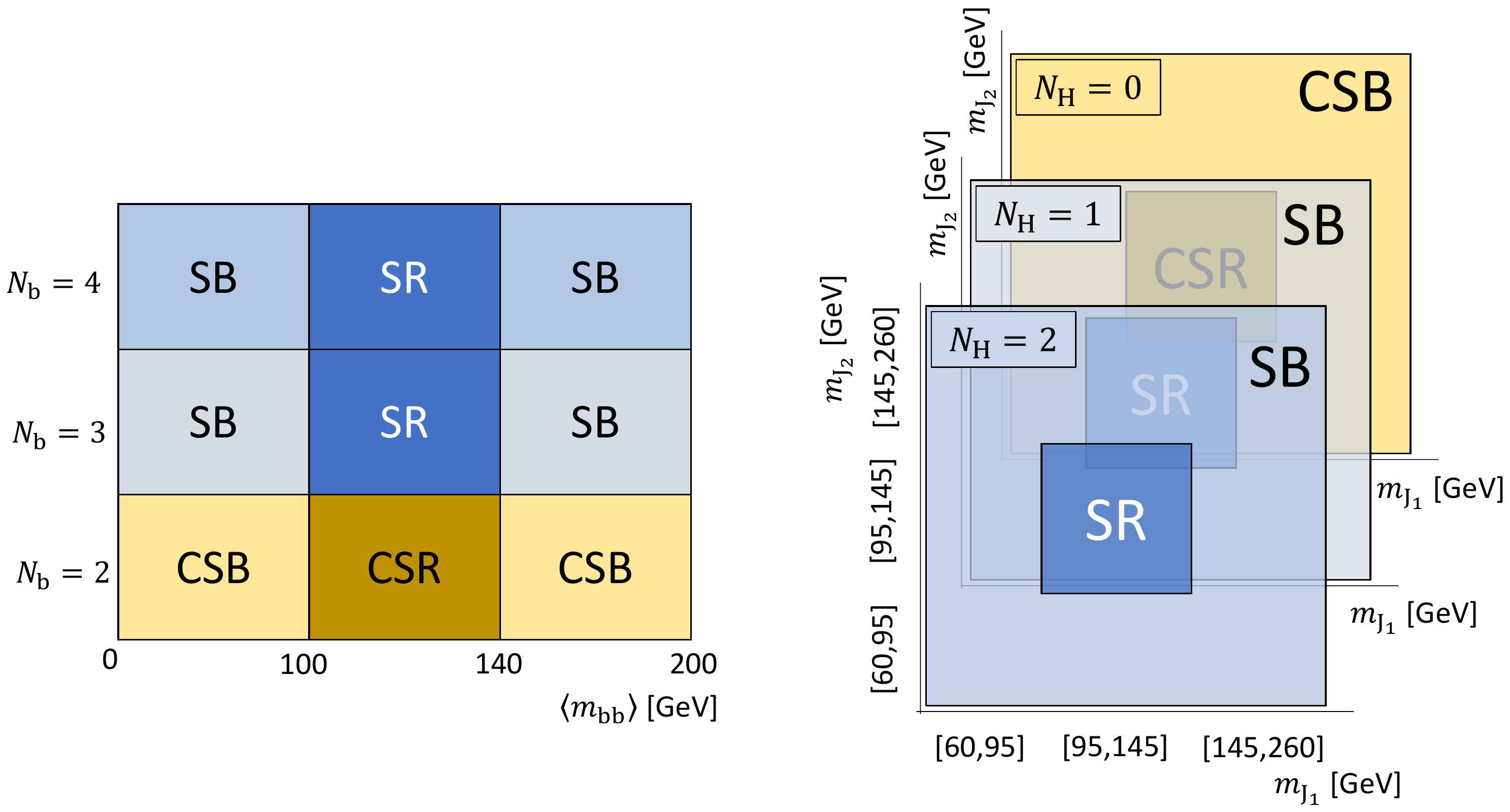}
  \caption{Configuration of the signal and control regions for the (left) resolved and (right) boosted signatures.  The patterns shown are repeated in each of several bins in kinematic or topological variables for improved sensitivity, as discussed in the text.}
  \label{fig:ABCDcartoon}
\end{figure}

After applying the baseline event selections for the resolved and boosted signatures described in Section~\ref{sec:evtsel}, we define
the analysis regions shown in Fig.~\ref{fig:ABCDcartoon}. The background in a given signal region (SR) is estimated using the event yields observed
in three types of neighboring regions: the sideband regions (SB), the control sideband regions (CSB), and the control signal region (CSR). 
For both signatures, these regions are constructed using two types of discriminating variables:   
\begin{enumerate}
\item  Variable(s) characterizing the masses of the \PH boson candidates. The resolved signature uses a single variable, \amjj,
while the boosted signature uses both ${\mjet}_1$ and ${\mjet}_2$, where $\text{J}_1$ and $\text{J}_2$ are the \pt-leading and subleading AK8 jets, respectively.
\item A {\PQb}-counting variable derived from the jet flavor tagging.  For the resolved signature, the variable is \nb, while for the boosted signature, it is \nH.
\end{enumerate}
For the resolved signature, signal events primarily have four \PQb-tagged jets, but they can also populate the region with three \PQb-tagged jets, so the signal regions are defined
by requiring \amjj to be in the Higgs boson mass window and either $\nb=4$ or $\nb=3$. For the boosted signature, signal events primarily have two identified double-\PQb-tagged jets, but they can also populate the region with only one. The signal regions are therefore defined by requiring both ${\mjet}_1$ and ${\mjet}_2$ to be in the Higgs boson mass window and either
$\nH = 1$ or $\nH=2$.  

The background event yield in a signal region is then estimated with the ``ABCD method,'' which uses the measured event yields in three background-dominated
regions to predict the background in a signal region. With the regions defined as shown in Fig.~\ref{fig:ABCDcartoon}, the estimated background yield is
\begin{linenomath}
\begin{equation}
  \nsrp = \kappa\frac{\ncsr}{\ncsb}\nsb,  
  \label{eq:ABCD}
\end{equation}
\end{linenomath}
where \nsb is the observed event yield in the \PH boson mass sideband for events satisfying the same {\PQb}-counting criteria as those applied in the SR, while \ncsr and \ncsb are, respectively, the event yields in regions corresponding to signal and sideband regions with respect to the Higgs candidate mass variable, but in the background regions with respect to the {\PQb}-counting criterion. The coefficient \kapa is a correction factor, typically near unity, that takes into account a potential correlation between the two types of discriminating variables. It is taken from simulation but is validated with separate control regions in data, as discussed in Section~\ref{sec:systematics}. 

For the resolved signature, Fig.~\ref{fig:ABCDcartoon} (left), the variable \nb was defined
in Section~\ref{sec:evtsel}, and
the \PH boson mass window for the SR or CSR is given by $100<\amjj<140\GeV$, with the events both above and below this window constituting the SB or CSB.
The SRs and SBs reside in \PQb tagging regions $\nb=3$ and $\nb=4$, while the $\nb=2$ region contains the CSR and CSB.
To increase the sensitivity to signal, these regions are further sorted into bins in two discriminating variables, 
\drmax (0--1.1 and 1.1--2.2) and \ptmiss (150--200, 200--300, 300--400, and ${>}400\GeV$).
The background determination by the ABCD method is performed separately for each of these eight bins, and for both the $\nb=3$ and $\nb=4$ SRs, for a total of 16 bins.

The corresponding ABCD regions for the boosted signature, Fig.~\ref{fig:ABCDcartoon} (right), are based on the number \nH of double-\PQb-tagged jets (defined in Section~\ref{sec:evtsel}), with events sorted into regions 0H, 1H, and 2H having $\nH=0$, 1, and 2, respectively.  The SRs and SBs lie within the 1H and 2H regions, and CSR and CSB in the 0H region.  The events in the SR and CSR have $95<\mjet<145\GeV$ for both AK8 jets, while those in the SB and CSB have at least one of the jets lying outside that mass window, as illustrated in the right-hand plot of Fig.~\ref{fig:ABCDcartoon}.  The background yield \tbkg is determined for each of the 1H and 2H SRs by the ABCD approach, Eq.~(\ref{eq:ABCD}).  The events in these SRs are further sorted into three \ptmiss bins (300--500, 500--700, and ${>}700\GeV$), for a total of six bins.  The distribution of the predicted background yield among these \ptmiss bins is discussed below.

\begin{figure}[tbh!]
  \includegraphics[width=\textwidth]{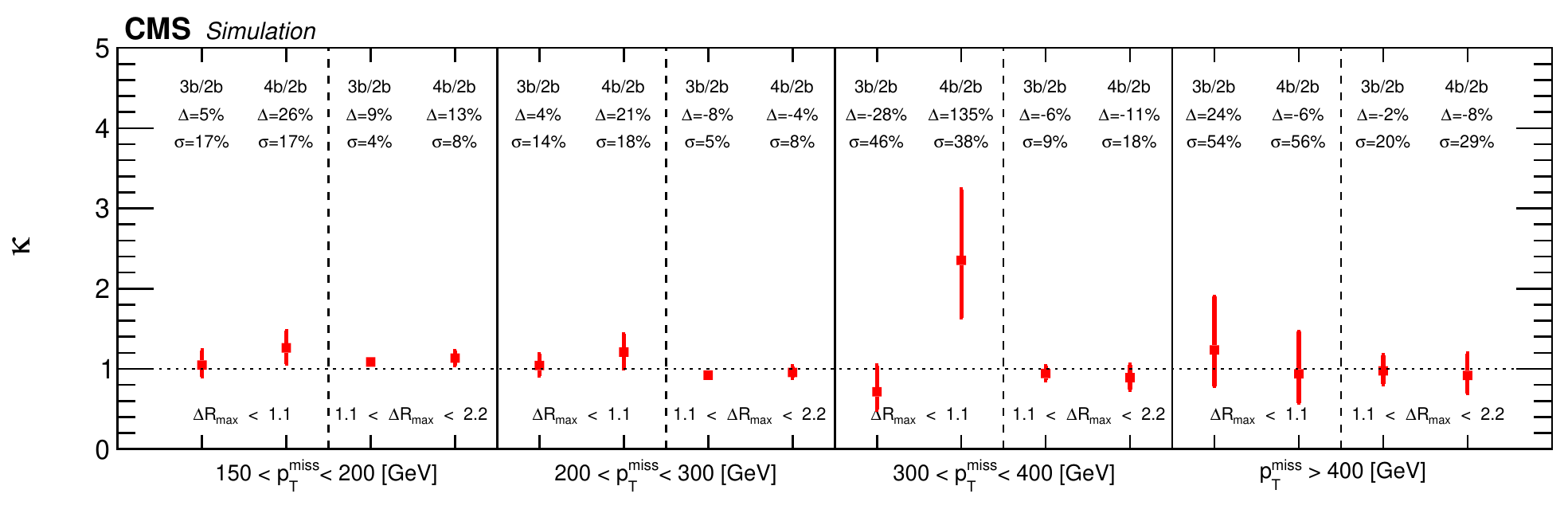}
    \caption{The double ratio $\kappa = (\nsr/\nsb)/(\ncsr/\ncsb)$ from the SM simulation for
    the $\nb=3$ and $\nb=4$
    search samples for each $(\ptmiss, \drmax)$ bin of the resolved signature.
    The value $\Delta$ gives the deviation of \kapa from unity, and $\sigma$
    its relative statistical uncertainty.
  }
  \label{fig:resolvedKappa}
\end{figure}

\begin{figure}[bth!]
\centering
\includegraphics[width=0.41\textwidth]{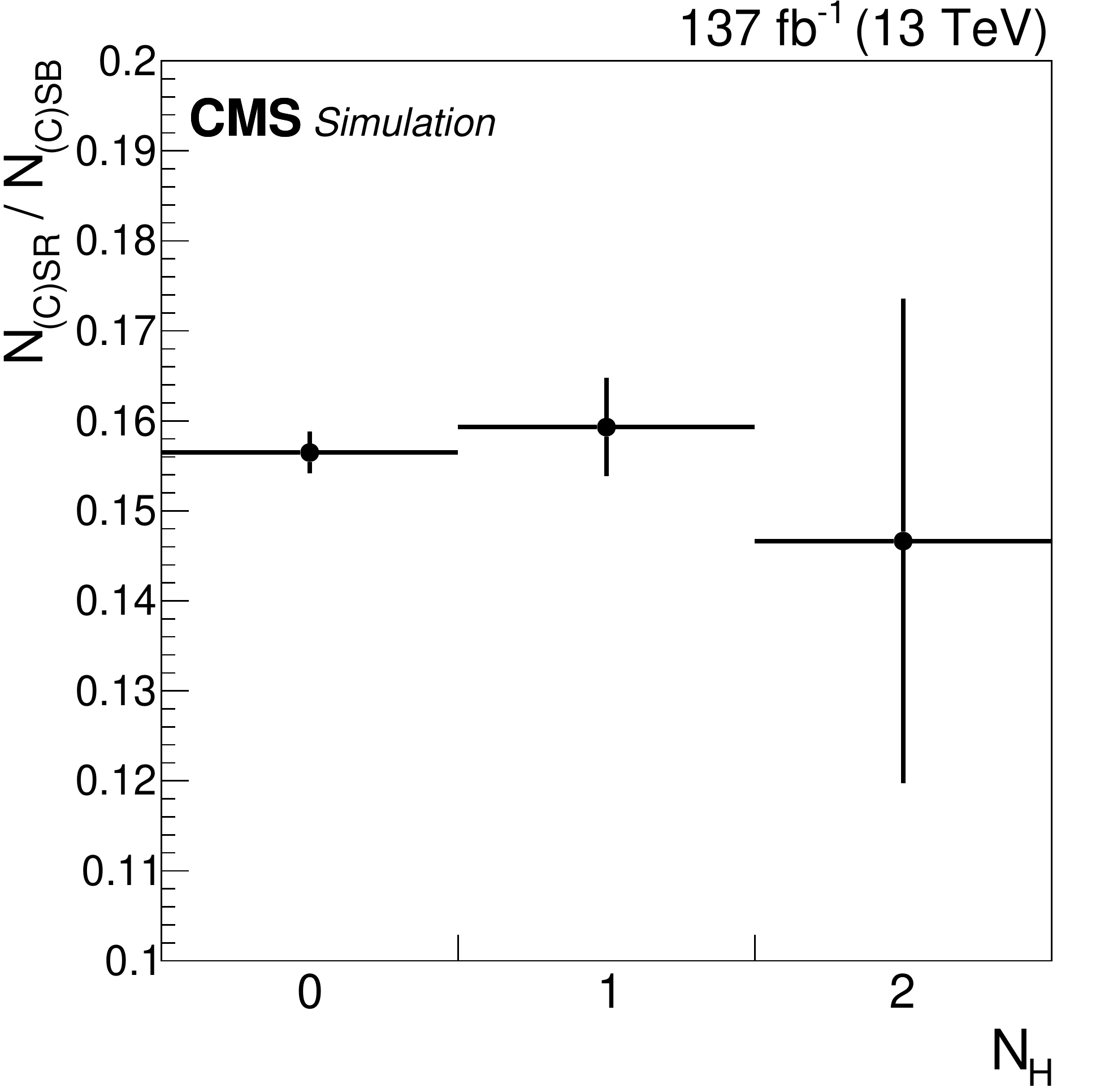}
\includegraphics[width=0.56\textwidth]{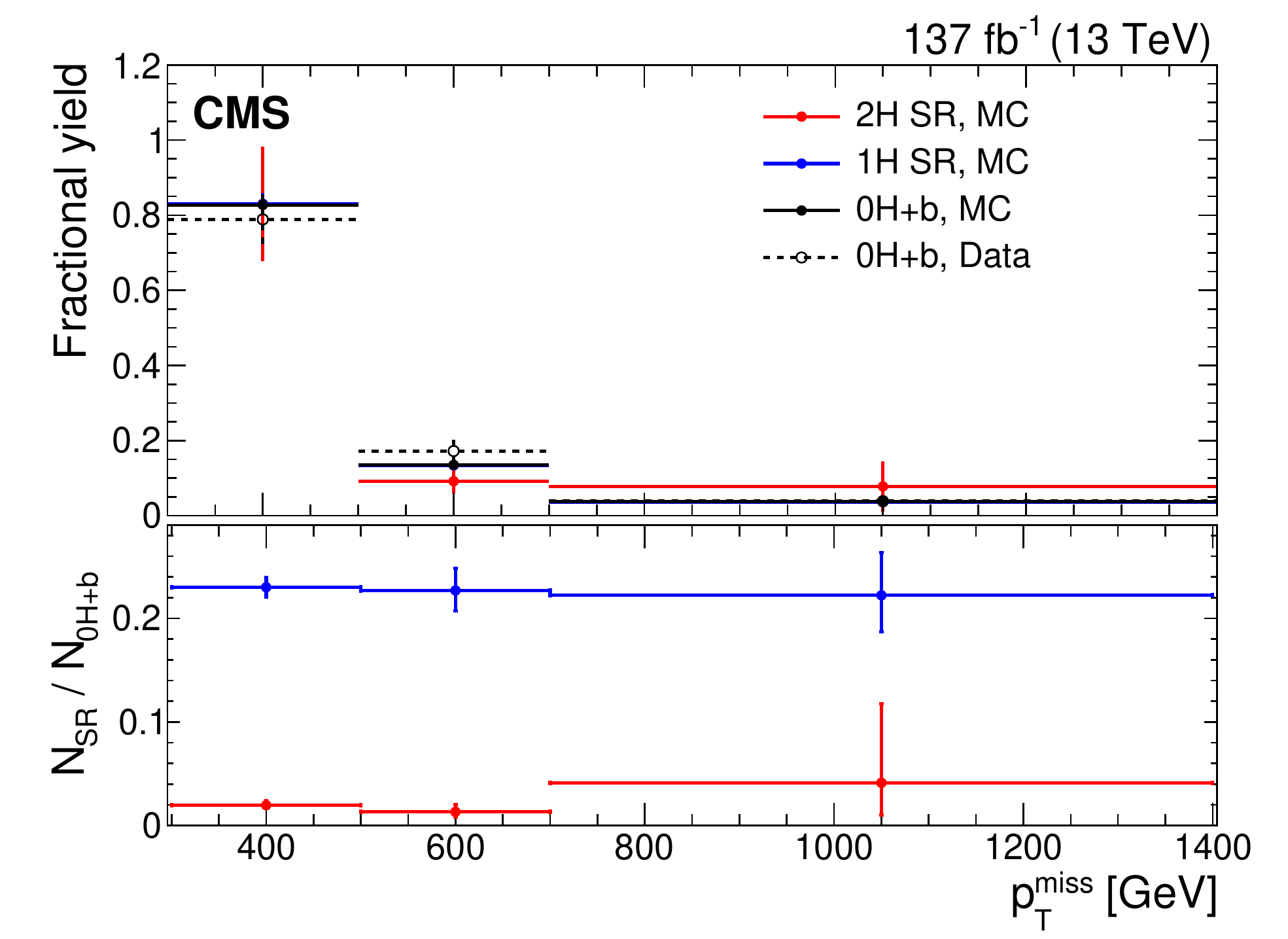}
\caption{%
(left) The yield ratios $\ncsr/\ncsb$ for the 0H region, and $\nsr/\nsb$ for 1H and 2H regions of the boosted signature, from the SM simulation.
(right) The $\ptmiss$ distributions of the 2H and 1H SR yields and the \crb control region yields, for SM simulation (MC, solid points), and for the control region in data (open black points).
In the right-hand plot the upper panel shows the distributions normalized to unit area.  The blue points, and in one bin the open point, are hidden under the solid black points.  The lower panel shows the (unnormalized) ratios to the \crb yields of the 1H (blue) and 2H (red) SR yields, for the simulation.  The error bars show the statistical uncertainties.  
}
\label{boostedSR-SBcomp}
\end{figure}

The factors $\kappa$ in Eq.~(\ref{eq:ABCD}) can differ from unity if the \PH candidate mass variable is correlated with the {\PQb}-counting variable, such that the $\nsr/\nsb$ and $\ncsr/\ncsb$ yield ratios differ. We thus evaluate \kapa as a double ratio
\begin{linenomath}
\begin{equation}
  \kappa = \frac{\nsr/\nsb}{\ncsr/\ncsb},
  \label{eq:kappa}
\end{equation}
\end{linenomath}
taking the yields in this case from simulation. A strength of this method is that many
systematic uncertainties cancel in this double ratio. Figure~\ref{fig:resolvedKappa} shows the values obtained for \kapa in each of the 16 analysis bins of the resolved signature.  These corrections, with their uncertainties, are applied to the background yield predictions, as indicated in Eq.~(\ref{eq:ABCD}).
The determination of the values of \kapa for the boosted signature is illustrated in Fig.~\ref{boostedSR-SBcomp} (left), which shows the
ratios from simulation $\ncsr/\ncsb$ for the 0H, and $\nsr/\nsb$ for the 1H and 2H analysis regions.
Dividing this ratio for the 1H (2H) region by the ratio for the 0H region we obtain the corresponding factor, with statistical uncertainties, $\kapa=1.02\pm0.04$ ($0.94\pm0.17$).  As these are consistent with unity, we set $\kappa=1$ in the calculation of the central values of the background yields, propagating the uncertainties in \kapa to these yields.
For both signatures, systematic contributions to the \kapa uncertainties are evaluated as discussed in Section~\ref{sec:systematics}.

For the boosted signature, the data yields are too small to permit a separate ABCD background calculation in each \ptmiss bin.  Instead we combine 
the \ptmiss-integrated estimate \tbkg for each of the 1H and 2H SRs with the binned \ptmiss distribution \fbkg obtained from a separate data control region.  This control region is a subset (\crb) of the 0H region, comprising both the CSR and CSB regions but satisfying the further requirement of one tight \PQb-tagged AK4 jet in addition to the two AK8 jets.
The \PQb tagging requirement serves to enrich this control region in \ttbar events, so as to approximate the SM content of the 1H and 2H regions.
We thus have for the predicted yield, \nsrpi, in \ptmiss bin $i$
\begin{linenomath}
  \begin{equation}
    \label{eq:bkeq}
          \nsrpi = \kapap_i\,\fbkgi\,\tbkg,
  \end{equation}
\end{linenomath}
where \tbkg is the predicted background yield, as calculated from Eq.~(\ref{eq:ABCD}) without any binning in \ptmiss, and \fbkgi is the fraction of events in the \crb region that fall in \ptmiss bin $i$.
Specifically,
\begin{linenomath}
  \begin{equation}
    \label{eq:fbkg}
    \fbkgi = \frac{N_{{\crb},i}}{N_{{\crb},1}+N_{{\crb},2}+N_{{\crb},3}},
  \end{equation}
\end{linenomath}
and \kapap is a correction factor obtained from simulation, as described below.

The yields $N_{{\crb},1,2,3}$ in the three \ptmiss bins of the \crb region are included in the likelihood function described in Section~\ref{sec:results} so as to propagate the statistical uncertainties of \fbkg to the signal yield.
We use the SM simulation to test the consistency of the \ptmiss distribution in the \crb region with those of the 1H and 2H regions.  This test is shown in
Fig.~\ref{boostedSR-SBcomp} (right).
The ratio distributions in the lower panel are seen to be uniform within the uncertainties of the simulation.
We thus set the central values of the \kapap factors to unity, accounting for their uncertainties as described in Section~\ref{sec:systematics}.
The figure also shows agreement between the simulation and the fractional distribution \fbkg as measured in the data (open points).

\clearpage
\section{Systematic uncertainties}
\label{sec:systematics}
Systematic uncertainties in the background estimates arise principally from the statistical uncertainties in control regions of the data, and from statistical and systematic uncertainties in the simulated samples used to measure \kapa and \kapap.
These quantities can be measured in SM-dominated control samples, providing a check on the procedure and estimates of the associated systematic uncertainties.  Uncertainties in the simulation also affect the signal predictions. 
Sections~\ref{sec:bkgsyst_resolved} and \ref{sec:bkgsyst_boosted} describe the evaluation of the background systematic uncertainties for the resolved and boosted signatures, respectively.  Section~\ref{sec:sigsyst} describes the systematic uncertainties in the signal model predictions.

\subsection{Background systematic uncertainties (resolved signature)}
\label{sec:bkgsyst_resolved}

We validate the \kapa factors
described in Section~\ref{sec:background} and displayed in Fig.~\ref{fig:resolvedKappa},
and assign systematic uncertainties, by comparing their values obtained in simulation with those from three control samples,
each of which is
enriched in a specific class of SM background process.
We compare the value of \kapa computed from the data with the one computed from the corresponding simulation.  Since for each control sample we find the comparison to be consistent across \ptmiss bins, we combine these bins, and take for the systematic uncertainty the larger of the data-simulation difference and the statistical uncertainty in the value obtained from simulation.

The \ttbar systematic uncertainty is measured in a \ttbar-dominated single-lepton control sample selected by replacing the lepton and charged particle vetoes of the search sample with the requirement of one electron or muon with $\pt>30\GeV$, and no additional muons or electrons.  The \dphi requirement is relaxed in the absence of QCD background, to improve the statistical precision, and a requirement $\mt<100\GeV$ is imposed to suppress signal contamination.  We
measure the relative systematic uncertainties in \kapa for the \ttbar background in the $\drmax<1.1$ bins to be
13\% and 19\% for $\nb=3$ and $\nb=4$, respectively; the corresponding numbers for the $1.1<\drmax<2.2$ bins are 2\% and 8\%.

The \vjets systematic uncertainty is measured in a \zjets dominated dilepton control sample in which we select events with two electrons or muons of opposite sign, at least one of which has $\pt>30\GeV$, dilepton mass between 80 and $100\GeV$, and $\ptmiss<50\GeV$.
We then compute the \pt of the dilepton system, $\pt(\PZ)$, and treat this quantity as the true \ptmiss.  That is, $\pt(\PZ)$ serves as a proxy for the \ptmiss that would be measured in an event with a \znn decay.
To increase the sample size, we relax the \dphi requirement and extend the range of ${\pt}({\PZ})$ down to zero.  To reduce \ttbar contamination, we take the CSR and CSB (SR and SB) regions to be defined by the \PQb jet selection $\nbm=0\,(1)$.  We
measure the relative systematic uncertainties in \kapa for the \vjets background to be 16\% and 5\% for the $\drmax<1.1$ and $1.1<\drmax<2.2$ bins, respectively.

The QCD multijet systematic uncertainty is measured in a QCD-enriched sample selected by inverting the \dphi requirement of the search sample. As with the dilepton control sample, the CSR and CSB (SR and SB) are defined by the \PQb jet selection $\nbm=0\,(1)$. We
measure the relative systematic uncertainties in \kapa for the QCD background to be 9\% and 7\% for the $\drmax<1.1$ and $1.1<\drmax<2.2$ bins, respectively.

Finally, these results for each SM process are weighted by the fraction of total background arising from that process in each of the 16 analysis bins to obtain the uncertainties given in Table~\ref{tab:resolvedBgsyst}.
This direct in situ comparison with data of the simulation affecting \kapa covers all of the background-related uncertainties associated with MC modeling. 

The effect of trigger efficiency modeling on \kapa has been evaluated and found to be negligible (${<}0.1\%$).

Table~\ref{tab:resolvedBgsyst} gives a summary of the uncertainties in the \kapa factors for the resolved signature.
These are propagated to the background prediction through Eq.~(\ref{eq:ABCD}).
The table shows that in most of the SRs the largest contribution to the systematic uncertainty in the background prediction is the statistical uncertainty in the simulation used to determine \kapa, which ranges from 4 to 56\%.

\begin{table}[tbp!]
  \renewcommand{\arraystretch}{1.1}
  \setlength{\tabcolsep}{10pt}
  \centering
  \topcaption{Summary of the uncertainties in \kapa for the resolved signature.
    The sources are statistical uncertainties in the determination of \kapa and the systematic uncertainties $\Delta\kappa(\text{data, MC})/\kappa$ derived from the comparison of simulation with data in the single-lepton, dilepton, and low-\dphi control samples, each weighted by the fraction of background arising from the associated process in each search bin.
}
 \label{tab:resolvedBgsyst}
  \begin{tabular}{llrrrrrrrr}
    \hline
\multicolumn{2}{l}{Source} & \multicolumn{7}{c}{Uncertainty [\%]}\\
    &$\met$ [\GeVns{}] & \multicolumn{2}{c}{\quad150--200}& \multicolumn{2}{c}{\quad200--300}& \multicolumn{2}{c}{\quad300--400}& \multicolumn{2}{c}{\quad${>}400$} \\
                    &$\nb$ & 3 &      4 &   3 &      4 &   3 &      4 &   3 &      4 \\
    \hline
    & & & & \multicolumn{4}{c}{$1.1 < \drmax < 2.2$} & & \\
   \multicolumn{2}{l}{\kapa stat. unc.}                &                 4 &       8 &       5 &       8 &       9 &      18 &      20 &      29 \\
 \multirow{3}{*}{$\Delta\kappa(\text{data, MC})/\kappa$} & \ttbar &      2 &       7 &       2 &       7 &       1 &       6 &       1 &       7 \\
                                              & \vjets &            ${<}1$ &  ${<}1$ &       1 &  ${<}1$ &       1 &       1 &       2 &       2 \\
                                              & QCD    &            ${<}1$ &  ${<}1$ &  ${<}1$ &       1 &  ${<}1$ &  ${<}1$ &  ${<}1$ &       1 \\ 
    \multicolumn{2}{l}{\kapa total syst. unc.}         &                 2 &       7 &       2 &       7 &       1 &       6 &       2 &       7 \\
    [\cmsTabSkip]
    & & & & \multicolumn{4}{c}{$\drmax < 1.1$} & & \\
   \multicolumn{2}{l}{\kapa stat. unc.}                &                  17 &      17 &      14 &      18 &      46 &      38 &      54 &      56 \\
  \multirow{3}{*}{$\Delta\kappa(\text{data, MC})/\kappa$} & \ttbar &      11 &      18 &      10 &      17 &       6 &      13 &       6 &       9 \\
                                              & \vjets &                   1 &       1 &       4 &       5 &      10 &       6 &       6 &      10 \\
                                              &   QCD  &                   1 &  ${<}1$ &  ${<}1$ &       2 &  ${<}1$ &  ${<}1$ &  ${<}1$ &  ${<}1$ \\
    \multicolumn{2}{l}{\kapa total syst. unc.}         &                  11 &      18 &      11 &      18 &      12 &      14 &       8 &      13 \\
   \hline
  \end{tabular}
\end{table}

\subsection{Background systematic uncertainties (boosted signature)}
\label{sec:bkgsyst_boosted}

Systematic uncertainties in the background yields are evaluated for each of the two steps (Eq.~(\ref{eq:ABCD}) and Eq.~(\ref{eq:bkeq})) in the background determination for the boosted signature, and are summarized in Table~\ref{tab:boostedBgsyst}.
For the first step, extraction of the \ptmiss-integrated yields \tbkg (Eq.~(\ref{eq:bkeq})), the systematic uncertainties are obtained by the method described above for the resolved signature. We construct a \ttbar and \wjets dominated single-lepton control sample by applying the baseline selection for the boosted-signature analysis region, but releasing the lepton and isolated track vetoes and the \dphi requirement, and instead requiring exactly one electron or muon with $\pt>30\GeV$ and $\mt<100\GeV$.  Taking the larger of the data-simulation difference and the statistical uncertainty in that comparison, we find systematic uncertainties of 9 and 13\% for the 1H and 2H SRs, respectively.
The statistical uncertainty from the calculation of \kapa in simulation is assigned as a further systematic uncertainty.

The second step in the background prediction for the boosted signature is the application of the \ptmiss distribution \fbkg measured with the \crb control region.
Systematic uncertainties arising from possible differences in the \ptmiss shape between the 0H+b control region and the SRs are contained in the correction factors \kapap in Eq.~(\ref{eq:bkeq}).  These uncertainties are measured as the deviations,
from the \ptmiss-averaged value, of a linear fit to the data in the lower panel of Fig.~\ref{boostedSR-SBcomp} (right); the values are 2, 3, and 9\% (13, 19, and 63\%) for the three \ptmiss bins in the 1H (2H) SR.
We obtain an additional uncertainty in the \ptmiss distribution, attributable to the background composition of the simulation, 
by varying the \zjets component by factors of two and one half.  Taking the larger effect of these on \kapap yields the uncertainties 3, 7, and 32\%.

We assign a systematic uncertainty covering the effects of MC mismodeling in both \tbkg and \fbkg from the following sources: {\PQb} and double-{\PQb} tagging, jet energy scale, jet energy resolution, ISR corrections, and renormalization (\rscale) and factorization (\fscale) scales. The effect of each of these contributions ranges from 0 to 5\%.  The total systematic uncertainty is calculated by taking the largest uncertainty for each source, and adding those in quadrature. The individual sources are described in Section~\ref{sec:evtreco} and their uncertainties in Section~\ref{sec:sigsyst}.

As seen in Table~\ref{tab:boostedBgsyst},
the leading contributions are statistical uncertainties in the SB, CSR, CSB, and \crb yields in the data, and in those of the MC samples that are used to measure $\kappa^\prime$.

\begin{table}[tbp!]
\centering
\topcaption{Summary of systematic uncertainties in the background prediction for the boosted signature.
Values in cells spanning the 1H and 2H columns
enter the yield calculations through the factor \fbkg, which is common to the 1H and 2H SRs.
The values from the ABCD measurement with the \pt-integrated sample appear in the rows labeled $\ptmiss>300\GeV$.  The row labeled $\Delta(\text{data, MC})$ gives the contribution derived from the comparison of simulation with data in the one-lepton control sample.}
\begin{tabular}{lcrr} \hline
  Source              & \ptmiss [\GeVns{}] & \multicolumn{2}{c}{Uncertainty [\%]} \\
                      &                & 1H          & 2H       \\ \hline
  SB, CSR, CSB statistics
                      & ${>}300$                  & 10 & 19 \\
  \multirow{3}{*}{\crb statistics}
                      & $[300,500]$             & \multicolumn{2}{c}{8}  \\
                      & $[500,700]$             & \multicolumn{2}{c}{18} \\
                      & ${>}700$                  & \multicolumn{2}{c}{38} \\
  [\cmsTabSkip]\kapa stat. unc.
                      & ${>}300$                  &  4 & 17 \\
  [\cmsTabSkip]
  \multirow{3}{*}{\ptmiss shape closure, \kapap}
                      & $[300,500]$             &  2 & 13 \\
                      & $[500,700]$             &  3 & 19 \\
                      & ${>}700$                  &  9 & 63 \\
  [\cmsTabSkip]$\Delta(\text{data, MC})$
                      & ${>}300$                  &  9 & 13 \\
  [\cmsTabSkip]
  \multirow{3}{*}{Bkg. composition}
                      & $[300,500]$             & \multicolumn{2}{c}{3} \\
                      & $[500,700]$             & \multicolumn{2}{c}{7} \\
                      & ${>}700$                  & \multicolumn{2}{c}{32} \\
  [\cmsTabSkip]
  \multirow{4}{*}{MC modeling}
                      & ${>}300$                  &  4 &  6 \\
                      & $[300,500]$             & \multicolumn{2}{c}{1} \\
                      & $[500,700]$             & \multicolumn{2}{c}{2} \\
                      & ${>}700$                  & \multicolumn{2}{c}{5} \\
\hline
\end{tabular}
\label{tab:boostedBgsyst}
\end{table}

\subsection{Signal systematic uncertainties}
\label{sec:sigsyst}
\begin{table}[thb]
\centering
\topcaption{
Sources of systematic uncertainties and their typical impact on the signal yields obtained from simulation. The range is reported as the median 68\% confidence interval among all signal regions for every signal mass
point considered.  Entries reported as 0 correspond to values smaller than 0.5\%.
}
\begin{tabular}{lcc}
\hline
\multirow{2}{*}{Source}  & \multicolumn{2}{c}{Relative uncertainty [\%]} \\
                         & Resolved      & Boosted  \\
\hline
MC sample size                           & 0--18 & 1--15 \\
ISR modeling                             & 0--2  & 0--18 \\
Renormalization and factorization scales $\rscale$ and $\fscale$ & 0--2 & 0--7  \\
Pileup corrections                       & 0--3 & 0--9 \\
Integrated luminosity                    & \multicolumn{2}{c}{1.6} \\
Jet energy scale                         & 0--7 & 0--12 \\
Jet energy resolution                    & 0--7 & 0--7 \\
Isolated track veto                      & 2--9 & 1--8 \\
Trigger efficiency                       & 1--12 & 0--4 \\
\mjet resolution                         & \NA & 0--9 \\
\PQb tagging efficiency                  & 2--6 & \NA\\
\PQb mistagging                          & 0--1 & \NA\\
$\PQb\PQb$ tagging efficiency            & \NA & 6--15 \\
\multicolumn{3}{c}{Uncertainties attributable to the fast simulation} \\
Jet quality requirements                 & \multicolumn{2}{c}{1} \\
\ptmiss modeling                         & 0--14 & 0--12 \\
\mjet resolution                         & \NA & 2--4 \\
\PQb tagging efficiency                  & 0--1 & \NA\\
\PQb mistagging                          & 0--1 & \NA\\
$\PQb\PQb$ tagging efficiency            & \NA & 0--1 \\
\hline
\end{tabular}
\label{tab:signalSyst}
\end{table}

Systematic uncertainties affecting the signal yields from simulation are listed in Table~\ref{tab:signalSyst}.
The systematic uncertainty in the ISR correction of the simulation of electroweak production models is evaluated by varying the correction factors described in Section~\ref{sec:evtreco} (18--22\%) by 100\% of their nominal value. For strong production models, the systematic uncertainty is evaluated by varying the correction factors (4--26\%) by 50\% of their nominal value.
To evaluate the uncertainty associated with \rscale and \fscale, we vary one or both by factors of 2.0 and 0.5 and take the combination (excluding the case of opposite variations of the two) that gives the largest effect on the yield~\cite{Kalogeropoulos:2018cke,Catani:2003zt,Cacciari:2003fi}.
The uncertainty associated with the pileup reweighting is evaluated by varying the value of the total inelastic cross section by 5\%~\cite{Sirunyan:2018nqx}.
The systematic uncertainty in the determination of the integrated luminosity varies between 1.2 and 2.5\%~\cite{CMS:2021xjt,CMS-PAS-LUM-17-004,CMS-PAS-LUM-18-002}, depending on the year of data collection; the total integrated luminosity for the three years has an uncertainty of 1.6\%.
The uncertainties related to the jet energy scale and jet energy resolution are calculated by varying jet properties in bins of \pt and~$\eta$, according to the uncertainties in the jet energy corrections.
The uncertainty in the \mjet resolution is derived by comparing the mean and width of the \PW boson peak in the \mjet spectrum between data and simulation in a \ttbar-enriched sample.
The efficiency of the isolated-track veto for signal events, as measured in simulation, is greater than 90\% except at the lowest NLSP masses.  Differences with respect to simulation measured in data control samples lie in the range 20--30\% of the inefficiency.  We assign a systematic uncertainty of 50\% of the inefficiency to account for the effect of this veto on the signal yields.

The uncertainty in the trigger efficiency modeling is derived from the statistical uncertainty, and variations in the kinematic selections, for the trigger measurement sample, and from variation in the independent trigger used to measure the efficiency.
This uncertainty is applied to each SR bin for each signal model based on its distribution in \ptmiss and \HT in that bin. For the resolved signature, typical values range from 1--13\%, though some bins have larger values.
For $\ptmiss>300\GeV$ the largest uncertainty is 4\%.

Systematic uncertainties in the {\PQb} tagging are obtained from the measurements in data of the {\PQb} and double-{\PQb} tagging efficiencies and mistag rates, described in Section~\ref{sec:evtreco}.  These uncertainties lie in the ranges 2--7\% for jets tagged as {\PQb} quarks and 6--15\% for the double-{\PQb}-tagged jets.

Systematic uncertainties associated with use of the fast simulation are taken to be 100\% of the differences with respect to the {\GEANTfour} simulation.  This amounts to 1\% for the jet quality requirements, 2--4\% for \mjet resolution, and 1\% for {\PQb} and double-{\PQb} tagging.

\section{Results and interpretation}
\label{sec:results}
\begin{figure}[tbp!]
\centering
\includegraphics[width=0.49\textwidth]{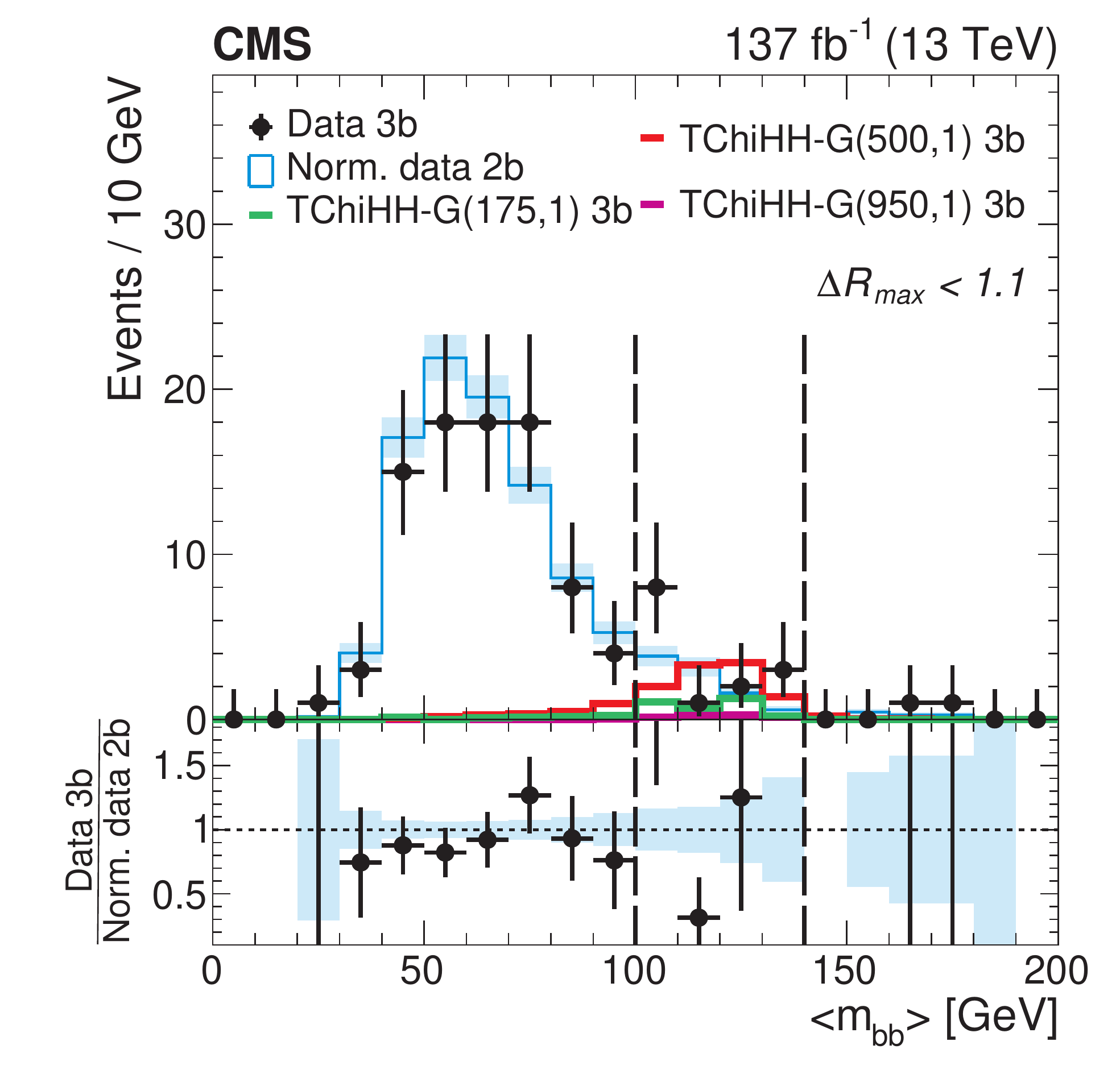}
\includegraphics[width=0.49\textwidth]{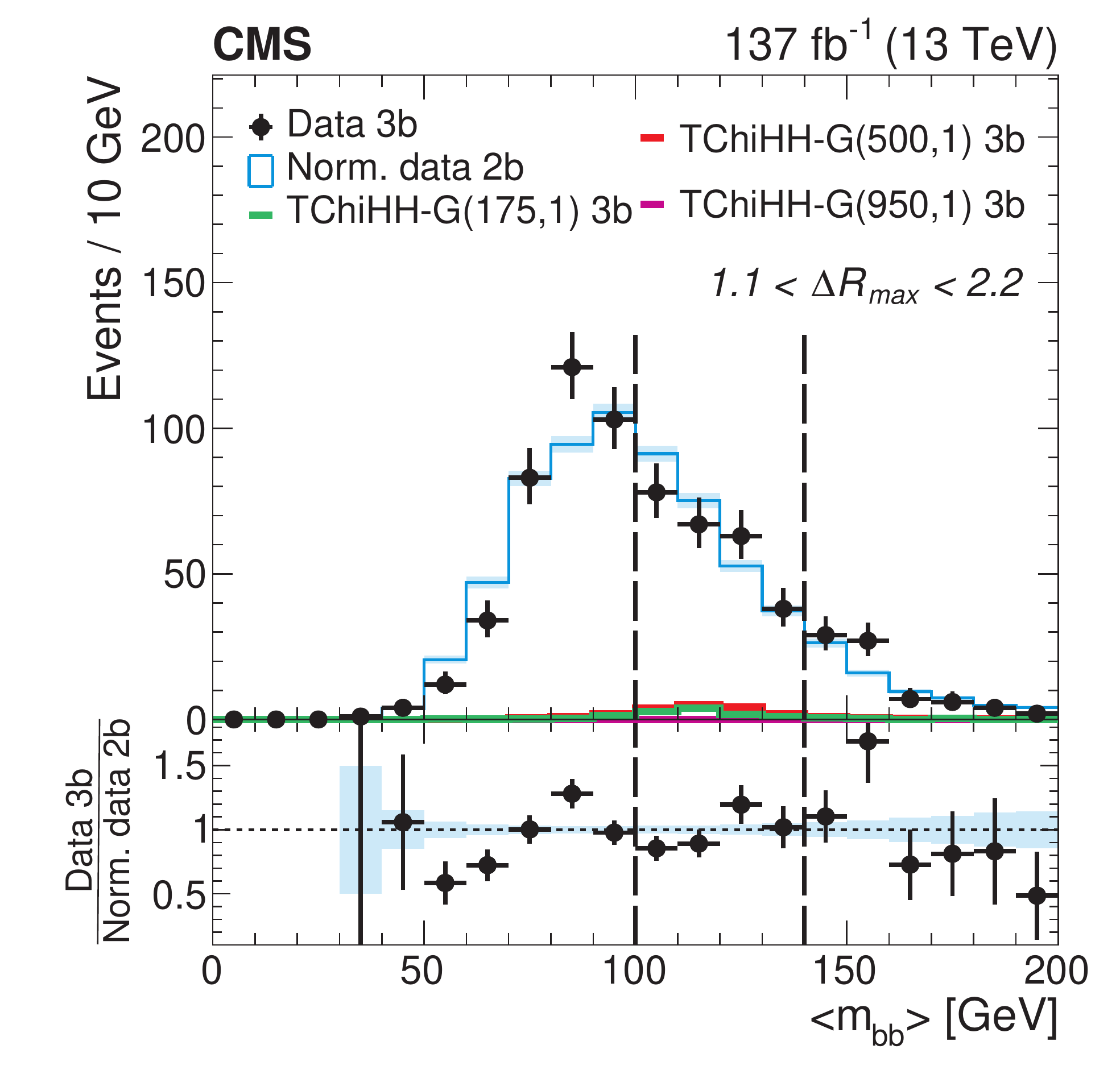}
\includegraphics[width=0.49\textwidth]{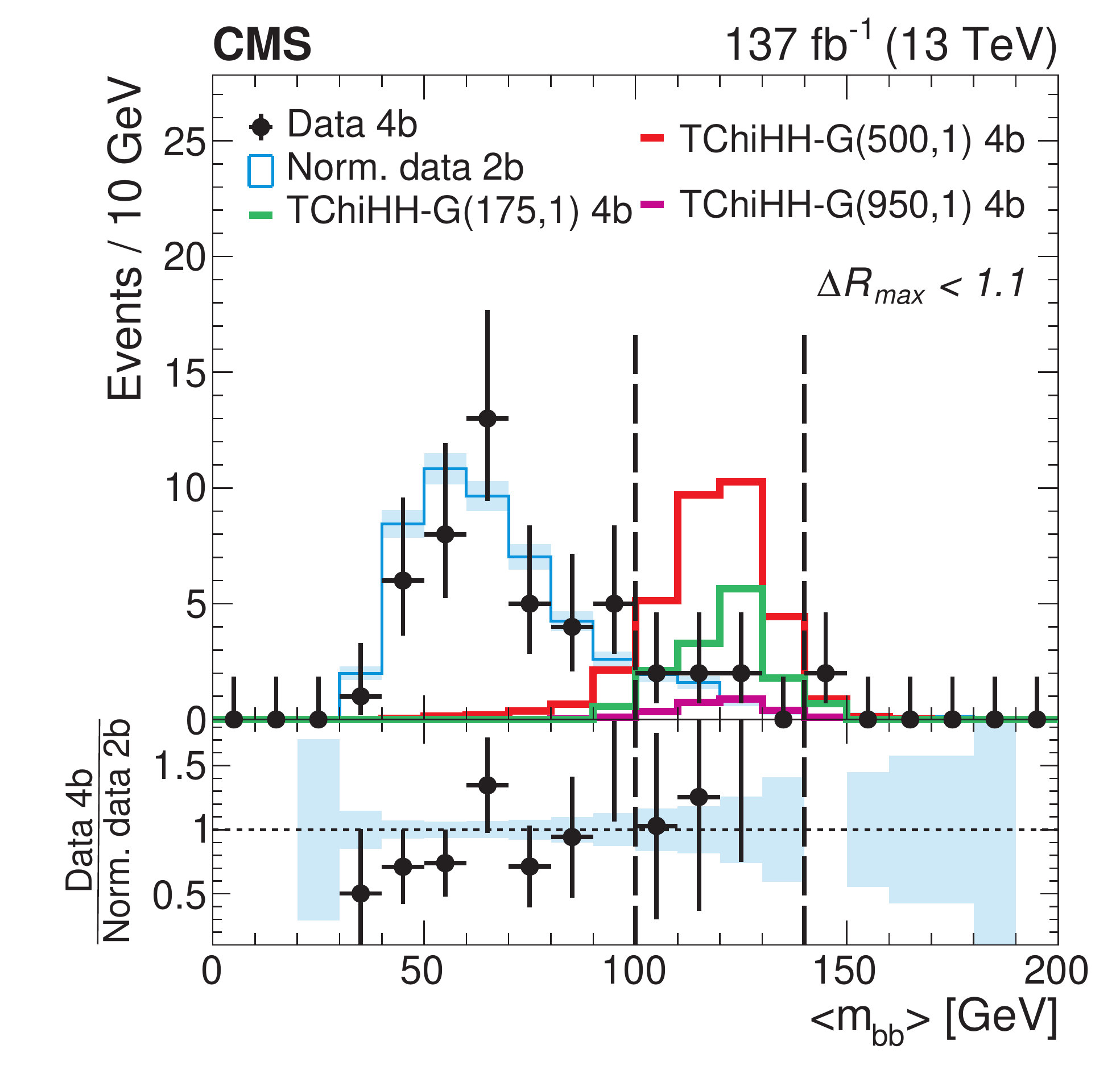}
\includegraphics[width=0.49\textwidth]{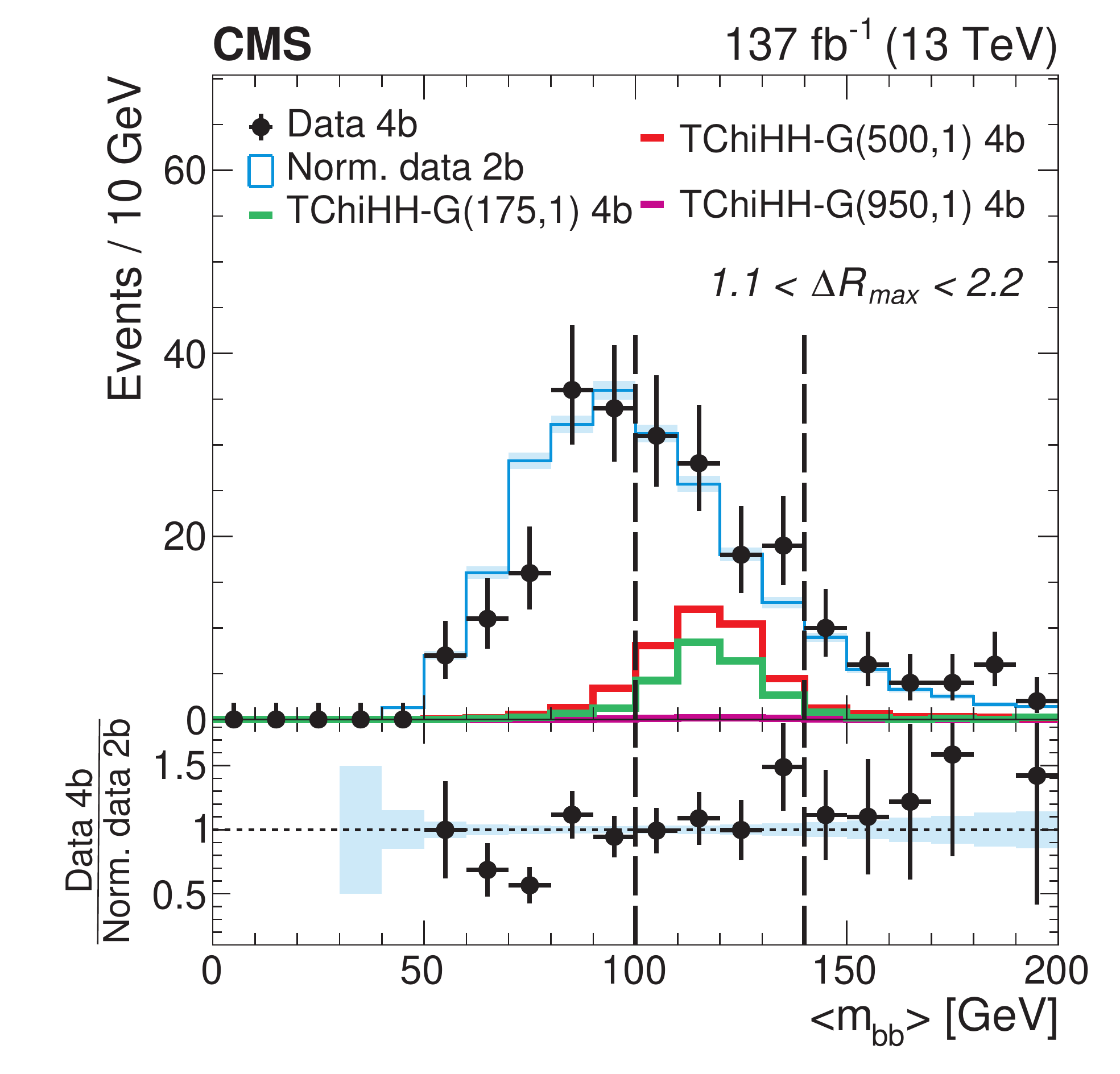}
\caption{Distributions in \amjj for the (upper row) $\nb=3$ and (lower row) $\nb=4$ data, denoted by black markers, with error bars indicating the statistical uncertainties.
The left-hand set of plots corresponds to $\drmax<1.1$, while the right-hand set corresponds to $1.1<\drmax<2.2$, in both cases integrated over \ptmiss.
The overlaid cyan histograms show the corresponding distributions of the $\nb=2$ data, reweighted with the appropriate \kapa values and scaled in area to the $\nb=3$ or $\nb=4$ distribution and with statistical uncertainties indicated by the cyan shading (absent in the 140--150 bin because of a vanishing $\nb=2$ yield there).
The ratio of these distributions appears in the lower panel.  The red, green, and violet histograms show simulations of representative signals, denoted $\text{TChiHH-G}(m(\chizone),m(\sGra)\,[\GeVns])$ in the legends.
}
\label{fig:resolvedResults:drmaxPlaneResults}
\end{figure}

\begin{figure}[tbp!]
\centering
\includegraphics[width=0.49\textwidth]{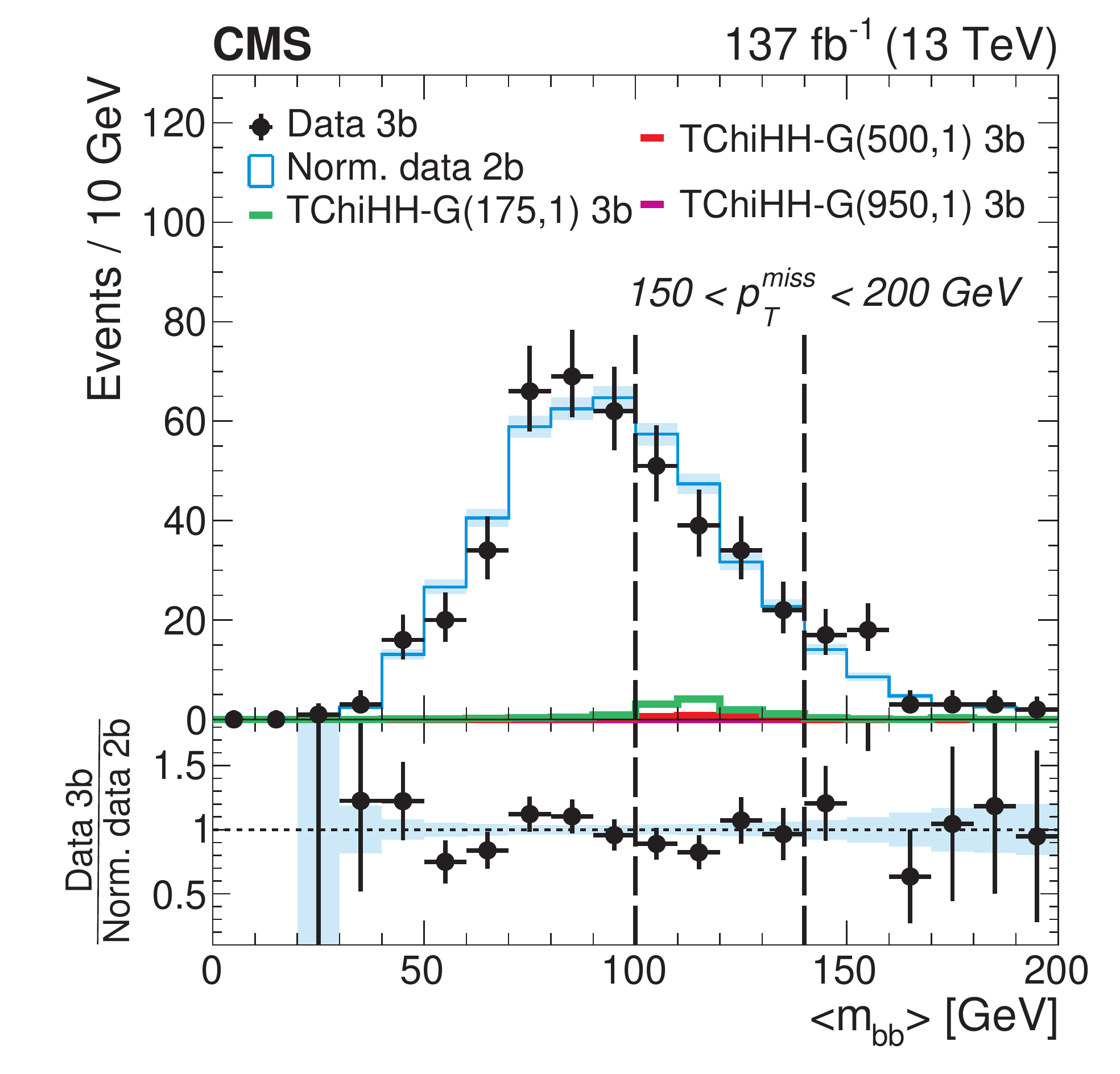}
\includegraphics[width=0.49\textwidth]{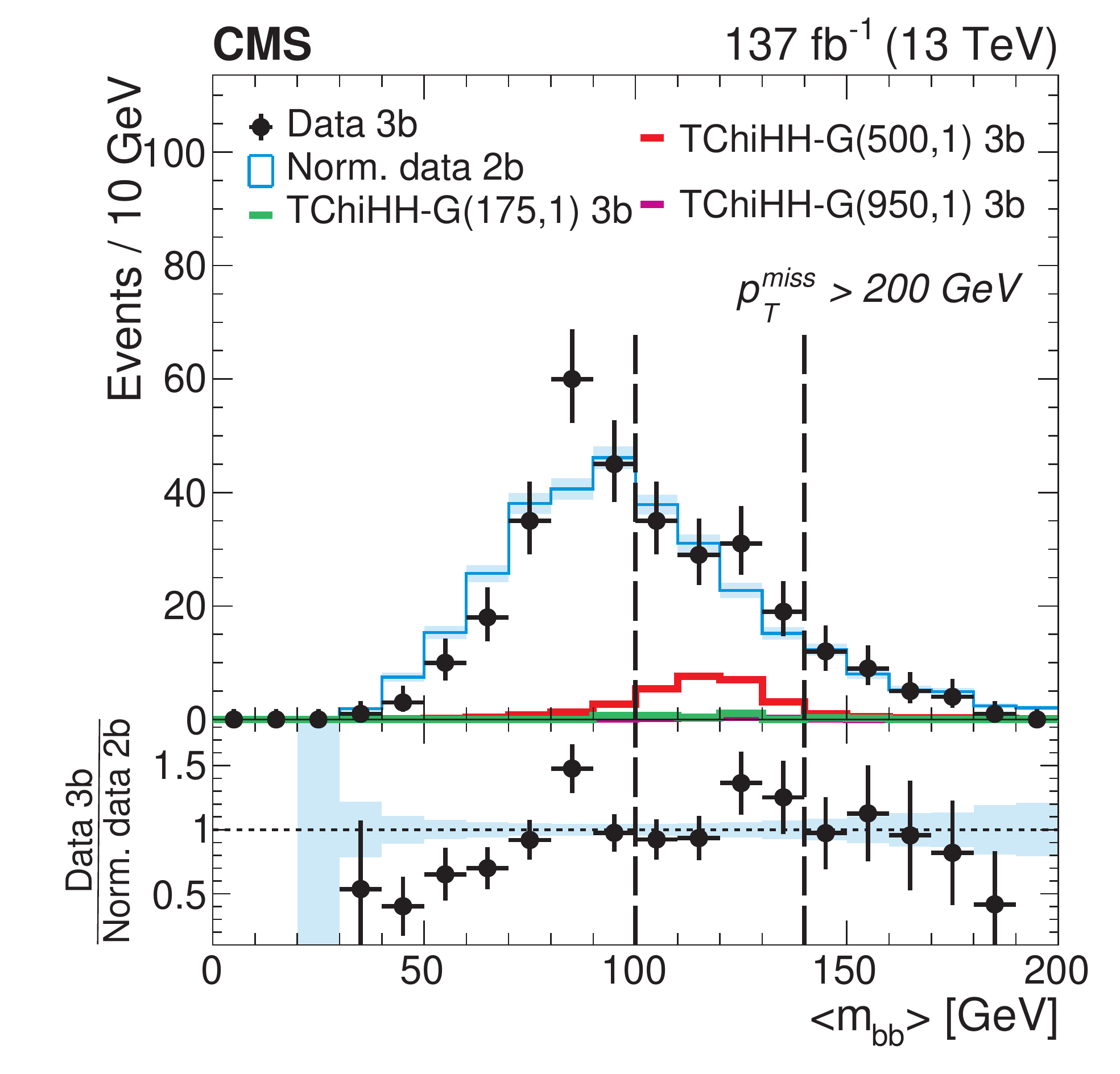}
\includegraphics[width=0.49\textwidth]{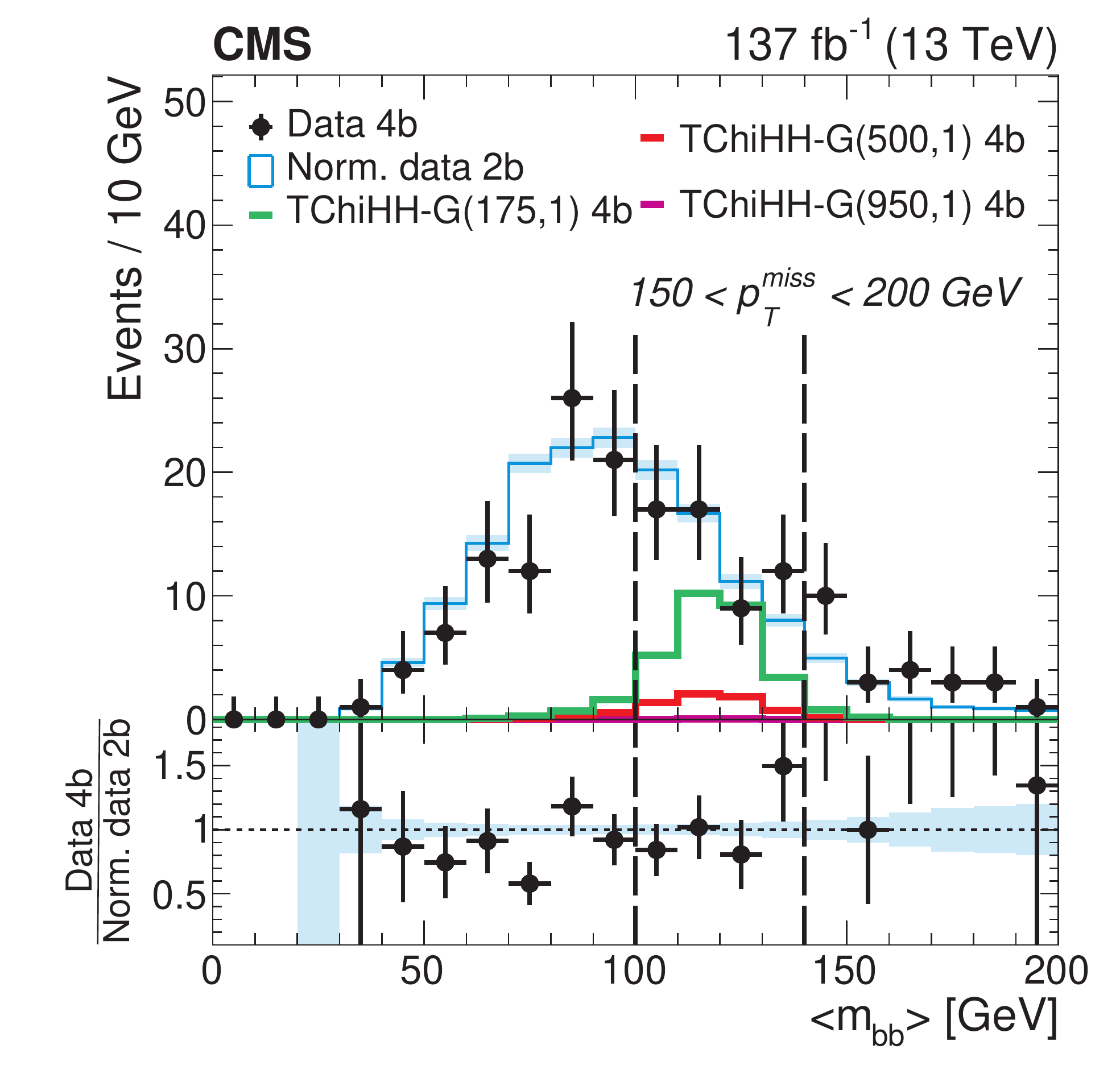}
\includegraphics[width=0.49\textwidth]{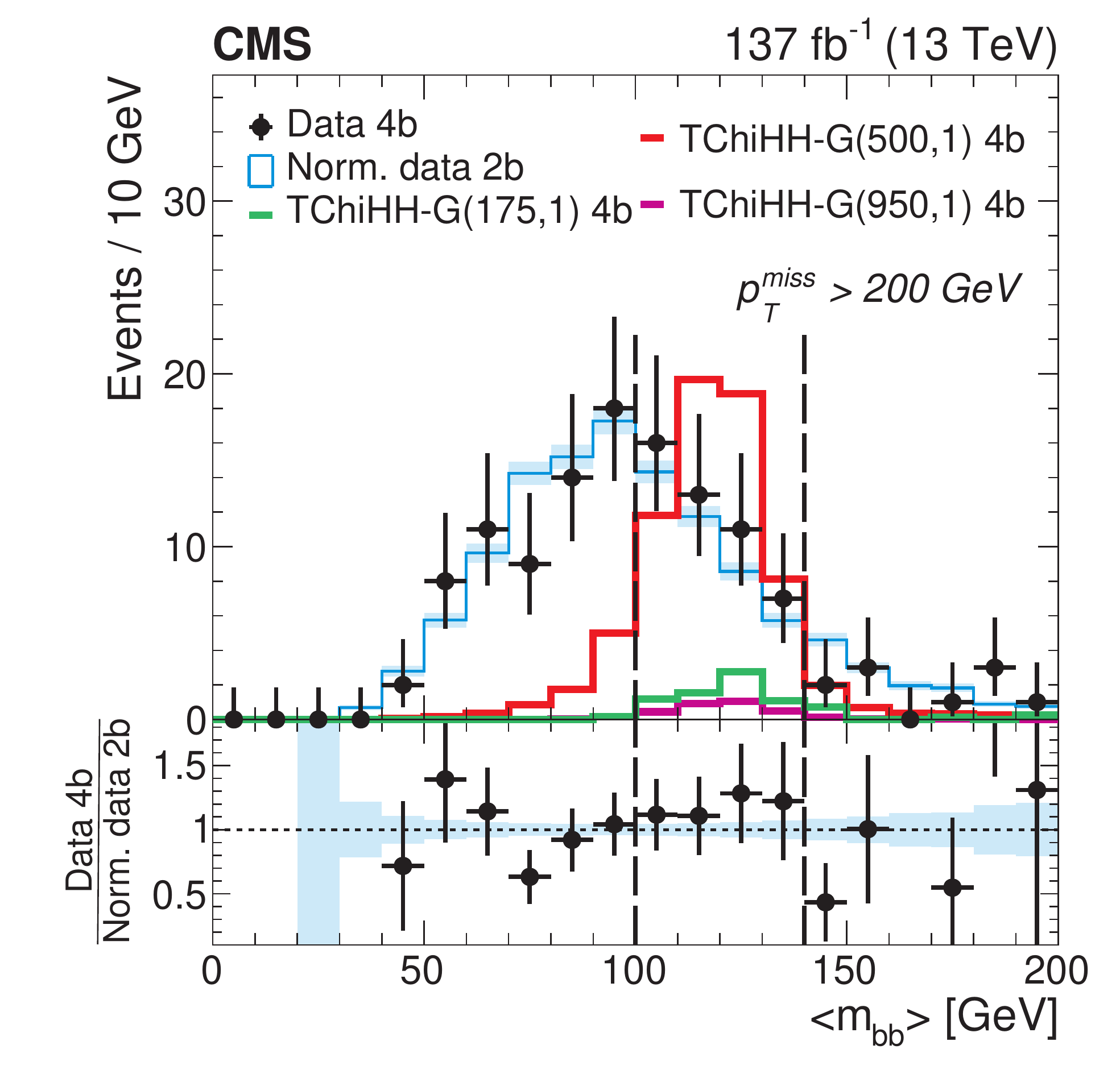}
\caption{Distributions in \amjj for the (upper row) $\nb=3$ and (lower row) $\nb=4$ data, denoted by black markers, with error bars indicating the statistical uncertainties.
The left-hand set of plots corresponds to $150<\met<200$, while the right-hand set corresponds to $\met>200\GeV$, in both cases integrated over \drmax.
The overlaid cyan histograms show the corresponding distributions of the $\nb=2$ data, reweighted with the appropriate \kapa values and scaled in area to the $\nb=3$ or $\nb=4$ distribution and with statistical uncertainties indicated by the cyan shading.
The ratio of these distributions appears in the lower panel.  The red, green, and violet histograms show simulations of representative signals, denoted $\text{TChiHH-G}(m(\chizone),m(\sGra)\,[\GeVns])$ in the legends.
}
\label{fig:resolvedResults:metPlaneResults}
\end{figure}

\begin{figure}[tbp!]
\centering
\includegraphics[width=0.45\textwidth]{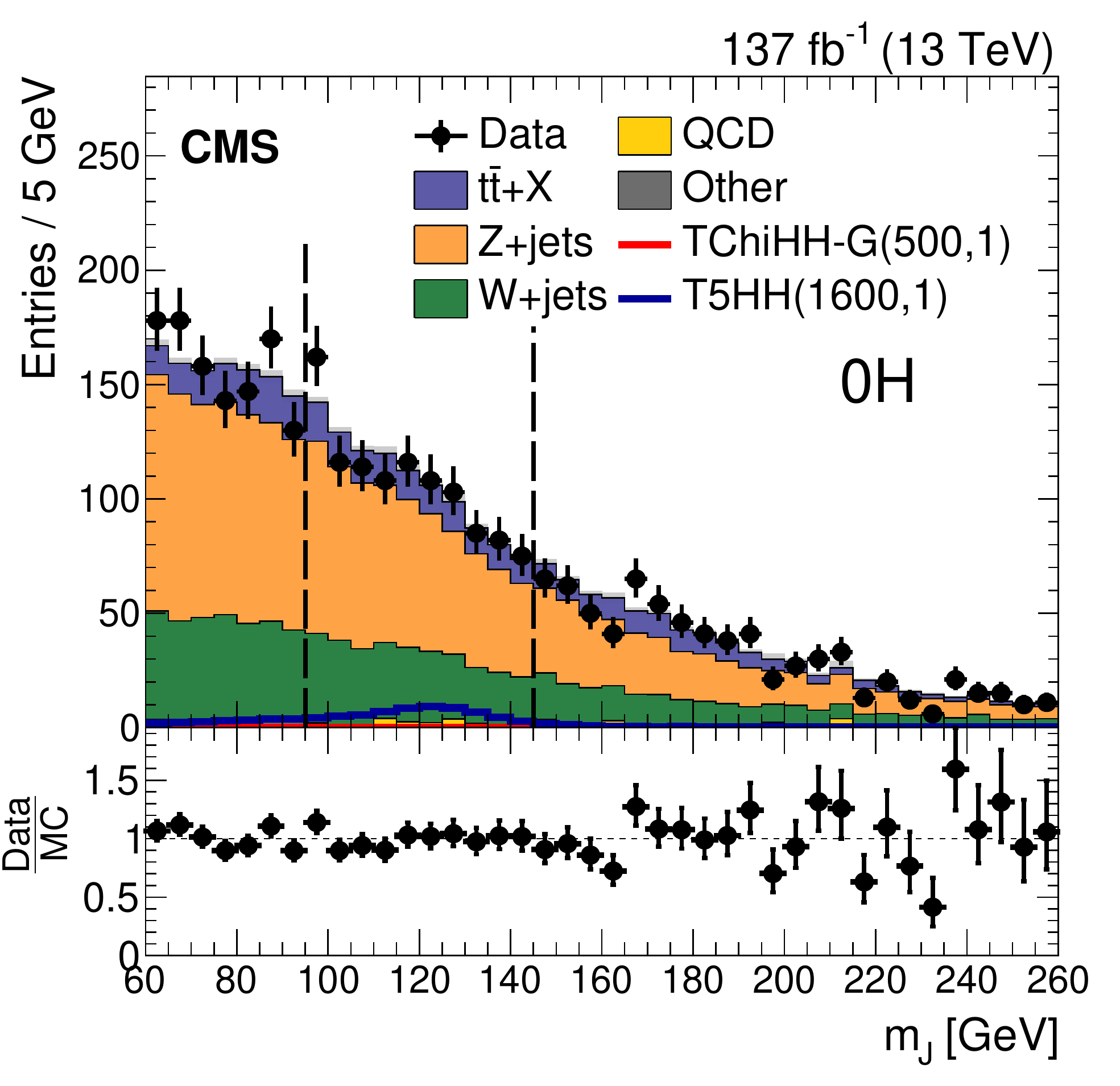}
\includegraphics[width=0.45\textwidth]{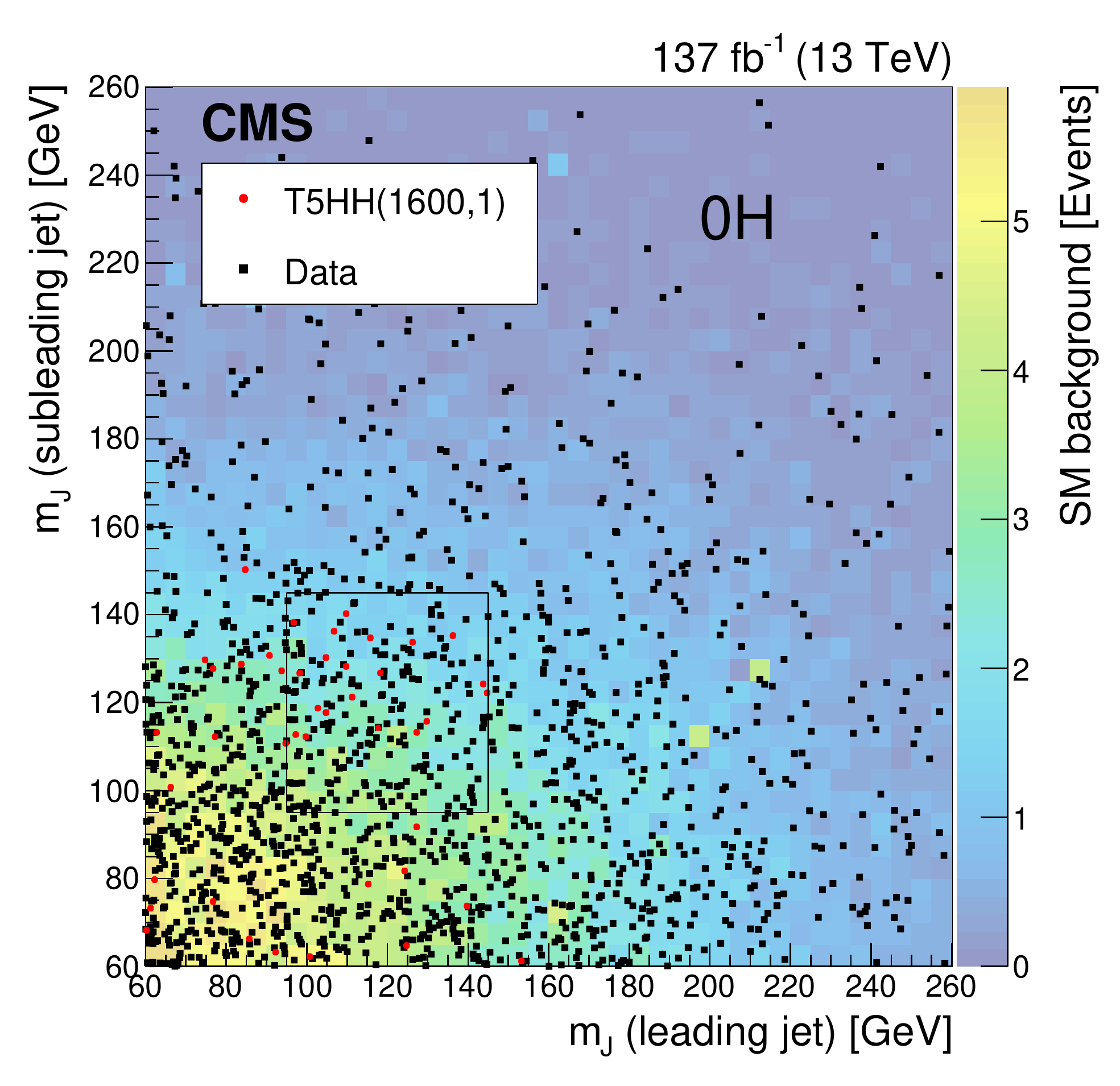}
\includegraphics[width=0.45\textwidth]{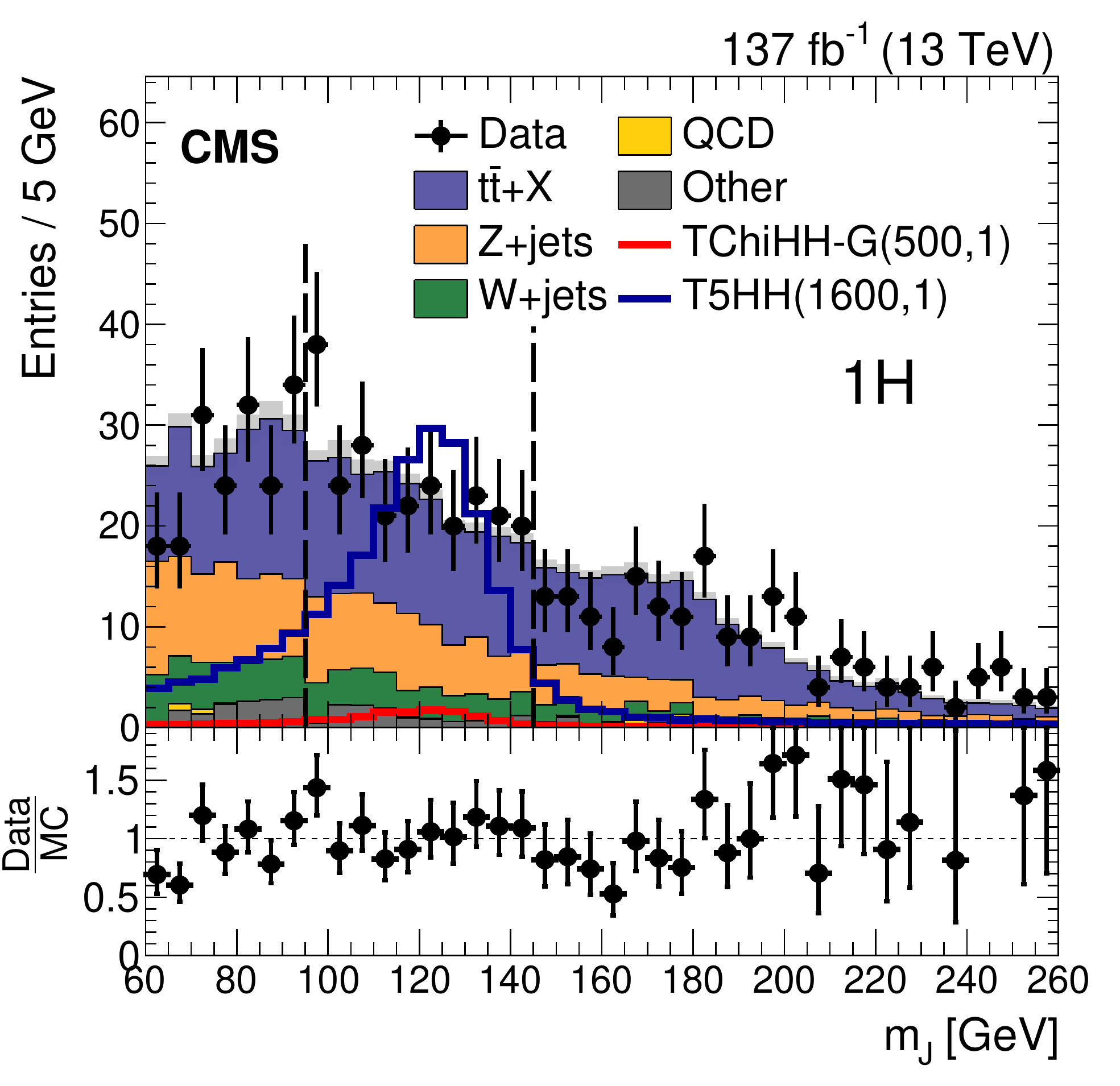}
\includegraphics[width=0.45\textwidth]{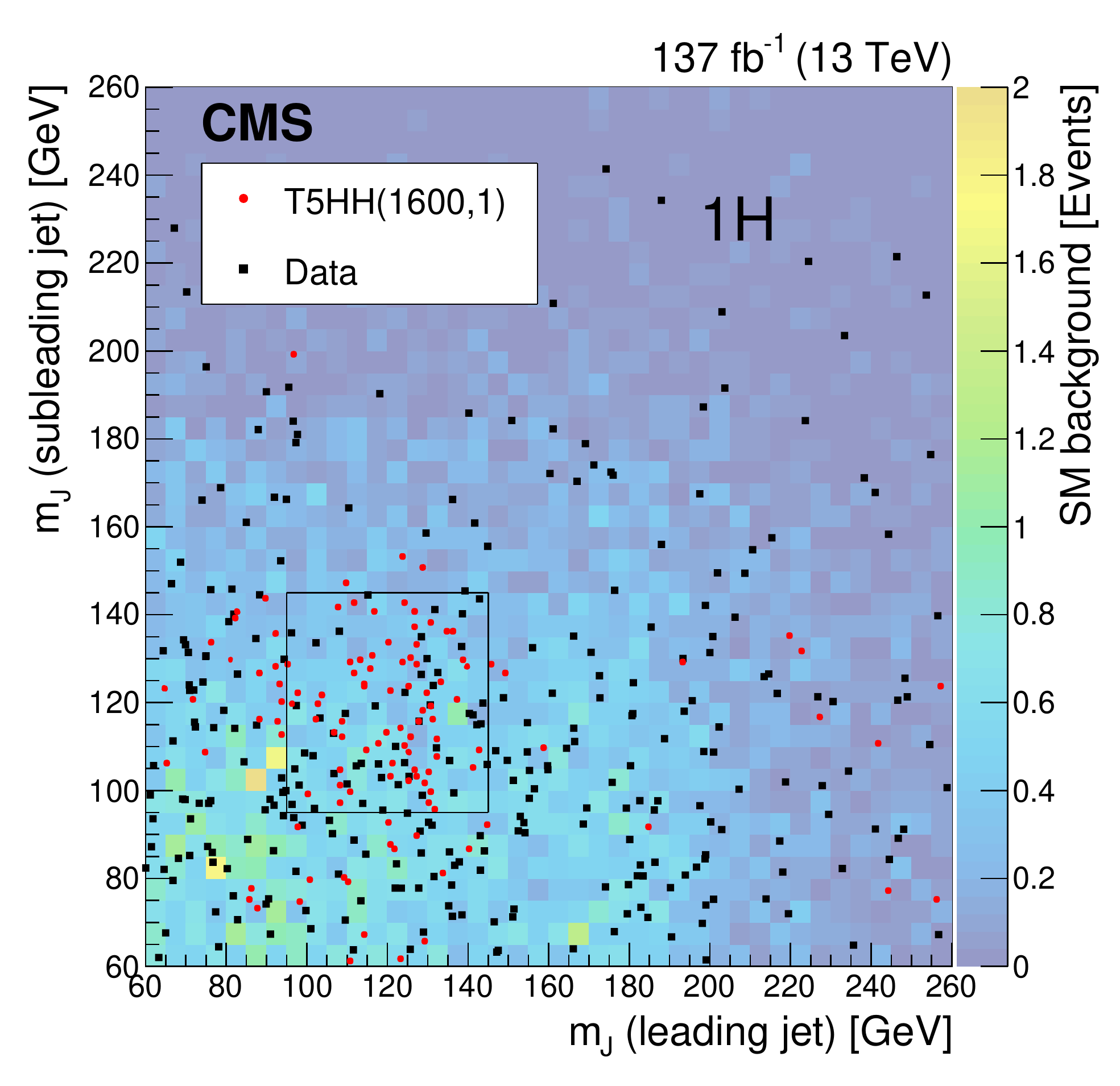}
\includegraphics[width=0.45\textwidth]{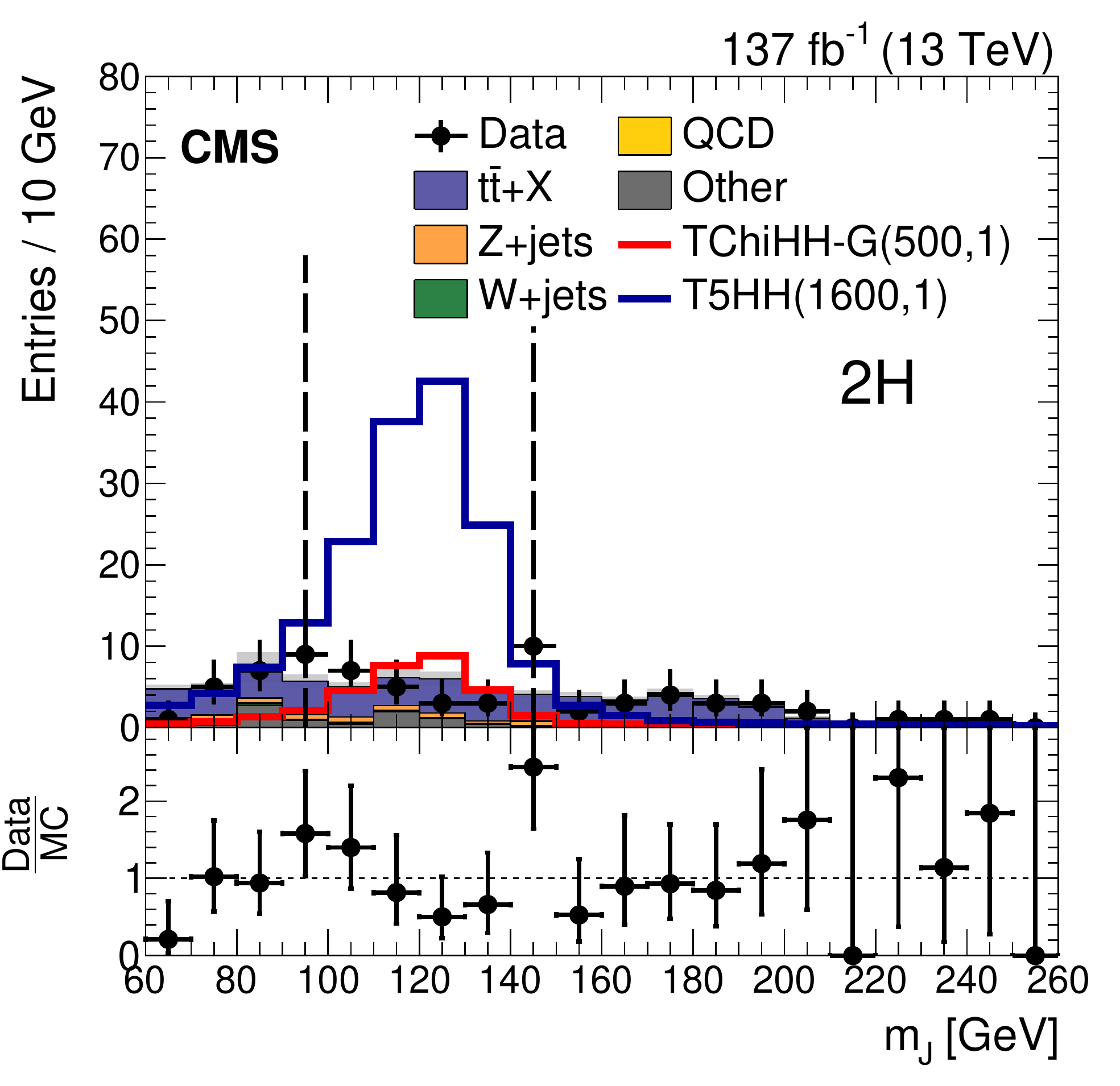}
\includegraphics[width=0.45\textwidth]{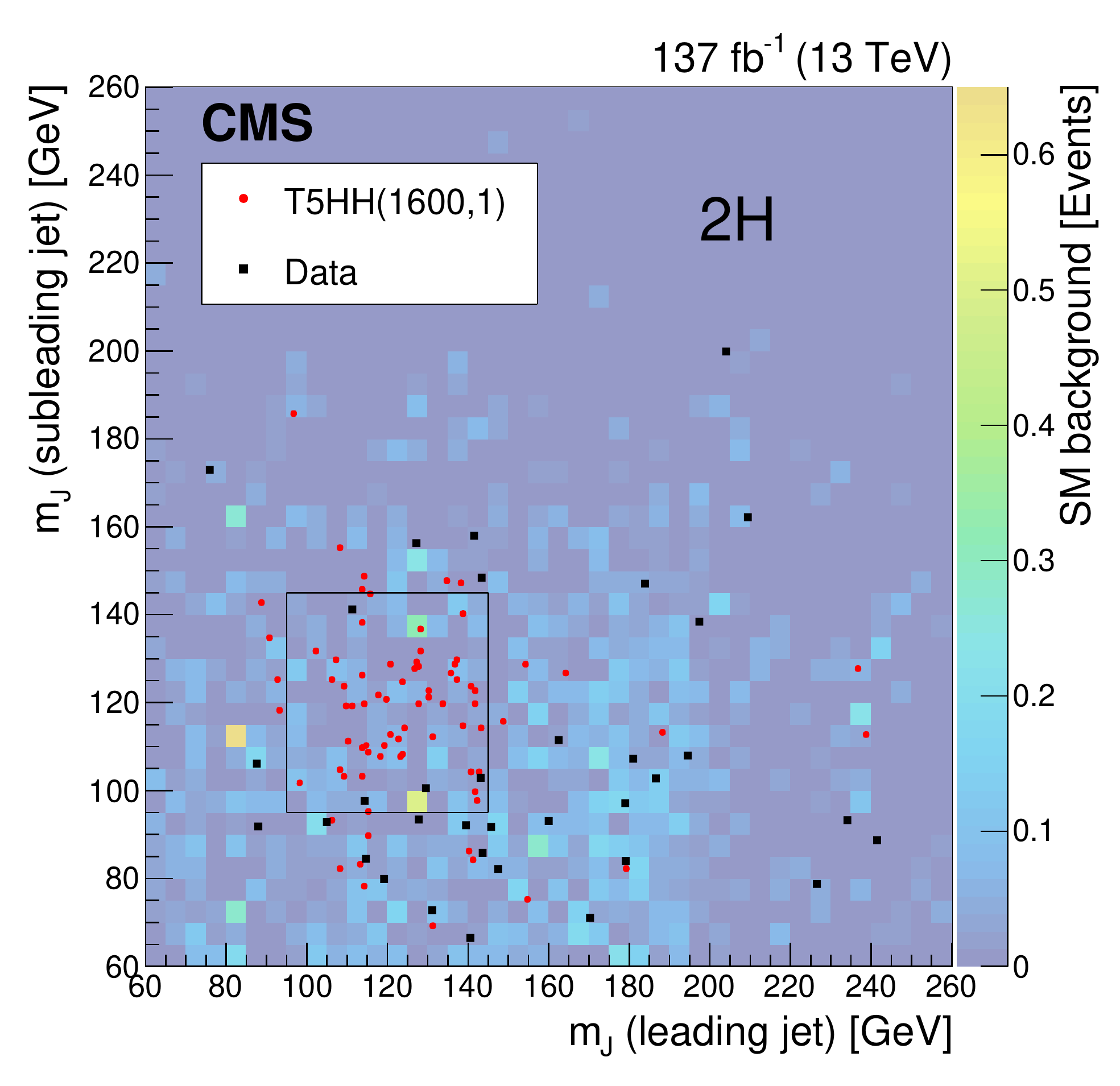}
\caption{%
Distributions in \mjet for the boosted signature, integrated in \ptmiss.
The projections (left column) contain two entries per event, with statistical uncertainties in the data and simulation given by the vertical bars and gray shading, respectively.  The SM yields are scaled to the data integral, by factors of 0.84, 0.92, and 1.13 in the 0H (upper), 1H (middle), and 2H (lower) plots, respectively.  The data to simulation ratio appears in each lower panel.
Simulated signals $\text{TChiHH-G}(m(\chizone),m(\sGra))$ and $\text{T5HH}(m(\gluino),m(\chizone)\,[\GeVns])$ are also shown.
In the correlation plots (right column) the color scale represents the SM background, black dots the data, and red dots the expected signal.  The dashed lines and boxes denote the boundaries of the SRs.
}
\label{fig:boostResults:mass}
\end{figure}

The observed distributions in the appropriate Higgs candidate mass quantity (\amjj for the resolved signature and \mjet for the boosted signature) are shown in
Figs.~\ref{fig:resolvedResults:drmaxPlaneResults}--\ref{fig:boostResults:mass}.
Figures~\ref{fig:resolvedResults:drmaxPlaneResults} and \ref{fig:resolvedResults:metPlaneResults} show the \amjj distributions for the resolved signature in the SR+SB regions ($\nb = 3$ and $\nb=4$) of the data, with the corresponding distributions for the CSR+CSB regions ($\nb = 2$) overlaid. The latter distributions are reweighted event-by-event with the appropriate values of \kapa and are normalized to the same overall event yield as observed in the SR+SB region. 
The distributions are presented (i) in two slices of \drmax but integrated in \ptmiss (Fig.~\ref{fig:resolvedResults:drmaxPlaneResults})
and (ii) in two slices of \ptmiss but integrated over \drmax (Fig.~\ref{fig:resolvedResults:metPlaneResults}).
Overall, we observe that the shapes of the \amjj distributions for $\nb=3$ and $\nb=4$ are
highly similar to those for $\nb=2$, and we do not observe a large excess at the Higgs-boson mass. In Fig.~\ref{fig:resolvedResults:metPlaneResults}, comparison of
the \amjj TChiHH-G distributions for $\nb=4$ at low and high values of \ptmiss illustrates how the
different \ptmiss regions provide sensitivity to different parts of the model space. 
For the boosted signature, Fig.~\ref{fig:boostResults:mass} shows the data distributions in \mjet, and in $({\mjet}_1,{\mjet}_2)$, with simulation overlaid, for the 0H CSR+CSB and the 1H and 2H SR+SB regions, integrated in \ptmiss. Again, we do not observe a large excess in the data in the \mjet distributions at the Higgs-boson mass.
Systematic uncertainties are not included in these figures, as those uncertainties are computed only for the total yields in the analysis regions.  The plots also show representative distributions for potential SUSY signals, with their clear peaks in the \PH boson mass window.

To extract the signal strength, $\mu$, we perform maximum likelihood fits to the data using a probability density function for each bin that represents the SM yield plus the yield predicted by a specific model for the signal.  The parameter $\mu$ is the ratio of the fitted to theoretical cross section.  The likelihood function is a product of Poisson distributions, one for each of the SRs of the resolved and boosted signatures and their corresponding SB, CSR, and CSB regions.  Additional Poisson variables represent the event yields in the three \ptmiss bins of each of the \crb control regions used for the boosted signature.  Equation (\ref{eq:ABCD}), and Eq.~(\ref{eq:bkeq}) for the boosted signature, are imposed as constraints on the
background component of the
Poisson means for each of these factors.  The measured values of \kapa, with their statistical uncertainties, are introduced as Gaussian constraints.  Other systematic uncertainties are implemented as additional free parameters with log-normal constraints.  Correlations among signal bins are taken into account.  Inclusion of the signal component in the functions for the Poisson means of the SB, CSR, CSB, and \crb regions accounts for potential signal contamination of these regions.

We determine confidence intervals for $\mu$ using the test statistic
$q_\mu =  - 2 \ln ( \mathcal{L}_\mu/\mathcal{L}_\text{max})$,
where $\mathcal{L}_\text{max}$ is the maximum likelihood
determined by allowing all parameters including
$\mu$ to vary,
and $\mathcal{L}_\mu$ is the maximum likelihood for a fixed signal strength.
Confidence ranges are set under the asymptotic approximation~\cite{Cowan:2010js},
with $q_\mu$ approximated with an Asimov data set~\cite{Cowan:2010js} and used
in conjunction with the \CLs criterion described in Refs.~\cite{Junk1999,bib-cls}.

\begin{table}[tbh!]
\topcaption{For each SR of the resolved signature, the MC correction factor \kapa, predicted background yield \nsrp, yield from the background-only ($\mu=0$) fit \nsrf, and observed yield \nsro.
The first and second uncertainties in the \kapa factors are statistical and systematic, respectively.  The uncertainties in \nsrp and \nsrf, extracted from the maximum likelihood fit, include both statistical and systematic contributions.
The interpretation of the results in bin 11
is discussed in the text.
}
\centering
\renewcommand{\arraystretch}{1.2}
\begin{tabular}{rrrccccr}
\hline
Bin & \drmax & \nb & \ptmiss [\GeVns{}] & \kapa  & \nsrp & \nsrf & \nsro \\
\hline
1 & \multirow{8}{*}{1.1--2.2} &
\multirow{4}{*}{3} & 150--200 &$1.09\pm0.04 \pm 0.02$ & $161^{+14}_{-13}$ & $149.7^{+8.9}_{-8.5}$ & $138$\\
2 &                  & & 200--300 &$0.92\pm0.04 \pm 0.02$ & $90.4^{+9.7}_{-9.0}$ & $91.5^{+6.9}_{-6.5}$ & $91$\\
3 &                  & & 300--400 &$0.94\pm0.09 \pm 0.01$ & $11.5^{+3.4}_{-2.7}$ & $12.8^{+2.6}_{-2.2}$ & $14$\\
\vspace{4pt}
4 &                  & &   ${>}400$ &$0.98^{+0.19}_{-0.16} \pm 0.02$ & $2.8^{+2.3}_{-1.4}$ & $2.8^{+1.4}_{-1.0}$ & $3$\\
5 & &\multirow{4}{*}{4} & 150--200 &$1.13\pm0.09 \pm 0.08$ & $53.5^{+8.8}_{-7.8}$ & $54.1^{+5.6}_{-5.2}$ & $54$\\
6 &                  & & 200--300 &$0.96\pm0.07 \pm 0.07$ & $28.3^{+5.6}_{-4.8}$ & $33.2^{+4.2}_{-3.9}$ & $38$\\
7 &                  & & 300--400 &$0.89^{+0.16}_{-0.15} \pm 0.05$ & $2.6^{+1.5}_{-1.1}$ & $3.2^{+1.3}_{-1.0}$ & $4$\\
\vspace{8pt}
8 &                  & &   ${>}400$ &$0.92^{+0.27}_{-0.22} \pm 0.07$ & $2.6^{+2.4}_{-1.4}$ & $1.27^{+0.98}_{-0.63}$ & $0$\\
9 & \multirow{8}{*}{${<}1.1$} &
\multirow{4}{*}{3} & 150--200 &$1.05^{+0.18}_{-0.15} \pm 0.12$ & $5.1^{+1.6}_{-1.3}$ & $5.9^{+1.4}_{-1.2}$ & $8$\\
10 &                  & & 200--300 &$1.04^{+0.14}_{-0.13} \pm 0.11$ & $2.17^{+0.79}_{-0.60}$ & $2.31^{+0.73}_{-0.57}$ & $2$\\
11 &                  & & 300--400 &$0.72^{+0.33}_{-0.22} \pm 0.08$ & $0.06^{+0.11}_{-0.04}$ & $0.72^{+0.53}_{-0.33}$ & $4$\\
\vspace{4pt}
12 &                  & &   ${>}400$ &$1.24^{+0.67}_{-0.45} \pm 0.10$ & $0.89^{+1.42}_{-0.60}$ & $0.52^{+0.65}_{-0.35}$ & $0$\\
13 & &\multirow{4}{*}{4} & 150--200 &$1.26^{+0.21}_{-0.20} \pm 0.23$ & $2.68^{+1.06}_{-0.79}$ & $2.58^{+0.85}_{-0.67}$ & $1$\\
14 &                  & & 200--300 &$1.21^{+0.22}_{-0.21} \pm 0.22$ & $1.26^{+0.62}_{-0.44}$ & $1.62^{+0.65}_{-0.48}$ & $3$\\
15 &                  & & 300--400 &$2.35^{+0.88}_{-0.72} \pm 0.34$ & $0.42^{+0.61}_{-0.27}$ & $1.16^{+0.87}_{-0.55}$ & $1$\\
\vspace{4pt}
16 &                  & &   ${>}400$ &$0.94^{+0.53}_{-0.36} \pm 0.13$ & $0.67^{+1.10}_{-0.46}$ & $0.78^{+0.76}_{-0.43}$ & $1$\\
\hline
\end{tabular}
\label{tab:resolvedYpredobs}
\end{table}

\begin{table}[tbhp!]
\topcaption{For each \nH SR of the boosted signature, the  total predicted background yield \tbkg, and for each \ptmiss bin the fraction \fbkg, both with their statistical uncertainties, the predicted background yield \nsrp, the yield from the background-only ($\mu=0$) fit \nsrf, and the observed yield \nsro.
The values of \nsrp and \nsrf are extracted from the maximum likelihood fit, with uncertainties that include both statistical and systematic contributions.}
\centering
\renewcommand{\arraystretch}{1.2}
\begin{tabular}{rrcccccr}
\hline
Bin & \nH & \ptmiss [\GeVns{}] & \tbkg & \fbkg & \nsrp & \nsrf & \nsro \\
\hline
17 & \multirow{3}{*}{1} & 300--500 & \multirow{3}{*}{$42.6\pm4.2$}                                                                                                               & $0.789\pm0.030$ & $33.6^{+ 6.1}_{-5.2}$      & $37.0^{+ 4.2}_{-4.0}$     & 42 \\
18 & & 500--700 &             & $0.172\pm0.028$ & $ 7.3^{+ 2.0}_{-1.6}$      & $ 7.2^{+ 1.5}_{-1.3}$     & 6 \\
19 & & ${>}700$   &             & $0.039\pm0.014$ & $ 1.65^{+ 1.04}_{-0.66}$   & $ 1.50^{+ 0.75}_{-0.53}$   & 1%
                                                                                    \vspace{4pt} \\
20 & \multirow{3}{*}{2} & 300--500 & \multirow{3}{*}{$5.1\pm1.0$}                                                                                                                & $0.789\pm0.030$ & $ 4.0^{+ 1.5}_{-1.1}$    & $ 4.0^{+ 1.2}_{-1.0}$   & 4 \\
21 & & 500--700 &             & $0.172\pm0.028$ & $ 0.88^{+ 0.40}_{-0.28}$    & $ 0.74^{+ 0.29}_{-0.21}$   & 0 \\
22 & & ${>}700$    &            & $0.039\pm0.014$ & $ 0.20^{+ 0.21}_{-0.10}$ & $ 0.14^{+ 0.13}_{-0.07}$ & 0 \\
\hline
\end{tabular}
\label{tab:boostedYpredobs}
\end{table}

\begin{figure}[tbp!]
  \includegraphics[width=0.99\textwidth]{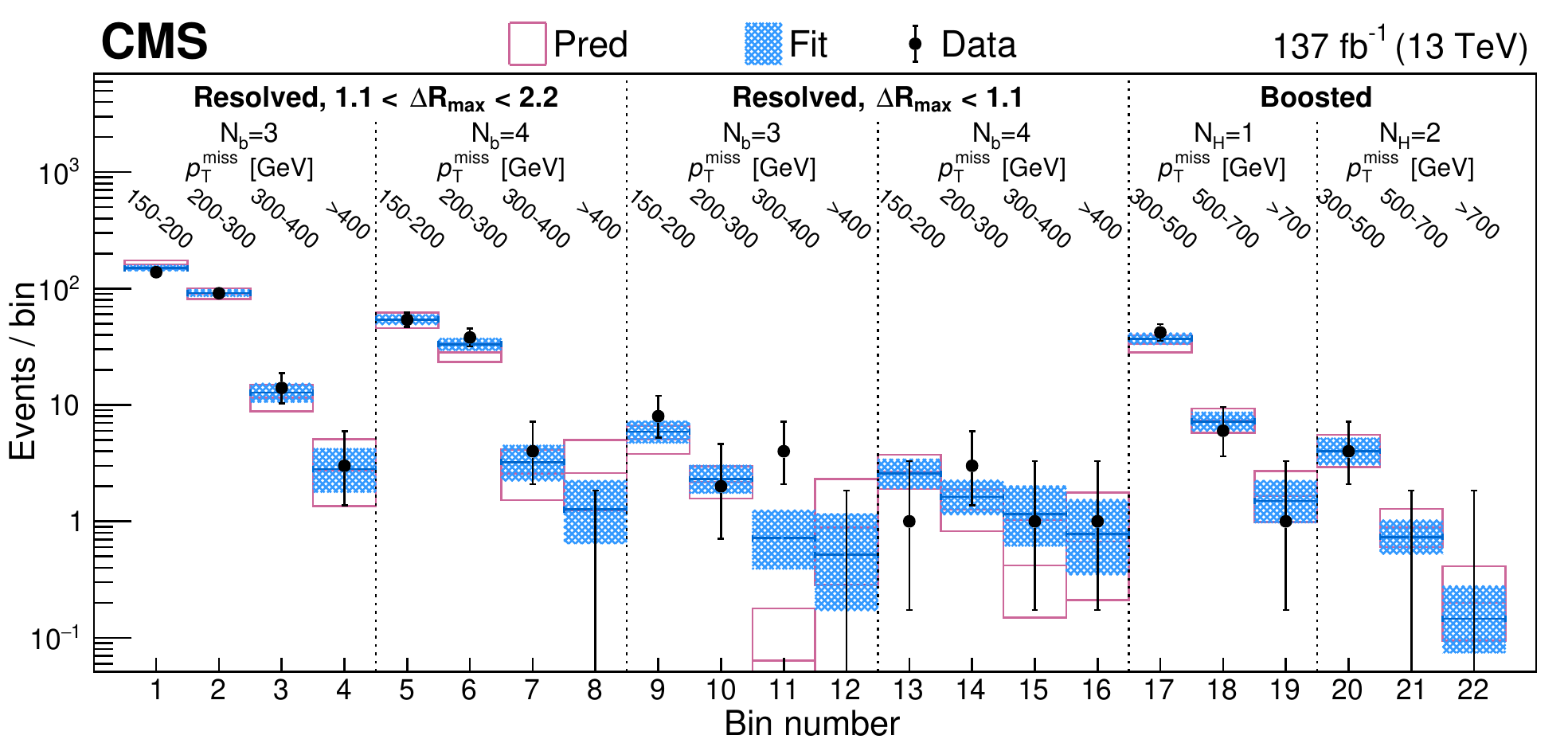}
  \caption{
    Observed and predicted yields in the search regions identified by the legend text.  The points with error bars represent the observed yields, the magenta outline bands the predicted background yields (``Pred'') with their total uncertainties derived as described in Sections \ref{sec:background} and \ref{sec:systematics}, and the blue shaded bands the values determined by the background-only fit (``Fit'').
  }
  \label{fig:Ypredobs}
\end{figure}

The observed yields, together with the predicted SM yields, are given in Tables~\ref{tab:resolvedYpredobs} and \ref{tab:boostedYpredobs} for the resolved and boosted signatures, respectively, and are
summarized in Fig.~\ref{fig:Ypredobs}. 
For 15 of the 16 bins of the resolved signature, and the 6 bins of the boosted signature, the distribution of deviations of the observations with respect to SM predictions is consistent with statistical fluctuations.
In the single bin 11,
corresponding to $\drmax<1.1$, $\nb=3$, $\ptmiss=300$--$400\GeV$ of the resolved signature,
the observed yield is 4 events while the prediction is $0.06^{+0.11}_{-0.04}$ (Table~\ref{tab:resolvedYpredobs}). That prediction is based on small SB and CSR counts (2 events each), leading to highly asymmetric confidence bands.
The probability of observing 4 or more events given this background estimate, as measured with a large sample of simulated pseudo-experiments, corresponds to 3.3 standard deviations (s.d.).
The pseudo-experiments were generated with a null model based on the
observed yields in the SB, CSR, and CSB regions and associated systematic uncertainties.
When the observed SR yield is included in the background-only fit (column \nsrf in Table~\ref{tab:resolvedYpredobs}) the background estimate becomes $0.72^{+0.53}_{-0.33}$.
We have examined the four observed events in detail and find no exceptional features.
The models we consider for potential contributions beyond the SM predict that events would be distributed over multiple bins, contrary to the observations here.
To account for the look-elsewhere effect~\cite{cowan:statLHC}, we perform pseudo-experiments with all 16 bins of the resolved signature.  We find that the probability of observing a 3.3 s.d. or larger excess in at least one bin
corresponds to a global significance of 2.1 s.d.

\begin{figure}[hbt!]
\centering
\includegraphics[width=0.85\textwidth]{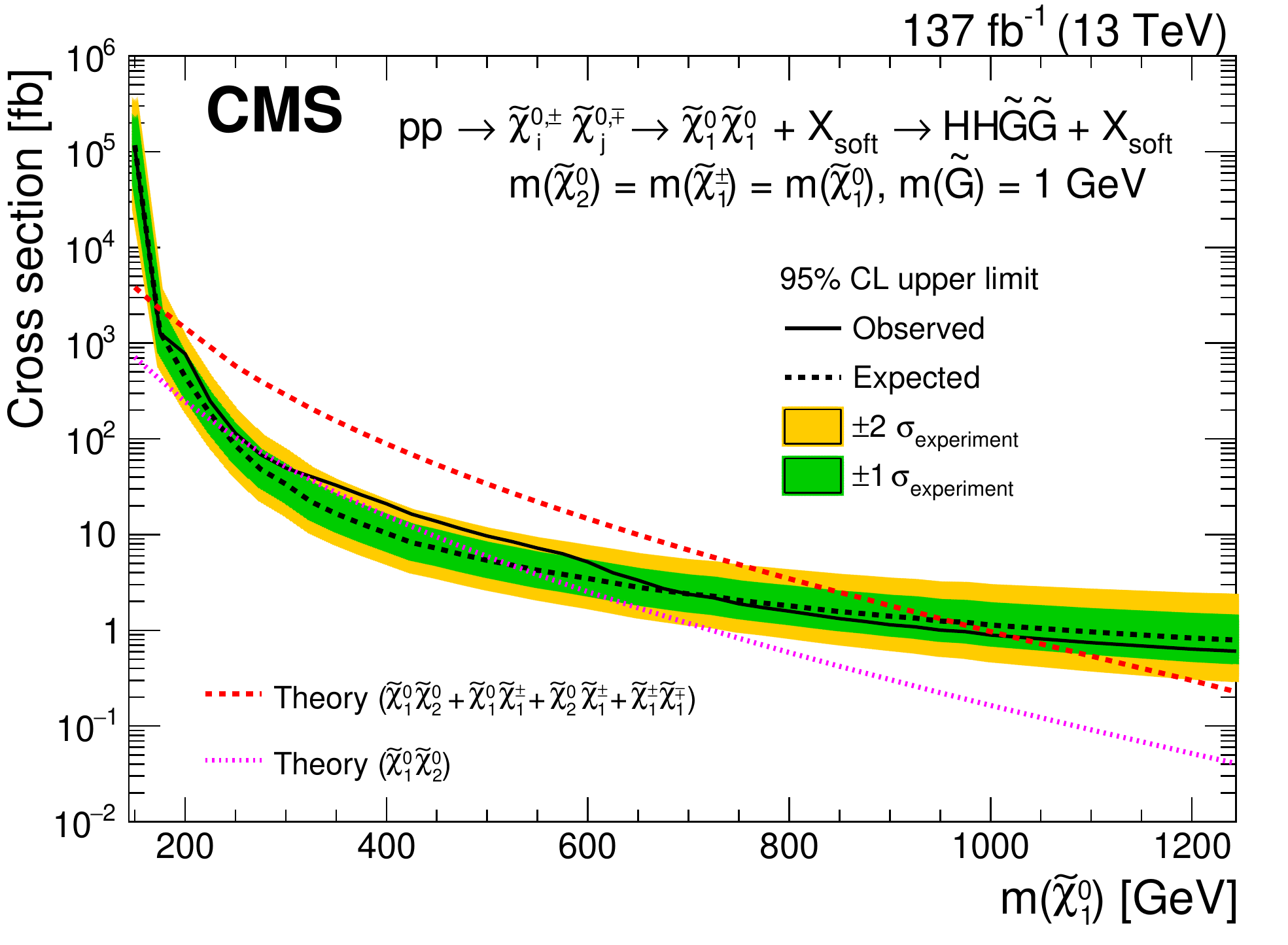}
  \caption{Observed and expected upper limits at 95\% \CL on the cross section for the GMSB-motivated simplified model TChiHH-G.
The symbol $\text{X}_{\text{soft}}$ in the legend represents low-energy particles emitted in the transitions to the \chizone NLSPs.  The dashed black line with green and yellow bands shows the expected limit with its 1- and 2-s.d. uncertainties, while the solid black line shows the observed limit.  The theoretical cross section is indicated by the dashed red line under the assumption that
the decay chains leading to the $\chizone\chizone$ intermediate state
include a degenerate set of all charginos and neutralinos, and by the dotted magenta line under the assumption that only the combination $\chizone\chiztwo$ contributes.
  }
  \label{fig:xseclim_GMSB}
\end{figure}

\begin{figure}[hbt!]
\centering
\includegraphics[width=0.85\textwidth]{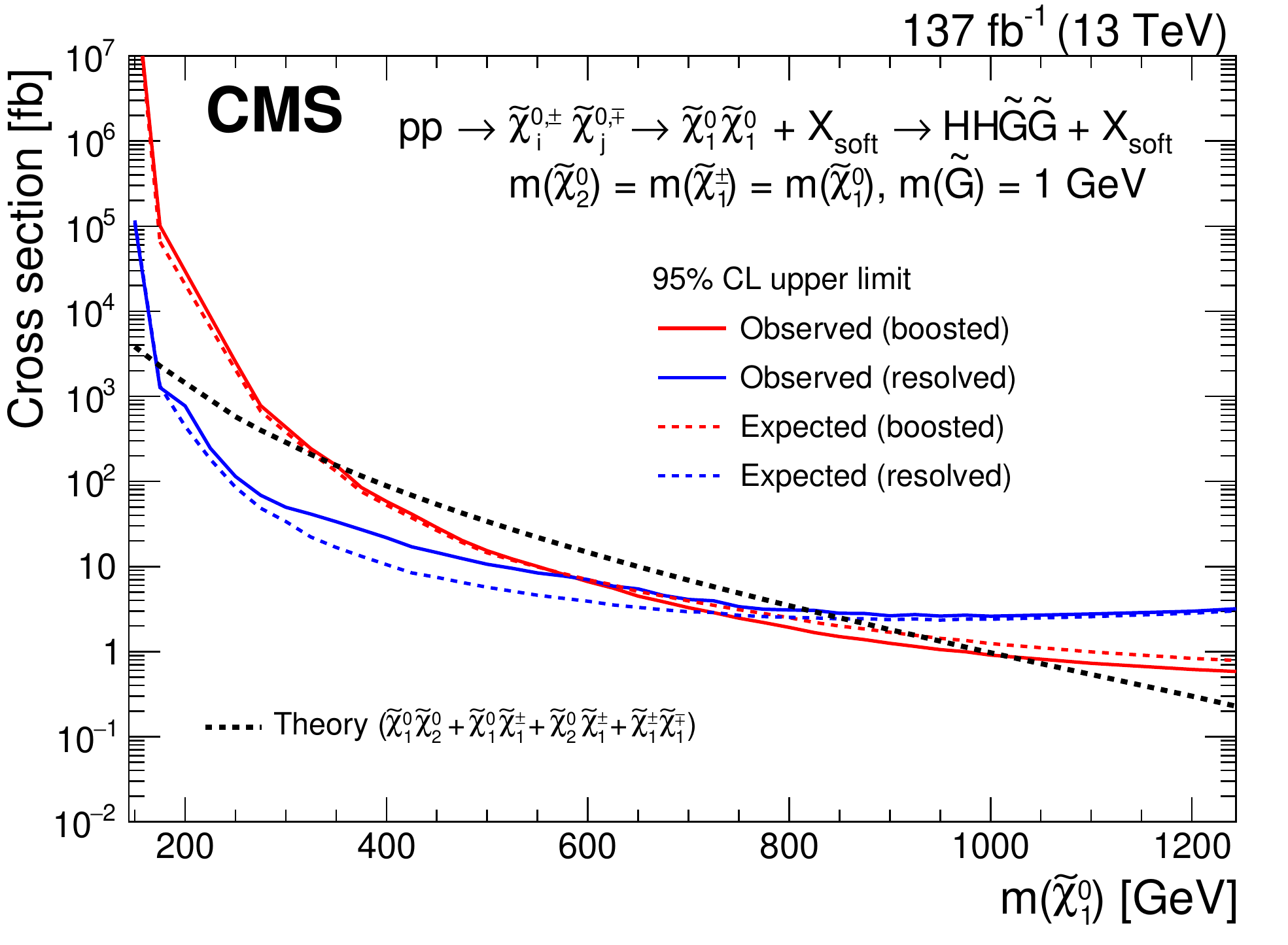}
  \caption{Separate resolved-signature (blue) and boosted-signature (red) contributions to the cross section limits for the GMSB-motivated simplified model TChiHH-G shown in Fig.~\ref{fig:xseclim_GMSB}.
Here no overlapping events were removed.
The symbol $\text{X}_{\text{soft}}$ in the legend represents low-energy particles emitted in the transitions to the \chizone NLSPs.
The solid and dashed lines, respectively, give the observed and expected upper limits at 95\% \CL.
The dashed black line shows the theoretical cross section computed under the assumption that
the decay chains leading to the $\chizone\chizone$ intermediate state
include a degenerate set of all charginos and neutralinos.
  }
  \label{fig:xseclim_GMSB_boostVRes}
\end{figure}

We interpret the data in terms of the three models discussed in Section~\ref{sec:introduction}.
In each case, the data from both resolved and boosted signatures are included in the fit.
Figure~\ref{fig:xseclim_GMSB} shows the upper limit on the cross section at 95\% \CL for the TChiHH-G model of Fig.~\ref{fig:susy_models} (left).
For the case of the production of the sum of all higgsino pairs, the data exclude masses of the \chizone in the range 175--1025\GeV.
This represents the best limit to date on this model, extending previous ATLAS results~\cite{Aaboud:2018htj}, which exclude mass ranges 130--230\GeV and 290--880\GeV, and previous CMS results~\cite{Sirunyan:2017obz}, which exclude 230--770\GeV.
If instead only $\chizone\chiztwo$ production is included in the model, the excluded mass range is 265--305\GeV.  Here the single-bin excess noted above leads to a limit less restrictive than the expected one.
The separate contributions of the resolved and boosted signatures are displayed in Fig.~\ref{fig:xseclim_GMSB_boostVRes}.  As expected, the resolved analysis provides most of the sensitivity at lower NLSP masses, where the energy available in the NLSP decay to LSP and \PH is limited.  At higher NLSP masses the Higgs bosons receive a larger Lorentz boost, and the daughter {\PQb} quarks tend to merge; the crossover where the boosted signature is more sensitive is about $800\GeV$ for the expected limits, and $600\GeV$ for the observed limits since the SRs of the boosted signature show no excess in the data.

\begin{figure}[btb!]
\centering
\includegraphics[width=0.85\textwidth]{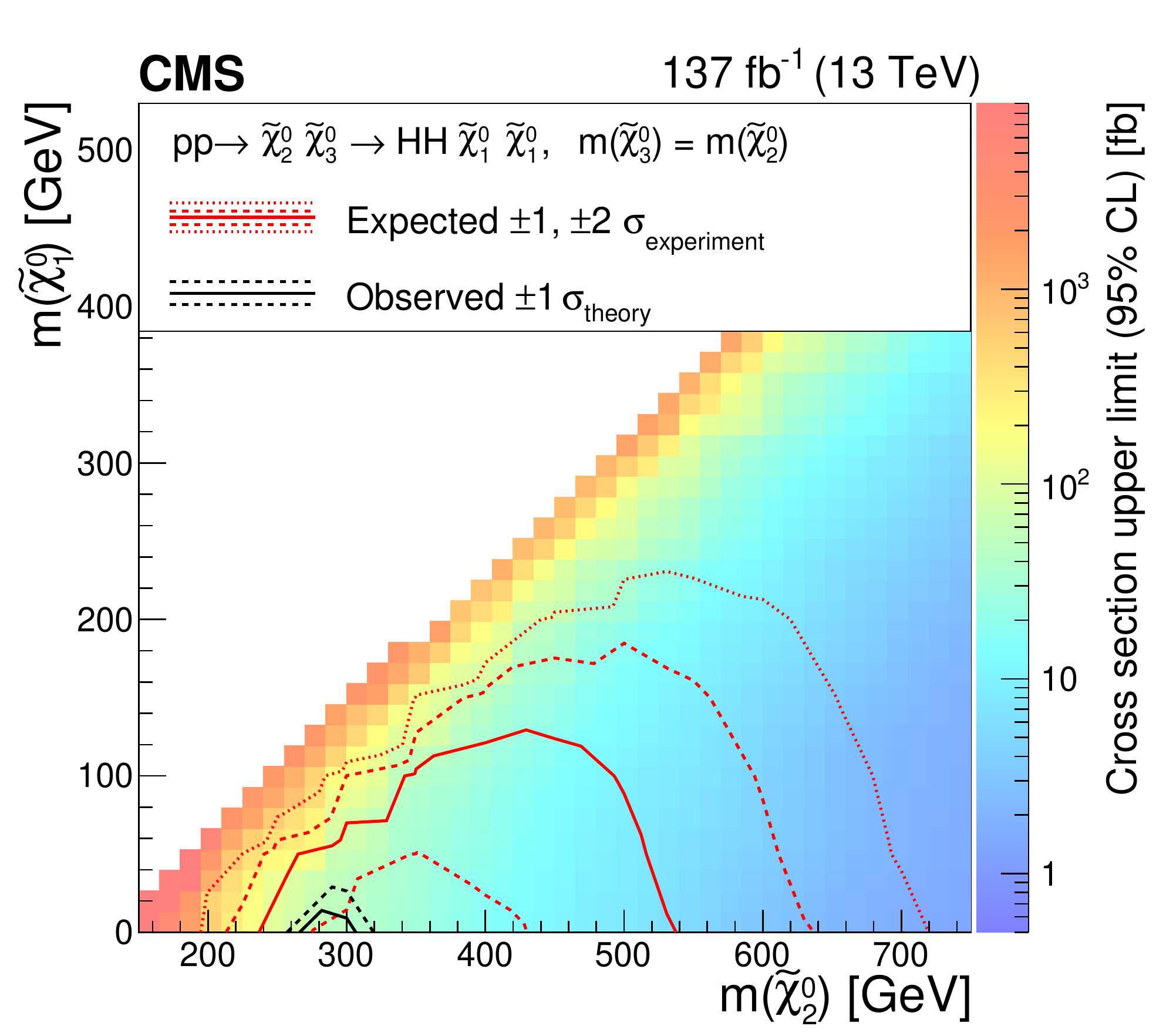}
  \caption{Limits at 95\% \CL on the cross section for the TChiHH signal model in which production of the intermediate state $\chiztwo\chizthree$ (assumed mass degenerate) is followed by the decay of each to $\chizone\PH$.  The color scale gives the cross section limit as a function of $(m(\chiztwo),m(\chizone))$.
The black solid and dashed curves show the observed excluded region and its 1-s.d. uncertainty.
The red solid and dashed (dotted) contours show the expected excluded region with its $1\,(2)$-s.d. uncertainty.
}
  \label{fig:xseclim_TchiHH}
\end{figure}

For the TChiHH model (Fig.~\ref{fig:susy_models}, center) the cross section limit is presented in Fig.~\ref{fig:xseclim_TchiHH} as a function of the independent masses $m(\chiztwo)$ and $m(\chizone)$. While the expected limit would exclude a substantial region of this plane,
the observed exclusion is limited to a small region where $m(\chizone)$ is less than 15\GeV, again because of the single-bin excess.  For $m(\chizone)=1\GeV$ the excluded range of $m(\chiztwo)$ is 265--305\GeV.

\begin{figure}[tbh!]
\centering
\includegraphics[width=0.85\textwidth]{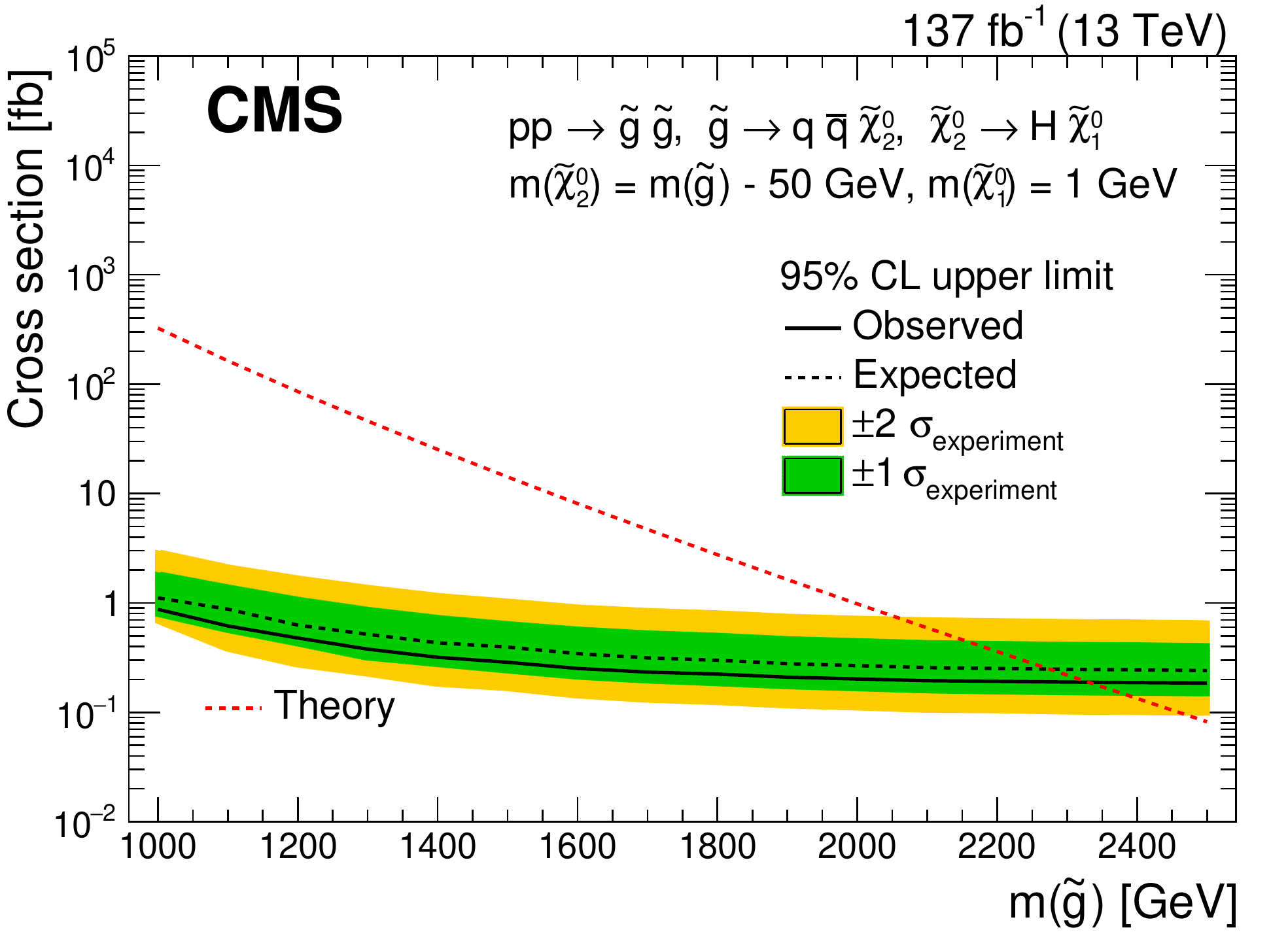}
  \caption{Observed and expected upper limits at 95\% \CL on the cross section for the simplified model T5HH, the strong production of a pair of gluinos each of which decays via a three-body process to quarks and a \chiztwo NLSP, which
subsequently decays to a Higgs boson and a \chizone LSP.  The dashed black line with green and yellow bands shows the expected limit with its 1- and 2-s.d. uncertainties; the solid black line shows the observed limit, and the dashed red line the theoretical cross section.
  }
  \label{fig:xseclim_T5HH}
\end{figure}

Figure~\ref{fig:xseclim_T5HH} shows
the cross section upper limit as a function of $m(\PSg)$ for the T5HH model (Fig.~\ref{fig:susy_models}, right) of gluino pair production with a \chiztwo NLSP slightly less massive than the gluino
and a light LSP.  Masses $m(\PSg) < 2330\GeV$ are excluded. This is the strongest mass limit for this model to date, extending a previous
CMS result~\cite{Sirunyan:2017bsh}, which excluded $m(\PSg) < 2010\GeV$.
Most of the sensitivity to this model is provided by the boosted signature.  This reflects the choice of large NLSP mass, which leads to energetic \PH boson daughters.

\clearpage
\section{Summary}
\label{sec:summary}
A search has been presented for physics beyond the standard model in channels leading to pairs of Higgs bosons and an imbalance of transverse momentum, produced in proton-proton collisions at $\sqrt{s}=13\TeV$.  The data sample, collected by the CMS experiment at the LHC, corresponds to an integrated luminosity of 137\fbinv.
The Higgs bosons are reconstructed via their decay to a pair of \PQb quarks, observed either as distinct \PQb quark jets (resolved signature), or as wide-angle jets that each contain the pair of \PQb quarks (boosted signature).
The observed event yields in 15 of the 16 analysis bins of the resolved signature, and all 6 bins of the boosted signature, are consistent with the background predictions based on SM processes.
In one bin of the resolved signature, an excess is observed, for which the global significance is 2.1 standard deviations when all 16 bins are considered.
A narrow, single-bin effect is not typical of the supersymmetry (SUSY) models under consideration.

These results are used to set limits on the cross sections for the production of particles predicted by SUSY, considering both the direct production of neutralinos and their production through intermediate states with gluinos.
For the electroweak production of nearly-degenerate higgsinos, each of whose decay cascades yields a {\chizone} that in turn decays through a Higgs boson to the lightest SUSY particle (LSP), a massless goldstino, \chizone masses in the range 175--1025\GeV are excluded at 95\% confidence level.  For a model with a mass splitting between the directly produced, degenerate higgsinos $\chiztwo\chizthree$, and a bino LSP,
a small region where $m(\chizone)$ is less than 15\GeV is excluded; for $m(\chizone)=1\GeV$ the excluded range of $m(\chiztwo)$ is 265--305\GeV.
For the strong production of gluino pairs decaying via a slightly lighter \chiztwo to a Higgs boson and a light \chizone LSP, gluino masses below 2330\GeV are excluded.  The bounds on masses found here extend previous limits for these models by about 150 and 320\GeV for the \chizone and gluino, respectively.

\begin{acknowledgments}
  We congratulate our colleagues in the CERN accelerator departments for the excellent performance of the LHC and thank the technical and administrative staffs at CERN and at other CMS institutes for their contributions to the success of the CMS effort. In addition, we gratefully acknowledge the computing  center s and personnel of the Worldwide LHC Computing Grid and other  center s for delivering so effectively the computing infrastructure essential to our analyses. Finally, we acknowledge the enduring support for the construction and operation of the LHC, the CMS detector, and the supporting computing infrastructure provided by the following funding agencies: BMBWF and FWF (Austria); FNRS and FWO (Belgium); CNPq, CAPES, FAPERJ, FAPERGS, and FAPESP (Brazil); MES and BNSF (Bulgaria); CERN; CAS, MoST, and NSFC (China); MINCIENCIAS (Colombia); MSES and CSF (Croatia); RIF (Cyprus); SENESCYT (Ecuador); MoER, ERC PUT and ERDF (Estonia); Academy of Finland, MEC, and HIP (Finland); CEA and CNRS/IN2P3 (France); BMBF, DFG, and HGF (Germany); GSRI (Greece); NKFIA (Hungary); DAE and DST (India); IPM (Iran); SFI (Ireland); INFN (Italy); MSIP and NRF (Republic of Korea); MES (Latvia); LAS (Lithuania); MOE and UM (Malaysia); BUAP, CINVESTAV, CONACYT, LNS, SEP, and UASLP-FAI (Mexico); MOS (Montenegro); MBIE (New Zealand); PAEC (Pakistan); MSHE and NSC (Poland); FCT (Portugal); JINR (Dubna); MON, RosAtom, RAS, RFBR, and NRC KI (Russia); MESTD (Serbia); MCIN/AEI and PCTI (Spain); MOSTR (Sri Lanka); Swiss Funding Agencies (Switzerland); MST (Taipei); ThEPCenter, IPST, STAR, and NSTDA (Thailand); TUBITAK and TAEK (Turkey); NASU (Ukraine); STFC (United Kingdom); DOE and NSF (USA).
  
  \hyphenation{Rachada-pisek} Individuals have received support from the Marie-Curie program  and the European Research Council and Horizon 2020 Grant, contract Nos.\ 675440, 724704, 752730, 758316, 765710, 824093, 884104, and COST Action CA16108 (European Union); the Leventis Foundation; the Alfred P.\ Sloan Foundation; the Alexander von Humboldt Foundation; the Belgian Federal Science Policy Office; the Fonds pour la Formation \`a la Recherche dans l'Industrie et dans l'Agriculture (FRIA-Belgium); the Agentschap voor Innovatie door Wetenschap en Technologie (IWT-Belgium); the F.R.S.-FNRS and FWO (Belgium) under the ``Excellence of Science -- EOS" -- be.h project n.\ 30820817; the Beijing Municipal Science \& Technology Commission, No. Z191100007219010; the Ministry of Education, Youth and Sports (MEYS) of the Czech Republic; the Deutsche Forschungsgemeinschaft (DFG), under Germany's Excellence Strategy -- EXC 2121 ``Quantum Universe" -- 390833306, and under project number 400140256 - GRK2497; the Lend\"ulet (``Momentum") Program  and the J\'anos Bolyai Research Scholarship of the Hungarian Academy of Sciences, the New National Excellence Program \'UNKP, the NKFIA research grants 123842, 123959, 124845, 124850, 125105, 128713, 128786, and 129058 (Hungary); the Council of Science and Industrial Research, India; the Latvian Council of Science; the Ministry of Science and Higher Education and the National Science Center, contracts Opus 2014/15/B/ST2/03998 and 2015/19/B/ST2/02861 (Poland); the Funda\c{c}\~ao para a Ci\^encia e a Tecnologia, grant CEECIND/01334/2018 (Portugal); the National Priorities Research Program by Qatar National Research Fund; the Ministry of Science and Higher Education, projects no. 0723-2020-0041 and no. FSWW-2020-0008, and the Russian Foundation for Basic Research, project No.19-42-703014 (Russia); MCIN/AEI/10.13039/501100011033, ERDF ``a way of making Europe", and the Programa Estatal de Fomento de la Investigaci{\'o}n Cient{\'i}fica y T{\'e}cnica de Excelencia Mar\'{\i}a de Maeztu, grant MDM-2017-0765 and Programa Severo Ochoa del Principado de Asturias (Spain); the Stavros Niarchos Foundation (Greece); the Rachadapisek Sompot Fund for Postdoctoral Fellowship, Chulalongkorn University and the Chulalongkorn Academic into Its 2nd Century Project Advancement Project (Thailand); the Kavli Foundation; the Nvidia Corporation; the SuperMicro Corporation; the Welch Foundation, contract C-1845; and the Weston Havens Foundation (USA).\end{acknowledgments}

\bibliography{auto_generated}
\cleardoublepage \appendix\section{The CMS Collaboration \label{app:collab}}\begin{sloppypar}\hyphenpenalty=5000\widowpenalty=500\clubpenalty=5000\cmsinstitute{Yerevan~Physics~Institute, Yerevan, Armenia}
A.~Tumasyan
\cmsinstitute{Institut~f\"{u}r~Hochenergiephysik, Vienna, Austria}
W.~Adam\cmsorcid{0000-0001-9099-4341}, J.W.~Andrejkovic, T.~Bergauer\cmsorcid{0000-0002-5786-0293}, S.~Chatterjee\cmsorcid{0000-0003-2660-0349}, K.~Damanakis, M.~Dragicevic\cmsorcid{0000-0003-1967-6783}, A.~Escalante~Del~Valle\cmsorcid{0000-0002-9702-6359}, R.~Fr\"{u}hwirth\cmsAuthorMark{1}, M.~Jeitler\cmsAuthorMark{1}\cmsorcid{0000-0002-5141-9560}, N.~Krammer, L.~Lechner\cmsorcid{0000-0002-3065-1141}, D.~Liko, I.~Mikulec, P.~Paulitsch, F.M.~Pitters, J.~Schieck\cmsAuthorMark{1}\cmsorcid{0000-0002-1058-8093}, R.~Sch\"{o}fbeck\cmsorcid{0000-0002-2332-8784}, D.~Schwarz, S.~Templ\cmsorcid{0000-0003-3137-5692}, W.~Waltenberger\cmsorcid{0000-0002-6215-7228}, C.-E.~Wulz\cmsAuthorMark{1}\cmsorcid{0000-0001-9226-5812}
% ...
\vskip\cmsinstskip
% ...
1:~Also at TU Wien, Wien, Austria\\
% ...
\end{sloppypar}
\end{document}